\documentclass[journal=jctcce,manuscript=article]{achemso}
\usepackage{chemformula} 
\usepackage[T1]{fontenc} 

\usepackage{graphicx}
\usepackage{subcaption}
\usepackage{amsmath}
\usepackage{amssymb}
\usepackage{braket}
\usepackage{bm}

\author{Paul Andrew Johnson}
\email{paul.johnson@chm.ulaval.ca}
\affiliation[Universit\'e Laval]
{D\'{e}partement de chimie, Universit\'{e} Laval, Qu\'{e}bec, Qu\'{e}bec, Canada}
\author{A. Eugene DePrince III}
\affiliation[Florida State University]{Department of Chemistry and Biochemistry, Florida State University, Tallahassee, FL 32306-4390}

\title{Single reference treatment of strongly correlated H$_4$ and H$_{10}$ isomers with Richardson-Gaudin states}

\abbreviations{IR,NMR,UV}

\begin{document}




\begin{abstract}
Richardson-Gaudin (RG) states are employed as a variational wavefunction ansatz for strongly correlated isomers of H$_4$ and H$_{10}$. In each case a single RG state describes the seniority-zero sector quite well. Simple natural orbital functionals offer a cheap and reasonable approximation of the outstanding weak correlation in the seniority-zero sector, while systematic improvement is achieved by performing a configuration interaction (CI) in terms of RG states. Other pair theories (e.g. generalized valence bond and pair-coupled-cluster doubles) can provide a good description of many of the geometries considered, but, at short distances, the wavefunctions for the 2D and 3D structures of H$_{10}$ take the form of an RG state that cannot be described by these other theories.
\end{abstract}

\section{Introduction}
Many problems in electronic structure theory can be treated as systems of weakly correlated electrons. In such cases, a single Slater determinant provides a qualitative description of the wavefunction and a short expansion in Slater determinants provides quantitative accuracy. For weakly-correlated systems Kohn-Sham density functional theory (DFT) and coupled-cluster (CC) theory\cite{cizek:1966,cizek:1971,bartlett_book,bartlett:2007} with singles and doubles\cite{purvis_1982} can be expected to predict physically correct results. 

Strongly-correlated systems are another story. The wavefunction has no single dominant Slater determinant, and even a qualitative description of the system can become complicated. The standard approach for dealing with strong correlation is the complete active space self-consistent field (CASSCF) approach\cite{roos:1980,siegbahn:1980,siegbahn:1981,roos:1987} which works well if a compact active space of chemically important orbitals can be identified. Even then, active spaces beyond 22 electrons in 22 orbitals\cite{vogiatzis:2017} are intractable, which has led to the development of a large number of configuration interaction (CI) based schemes as approximate CASSCF solvers,\cite{olsen:1988,malmqvist:1990,fleig:2001,ma:2011,manni:2013,thomas:2015,manni:2016,schriber:2016,levine:2020} as well as alternative representations of the electronic structure of the active space that abandon the CI framework altogether.\cite{ghosh:2008,yanai:2009,wouters:2014,sun:2017,ma:2017,gidofalvi:2008,fosso-tande:2016}

It has long been understood\cite{fock:1950} that two-electron functions, called geminals, describing pairs of weakly-interacting electrons, can provide a better basis for strongly-correlated electrons than Slater determinants of single-particle orbitals. For systems of paired electrons, an excellent description is obtained with the antisymmetrized product of interacting geminals (APIG), but this treatment is completely intractable in practice.\cite{silver:1969,silver:1970a,silver:1970b,silver:1970c,paldus:1972a,paldus:1972b,moisset:2022a} For attractive pairing interactions, such as in the Bardeen-Cooper-Schrieffer (BCS) mechanism\cite{bardeen:1957a,bardeen:1957b,schrieffer_book} and in nuclear structure, the antisymmetrized geminal power (AGP) gives qualitatively correct results for a meagre cost.\cite{coleman:1965,ortiz:1981,sarma:1989,coleman:1997,henderson:2019,khamoshi:2019,dutta:2020,khamoshi:2021,dutta:2021} For repulsive interactions, as arise in chemistry, AGP is qualitatively incorrect and requires the use of Jastrow factors to be size-consistent.\cite{neuscamman:2012,neuscamman:2013,neuscamman:2016} Correct understanding of repulsive systems is possible with the antisymmetrized product of strongly-orthogonal geminals (APSG)\cite{hurley:1953,kutzelnigg:1964} and in particular generalized valence bond/perfect pairing (GVB).\cite{goddard:1967,hay:1972,hunt:1972,goddard:1973,kutzelnigg:2010,kobayashi:2010,kutzelnigg:2012,surjan:2012,zoboki:2013,pernal:2014,jeszenszki:2014,pastorczak:2015,margocsy:2018,pernal:2018,pastorczak:2018,pastorczak:2019,piris:2011} These last two approaches require splitting the orbitals into disjoint subspaces, which can be done quite easily by looking at the corresponding unrestricted Hartree-Fock (UHF) orbitals.\cite{wang:2019} Finally, the antisymmetrized product of 1-reference orbital geminals (AP1roG),\cite{limacher:2013} which is equivalent to pair-coupled-cluster doubles (pCCD)\cite{stein:2014} has shown good results for ground state properties,\cite{limacher:2014a,limacher:2014b,henderson:2014a,henderson:2014b,boguslawski:2014a,boguslawski:2014b,boguslawski:2014c,tecmer:2014,boguslawski:2015,boguslawski:2016a,boguslawski:2016b,boguslawski:2017,boguslawski:2019,nowak:2019,boguslawski:2021,nowak:2021} and even excited state energies provided the orbitals are correctly optimized.\cite{marie:2021,kossoski:2021} However, AP1roG/pCCD is not a variational theory, so its wavefunction parameters must be solved by projection in a state-specific manner. 

The eigenvectors of the reduced BCS Hamiltonian, which we call Richardson-Gaudin (RG)\cite{richardson:1963,richardson:1964,richardson:1965,gaudin:1976} states, have shown excellent results for 1-dimensional strongly-correlated model systems. RG states are mean-field geminal wavefunctions that can be optimized with mean-field cost. What sets RG states apart from GVB, APSG, and AP1roG/pCCD is that they form a basis for the Hilbert space. Thus, to account for the missing weak correlation, \emph{single}-reference methods can be built from RG states in the same manner as for weakly-correlated systems in terms of Slater determinants. 

Previous applications\cite{johnson:2020,fecteau:2022} of RG states have been limited to small linear chains of hydrogen atoms, which are well described by GVB, and therefore APSG and AP1roG/pCCD. Hence, a natural next step in validating the efficacy of RG states is to consider more challenging model systems that remain small enough for an exact treatment. For this purpose, we consider a classic family of multi-reference systems, the Paldus H$_4$ isomers,\cite{paldus:1993} as well as more recently studied\cite{stair:2020} isomers of H$_{10}$, which resemble finite-sized Hubbard models with different connectivity patterns. 

This paper is outlined as follows. In section \ref{sec:theory} we summarize the structure of the reduced BCS Hamiltonian and RG states. In particular, the non-linear equations to be solved for each RG state, their reduced density matrix (RDM) elements, and transition density matrix (TDM) elements between RG states are presented. In section \ref{sec:numbers} we demonstrate that the Paldus isomers of H$_4$ and the Stair-Evangelista isomers of H$_{10}$ can, for the most part, be described qualitatively by orbital-optimized (OO) doubly-occupied configuration interaction (DOCI). In each case, a single RG state is energetically similar to OO-DOCI, while a configuration interaction (CI) (pair-) singles expansion in terms of RG states matches OO-DOCI with quantitative accuracy. Simple functionals of the occupation numbers are tested on top of a single RG state, and the results are not bad considering how computationally inexpensive they are.

\section{Theory} \label{sec:theory}
In this section we will briefly summarize the minimum of information required to compute with RG states. The development of the matrix elements is not complicated, but rather long and irrelevant for the present purpose. We refer the interested reader to refs.\cite{gorohovsky:2011,fecteau:2020,johnson:2021,fecteau:2021,moisset:2022a,faribault:2022} where \emph{exhaustive} detail is presented.

\subsection{Reduced BCS Hamiltonian and RG States}
Starting from second quantized operators $a^{\dagger}_{i\uparrow}$ and $a^{\dagger}_{i\downarrow}$ which create up and down spin electrons in spatial orbital $i$, pairs of electrons are described by the objects
\begin{align} \label{eq:su2_objects}
S^+_i = a^{\dagger}_{i\uparrow} a^{\dagger}_{i\downarrow}, \quad S^-_i = a_{i\downarrow} a_{i\uparrow}, \quad
S^z_i = \frac{1}{2} \left( a^{\dagger}_{i\uparrow}a_{i\uparrow} + a^{\dagger}_{i\downarrow}a_{i\downarrow} -1 \right)
\end{align}
which locally have the structure of the Lie algebra su(2)
\begin{subequations} \label{eq:su2_structure}
\begin{align} 
[S^+_i , S^-_j] &= 2 \delta_{ij} S^z_i \\
[S^z_i , S^{\pm}_j] &= \pm \delta_{ij} S^{\pm}_i.
\end{align}
\end{subequations}
With $N$ spatial orbitals, there are $N$ copies of these objects. It is convenient to use the number operator
\begin{align}
\hat{n}_i = 2S^z_i + 1,
\end{align}
which counts the number of individual electrons in the spatial orbital $i$. The vacuum $\ket{\theta}$ is destroyed by the pair removal operators
\begin{align}
S^-_i \ket{\theta} = 0, \quad \forall i.
\end{align}
With these objects, the reduced BCS Hamiltonian is written
\begin{align} \label{eq:bcs_ham}
\hat{H}_{BCS} = \frac{1}{2} \sum_i \varepsilon_i \hat{n}_i - \frac{g}{2} \sum_{ij} S^+_i S^-_j
\end{align}
and describes a system in which there is competition between filling the spatial orbitals (with single-particle energies $\{\varepsilon\}$) and a constant-strength pair-scattering $g$ between them. The interaction is attractive when $g$ is positive and repulsive when $g$ is negative, and this Hamiltonian is exactly solvable in all cases. RG states \eqref{eq:rg_states} are wavefunctions representing weakly-correlated pairs of electrons
\begin{align} \label{eq:rg_states}
\ket{\{u\}} = S^+(u_1) S^+(u_2) \dots S^+(u_M) \ket{\theta},
\end{align}
with the pair creators
\begin{align}
S^+(u) = \sum_i \frac{S^+_i}{u - \varepsilon_i}
\end{align}
defined by a set of complex numbers $\{u\}$ called the \emph{rapidities}. RG states are eigenvectors of the reduced BCS Hamiltonian provided that their rapidities are solutions of Richardson's equations
\begin{align} \label{eq:rich_eq}
\frac{2}{g} + \sum_i \frac{1}{u_a - \varepsilon_i} + \sum_{b (\neq a)} \frac{2}{u_b - u_a} = 0, \quad \forall a = 1,\dots,M.
\end{align}
The eigenvalue problem for the Hamiltonian \eqref{eq:bcs_ham} is thus reduced to solving the non-linear equations \eqref{eq:rich_eq}. Solving Richardson's equations for the rapidities is possible, but difficult and expensive as they possess divergent critical points where rapidities coincide with single-particle energies $\{\varepsilon\}$.\cite{rombouts:2004,guan:2012,pogosov:2012,debaerdemacker:2012} In terms of the eigenvalue-based variables (EBV), 
\begin{align}
U_i = \sum_a \frac{1}{\varepsilon_i - u_a},
\end{align}
Richardson's equations are equivalent to the non-linear equations
\begin{align} \label{eq:ebv_eq}
U^2_i - 2 U_i - g \sum_{k \neq i} \frac{U_k - U_i}{\varepsilon_k - \varepsilon_i} = 0, \quad \forall i = 1,\dots,N,
\end{align}
which are \emph{much} easier to solve numerically.\cite{faribault:2011,elaraby:2012} Rapidities could then be found with a root-finding procedure, but this is now completely unnecessary: all of the required RDM and TDM elements can be computed directly from the EBV.\cite{faribault:2022} However, the RG states do not have a direct representation in terms of the EBV so we will continue to label them as $\ket{\{u\}}$.

The EBV equations \eqref{eq:ebv_eq} decouple at $g=0$, where the eigenvectors of the reduced BCS Hamiltonian are Slater determinants. These states are thus well-defined by sites that are occupied and sites that are empty: they may be represented by a string of 1s and 0s that we have referred to as a \emph{bitstring}. At non-zero $g$, the RG states are \emph{not} Slater determinants, but evolve continuously and uniquely from the $g=0$ states. It is therefore unambiguous to label RG states at any $g$ as a bitstring based on the Slater determinant from which it evolves at $g=0$. At any $g$ the ground state of the reduced BCS Hamiltonian is always represented by the bitstring of $M$ 1s followed by $(N-M)$ 0s, and the highest excited state of the reduced BCS Hamiltonian is always the bitstring of $(N-M)$ 0s followed by $M$ 1s. The other RG states can and do cross, but the evolution from $g=0$ is unique. Note that, for Coulomb Hamiltonians, which are the Hamiltonians we would ultimately like to treat, the ground state of the reduced BCS Hamiltonian is not necessarily the variationally optimal state.

At $g=0$ we can solve the equations \eqref{eq:ebv_eq} explicitly. The solution is then evolved iteratively to a final target $g$. First, a step in $g$ toward the target is defined, and a fourth-order Taylor update to the EBV is computed. The EBV equations are then solved at the new $g$ by a Newton-Raphson procedure. If the terms in the Taylor series grow, or if the Newton-Raphson procedure causes a change in the norm of the EBV by more than 25\%, the step is rejected and reattempted with half the step-size. We have seen that if we begin by attempting as large a step as possible, the number of steps required seems to scale logarithmically with the pairing strength $g$. The Taylor update allows \emph{much} larger steps to be taken with no drawback. For details of the procedure, see refs. \cite{faribault:2011,elaraby:2012,claeys:2015,fecteau:2022}

The Coulomb Hamiltonians we wish to solve are written
\begin{align} \label{eq:C_ham}
\hat{H}_C = \sum_{ij} h_{ij} \sum_{\sigma} a^{\dagger}_{i \sigma} a_{j \sigma} + \frac{1}{2} \sum_{ijkl} V_{ijkl} \sum_{\sigma \tau} a^{\dagger}_{i \sigma} a^{\dagger}_{j \tau} a_{l \tau} a_{k \sigma}
\end{align}
with spin labels $\sigma$ and $\tau$ and a set of integrals computed in a given orbital basis $\{\phi\}$
\begin{align}
h_{ij} &= \int d\mathbf{r} \phi^*_i (\mathbf{r}) \left( - \frac{1}{2} \nabla^2 - \sum_I \frac{Z_I}{| \mathbf{r} - \mathbf{R}_I |} \right) \phi_j (\mathbf{r}) \\
V_{ijkl} &= \int d\mathbf{r}_1 d\mathbf{r}_2 \frac{\phi^*_i(\mathbf{r}_1)  \phi^*_j(\mathbf{r}_2)  \phi_k(\mathbf{r}_1)  \phi_l(\mathbf{r}_2)  }{| \mathbf{r}_1 - \mathbf{r}_2|}.
\end{align}
RG states can be used as a variational ansatz by minimizing
\begin{align}
E_{RG} = \min_{\{\varepsilon\},g} \frac{\braket{\{u\}|\hat{H}_C|\{u\}}}{\braket{\{u\}|\{u\}}}
\end{align}
with respect to $\{\varepsilon\}$ and $g$. Our first study of RG states in chemical problems\cite{johnson:2020}, in particular dissociations of linear hydrogen chains and molecular nitrogen, showed quite conclusively that the RG ground state 1...10...0 is \emph{not} representative of bond-breaking correlation. The RG ground state provided an acceptable description near the equilibrium geometry, and an exact description at dissociation, but the intermediate re-coupling region was not at all well described. In a later study,\cite{fecteau:2022} we found that another RG state, specifically one labelled 1010...10 which we have referred to as the N\'{e}el RG state, was actually the variationally optimal state. For linear H$_{8}$, for example, the optimal single-particle energies $\{\varepsilon\}$ arranged themselves in a 2-2-2-2 pattern: they form four sets of two $\{\varepsilon\}$ that are close in energy. The energy difference between each set of two $\{\varepsilon\}$ is much larger than the pairing strength $g$. The bitstring 10101010 places one rapidity in between each of the near-degenerate pairs of $\{\varepsilon\}$ so that each pair $S^+(u)$ is dominated by two sites. This can be summarized in the geminal coefficient matrix 
\begin{align} \label{eq:gvb_matrix}
G_{ai} = \frac{1}{u_a - \varepsilon_i} \approx
\begin{pmatrix}
* & * & 0 & 0 & 0 & 0 & 0 & 0 & 0 & 0 \\
0 & 0 & * & * & 0 & 0 & 0 & 0 & 0 & 0 \\
0 & 0 & 0 & 0 & * & * & 0 & 0 & 0 & 0 \\
0 & 0 & 0 & 0 & 0 & 0 & * & * & 0 & 0 \\
0 & 0 & 0 & 0 & 0 & 0 & 0 & 0 & * & * 
\end{pmatrix},
\end{align}
in which each row represents a pair while each column represents an orbital. The elements marked with zeroes in equation \eqref{eq:gvb_matrix} are not numerically zero, but they are much smaller than those marked with asterisks. In the dissociation limit, the N\'{e}el RG state explicitly becomes the GVB state
\begin{align}
\ket{\text{GVB}} = (S^+_1 - S^+_2)(S^+_3 - S^+_4)\dots (S^+_{2N-1} - S^+_{2N}) \ket{\theta}.
\end{align}
For the dissociation of molecular nitrogen, we found the optimal RG state to be 1111101010 with $\{\varepsilon\}$ arranged in a 1-1-1-1-2-2-2 pattern: there were four individual $\{\varepsilon\}$ along with three pairs of near-degenerate $\{\varepsilon\}$.

A single RG state optimized as described above will provide a reasonable, but not exact, description of the seniority-zero sector of the true wavefunction. For an improved description of this sector, we can perform a CI in the basis of RG states. For Slater determinants, the Slater-Condon rules ensure that a given reference only couples with single- and double-excitations through the Coulomb Hamiltonian. For RG states however, there are unfortunately no such rules. On paper, \emph{each} RG state will couple with \emph{each other} RG state. Fortunately, we have found numerically that these couplings go to zero quite rapidly.\cite{johnson:2021} With a given \emph{reference} bitstring, we will call a single-pair excitation one that differs from the reference by a single 1 and a single 0, a double-pair excitation one the differs from the reference by two 1s and two 0s, etc. Magnitudes of couplings decrease with excitation level, and are negligible past doubles. Thus, we will variationally optimize a single RG state, then solve the CI problem with its singles (RGCIS), and singles and doubles (RGCISD). In particular, for a variationally optimized set of $\{\varepsilon\}$ and $g$ for a particular RG bitstring, we will \emph{compute} the EBV for the $M(N-M)$ singles and $\binom{M}{2}\binom{N-M}{2}$ doubles. The Coulomb Hamiltonian is built in this basis and diagonalized. This procedure requires both RDM elements for each RG state and transition density matrix (TDM) elements between RG states.

\subsection{RDMs and TDMs}
Evaluating the energy of \eqref{eq:C_ham} with an RG state yields only seniority-zero contributions
\begin{align} \label{eq:e_functional}
E_{RG}[\{\varepsilon\},g] = 2\sum_k h_{kk} \gamma_k + \sum_{kl} (2V_{klkl}-V_{kllk})D_{kl} + \sum_{kl} V_{kkll} P_{kl}.
\end{align}
The 1-RDM $\gamma_k$ is diagonal 
\begin{align}
\gamma_k = \frac{1}{2} \frac{\braket{\{u\}|\hat{n}_k|\{u\}}}{\braket{\{u\}|\{u\}}},
\end{align}
and there are only $\mathcal{O}(N^2)$ non-zero elements from the 2-RDM arranged in what we call the diagonal-correlation function $D_{kl}$
\begin{align}
D_{kl} = \frac{1}{4} \frac{\braket{\{u\}|\hat{n}_k \hat{n}_l|\{u\}}}{\braket{\{u\}|\{u\}}},
\end{align}
and the pair-correlation function $P_{kl}$
\begin{align}
P_{kl} = \frac{\braket{\{u\}|S^+_k S^-_l | \{u\}}}{\braket{\{u\}|\{u\}}}.
\end{align}
In the pair representation, the diagonal elements $P_{kk}$ and $D_{kk}$ refer to the same matrix element, so to avoid double-counting we assign it to $P_{kk} = \gamma_k$ and set $D_{kk}=0$.

Expressions for the density matrix elements of RG states are known in terms of rapidities,\cite{faribault:2008,faribault:2010,gorohovsky:2011,fecteau:2020} and more recently directly in terms of EBV.\cite{faribault:2022} We will present only the results as the development is incredibly tedious. TDM elements are presented first as RDM elements are a special case.

The expressions for the 1- and 2-TDM elements require the 1st and 2nd co-factors of the matrix $J$,
\begin{align}
J_{ij} = \begin{cases}
U_i + V_i -\frac{2}{g} + \sum_{k(\neq i)} \frac{1}{\varepsilon_k - \varepsilon_i}, &\quad i = j \\
\frac{1}{\varepsilon_i - \varepsilon_j}, &\quad i \neq j
\end{cases}
\end{align}
where $\{U\}$ and $\{V\}$ are the EBV for the two RG states. First co-factors are understood
\begin{align}
A[J]^{i,j} = (-1)^{i+j} \det J^{i,j}
\end{align}
where $J^{i,j}$ is the matrix $J$ without the $i$th row and $j$th column. To correctly account for the sign, second co-factors require using a Heaviside function
\begin{align}
h(x) = \begin{cases}
1 & x > 0 \\
0 & x \leq 0
\end{cases}
\end{align}
in their definition
\begin{align}
A[J]^{ij,kl} = (-1)^{i+j+k+l+h(i-j)+h(k-l)} \det J^{ij,kl}.
\end{align}
Here $J^{ij,kl}$ is the matrix $J$ without the $i$th and $j$th rows, and the $k$th and $l$th columns.

For two distinct RG states, the \emph{non-normalized} elements of the 1-TDM are
\begin{align} \label{eq:ebv_1tdm}
\gamma^{UV}_k := \frac{1}{2}\braket{\{u\}|\hat{n}_k|\{v\}} = \eta \sum_l V_l A[J]^{l,k}
\end{align}
where the pre-factor is
\begin{align}
\eta = (-1)^{N-M}\left( \frac{g}{2} \right)^{N-2M}.
\end{align}
With
\begin{align}
K_{kl} = V_k V_l + \frac{V_k - V_l}{\varepsilon_k - \varepsilon_l},
\end{align}
the non-normalized elements of the 2-TDM are
\begin{align} \label{eq:ebv_t2d}
\frac{1}{\eta} D^{UV}_{kl} :&= \frac{1}{\eta} \frac{1}{4} \braket{\{u\}|\hat{n}_k\hat{n}_l|\{v\}} \\
&= K_{kl} A[J]^{kl,kl} + \sum_{i (\neq k,l)} K_{il} A[J]^{il,kl} + \sum_{i (\neq k,l)} K_{ik} A[J]^{ki,kl} \nonumber \\
&+ \sum_{i < j (\neq k,l)} \frac{(\varepsilon_k - \varepsilon_i)(\varepsilon_l-\varepsilon_j)+(\varepsilon_k - \varepsilon_j)(\varepsilon_l-\varepsilon_i)}{(\varepsilon_k-\varepsilon_l)(\varepsilon_j-\varepsilon_i)} K_{ij} A[J]^{ij,kl},
\end{align} 
and
\begin{align} \label{eq:ebv_t2p}
\frac{1}{\eta} P^{UV}_{kl} :&= \frac{1}{\eta} \braket{\{u\}|S^+_k S^-_l | \{v\}} \\
&= \left( V_l + (\varepsilon_k - \varepsilon_l)(V_l V_l - V_l J_{ll}) \right) A[J]^{l,k} + \sum_{i (\neq k,l)}\frac{\varepsilon_i-\varepsilon_k}{\varepsilon_i-\varepsilon_l} V_i A[J]^{i,k} \nonumber \\
&- 2 \sum_{i (\neq k,l)} \frac{\varepsilon_k - \varepsilon_i}{\varepsilon_l - \varepsilon_i} K_{i l}
A[J]^{i l,kl} 
- 2\sum_{\substack{ i < j  \\ (\neq k,l) } } \frac{(\varepsilon_k - \varepsilon_i)(\varepsilon_k - \varepsilon_j)}{(\varepsilon_k - \varepsilon_l)(\varepsilon_j-\varepsilon_i)}
K_{ij} A[J]^{ij,kl}.
\end{align}
In this contribution we evaluated the TDM elements by first computing the $\binom{N}{2}\binom{N}{2}$ second co-factors of the matrix $J$ for each pair of states. Such an approach is quite expensive and certainly could be improved by further studying the matrix elements. However, one of the points of the present contribution is to determine whether it is worth working it out. 

The square of the norm of an RG state is
\begin{align}
\braket{\{u\}|\{u\}} = \eta \det \bar{J}
\end{align}
where $\bar{J}$ is the matrix $J$ where both sets of EBV are the same. \emph{Normalized} RDM elements of RG states are particularly simple to compute: scaled second co-factors are writeable directly as a $2\times 2$ determinant of first co-factors 
\begin{align}
\frac{A[\bar{J}]^{ij,kl}}{\det \bar{J}} = \frac{A[\bar{J}]^{i,k}}{\det \bar{J}}\frac{A[\bar{J}]^{j,l}}{\det \bar{J}} -  \frac{A[\bar{J}]^{i,l}}{\det \bar{J}}\frac{A[\bar{J}]^{j,k}}{\det \bar{J}}.
\end{align}
This extends generally as a theorem of Jacobi\cite{vein_book} states that a $k$th-order scaled co-factor is a $k \times k$ determinant of scaled first co-factors. Next, the adjugate formula for the matrix inverse gives the elements of the inverse as
\begin{align}
\bar{J}^{-1}_{ij} = \frac{A[\bar{J}]^{j,i}}{\det \bar{J}}.
\end{align}
The 1- and 2-RDMs can therefore easily be computed once the matrix $\bar{J}$ is inverted. We have seen,\cite{faribault:2022} that so long as the set of $\{\varepsilon\}$ is non-degenerate, $\bar{J}$ is well-conditioned and it is reasonable to invert numerically. The normalized 1-RDM elements are
\begin{align} \label{eq:ebv_1dm}
\gamma_{k} = \sum_l U_l \bar{J}^{-1}_{kl}
\end{align}
while the 2-RDM elements are
\begin{align} \label{eq:ebv_d2d}
D_{kl} &= K_{kl} (\bar{J}^{-1}_{kk}\bar{J}^{-1}_{ll}-\bar{J}^{-1}_{lk}\bar{J}^{-1}_{kl}) 
+ \sum_{j (\neq l)} K_{jk} (\bar{J}^{-1}_{kk}\bar{J}^{-1}_{lj}-\bar{J}^{-1}_{lk}\bar{J}^{-1}_{kj}) 
+ \sum_{j (\neq k)} K_{jl} (\bar{J}^{-1}_{kj}\bar{J}^{-1}_{ll}-\bar{J}^{-1}_{lj}\bar{J}^{-1}_{kl})  \nonumber \\
&+ \sum_{\substack{ i < j  \\ (\neq k,l) } } \frac{(\varepsilon_k - \varepsilon_i)(\varepsilon_l - \varepsilon_j)+(\varepsilon_k - \varepsilon_j)(\varepsilon_l - \varepsilon_i)}{(\varepsilon_k - \varepsilon_l)(\varepsilon_j - \varepsilon_i)} K_{ij} (\bar{J}^{-1}_{ki}\bar{J}^{-1}_{lj}-\bar{J}^{-1}_{li}\bar{J}^{-1}_{kj}) 
\end{align}
and
\begin{align} \label{eq:ebv_d2p}
P_{kl} &= \left(2 U_l + \sum_{i (\neq k,l)} \frac{\varepsilon_i - \varepsilon_k}{\varepsilon_i - \varepsilon_l} U_i - \frac{2M}{g} \right) \bar{J}^{-1}_{kl}  \nonumber \\
&+ \sum_{i (\neq k,l)} \frac{\varepsilon_i - \varepsilon_k}{\varepsilon_i - \varepsilon_l} (U_i \bar{J}^{-1}_{ki} -2 K_{il} (\bar{J}^{-1}_{ki}\bar{J}^{-1}_{ll}-\bar{J}^{-1}_{li}\bar{J}^{-1}_{kl}) ) \nonumber \\
&-2 \sum_{\substack{ i < j  \\ (\neq k,l) } } \frac{(\varepsilon_k - \varepsilon_i)(\varepsilon_k - \varepsilon_j)}{(\varepsilon_k - \varepsilon_l)(\varepsilon_j - \varepsilon_i)}
K_{ij} (\bar{J}^{-1}_{ki}\bar{J}^{-1}_{lj}-\bar{J}^{-1}_{li}\bar{J}^{-1}_{kj}).
\end{align}
Unless it turns out to be possible to compute the RDM elements without linear algebra operations, these expressions are optimal: for each RG state, its 2-RDM can be constructed with $\mathcal{O}(N^4)$ cost.

\subsection{Weak correlation functionals}
The natural orbital functionals of Piris are closely related to geminal wavefunctions. In particular, PNOF5\cite{piris:2011} is equivalent to APSG,\cite{pernal:2013} while PNOF7\cite{piris:2017,piris:2019,mitxelena:2020a,mitxelena:2020b,rodriguez:2021} can be roughly understood as intra-pair contributions treated as APSG and inter-pair treated like AGP. Both are treatments of strong correlation in the seniority-zero channel. More recently, GNOF\cite{piris:2021,mitxelena:2022} tries to include weak correlation by updating the $P_{kl}$ elements with the ``dynamical'' part of the occupation number $d_i$. The energy of the update is of the form
\begin{align} \label{eq:Ewc_piris}
E_{WC} = - \sum_{ij}{}^{'} \sqrt{d_id_j} V_{iijj}
\end{align}
where the primed summation leaves out: i) diagonal terms, ii) terms arising from the same APSG subspace, iii) terms for which $\gamma_i$ and $\gamma_j$ are both near 1. Piris suggests an ansatz for $d_i$ as the occupation number $\gamma_i$ scaled by a Gaussian function. In this contribution, we considered a few variants applied only to the N\'{e}el RG state. This means in particular that each pair is considered to be dominated by 2 spatial orbitals. The same three conditions are used to omit terms. 

One can include either only contributions from occupation numbers near zero ($d^h$) or contributions from occupations near zero \textbf{and} one ($d^{hp}$)
\begin{align}
d^h_i &= \gamma_i f(\gamma_i) \\
d^{hp}_i &= \gamma_i f(\gamma_i) + (1 - \gamma_i) f(1-\gamma_i) 
\end{align} 
while also considering other decaying functions $f$. We considered both a Gaussian and a Slater decay
\begin{align}
f^G (\gamma_i) &= \exp \left( - \left( \frac{\gamma_i}{0.02\sqrt{2}} \right)^2  \right) \\
f^S (\gamma_i) &= \exp \left( - \left( \frac{\gamma_i}{0.02} \right)  \right)
\end{align}
with maxima at 0.02. Thus, four possible functionals were considered and we refer to them as: $E^{Gh}_{WC}$, $E^{Ghp}_{WC}$, $E^{Sh}_{WC}$, and $E^{Shp}_{WC}$. We also considered the same group of four functionals with maxima at 0.01, but they were nearly always inferior to the functionals with maxima at 0.02. 

\section{Numerical Results} \label{sec:numbers}
\subsection{Preliminaries}
Potential energy curves for the Paldus isomers of H$_4$ and the Stair-Evangelista isomers of H$_{10}$ were computed in the minimal basis STO-6G. Full CI (FCI) results were computed for the Paldus systems using the \textsc{Psi4} quantum chemistry package,\cite{smith:2020} while for the H$_{10}$ isomers the results were taken from Ref.\cite{stair:2020} RG states are strictly variational approximations to OO-DOCI so the results are computed in the basis of OO-DOCI orbitals. OO-DOCI calculations were performed using the implementation in \texttt{hilbert},\cite{hilbert} which is a plugin to the \textsc{Psi4} package. It is well known that the seniority-zero orbital landscape is prone to many local minima, so multiple OO-DOCI calculations were performed at each geometry with different initial starting conditions for the orbitals, and the lowest-energy solutions obtained are reported here.

The variational RG optimization remains proof of principle: the variables $\{\varepsilon\}$ and $g$ are pre-conditioned with the covariance matrix adaptation evolution strategy (CMA-ES)\cite{hansen:2001} before being optimized with the Nelder-Mead\cite{nelder:1965} simplex algorithm. Three consistency conditions are verified at each iteration of the variational optimization. First, $\gamma_k$ and $D_{kl}$ satisfy the sum rules
\begin{align} \label{eq:diag_consistency}
\sum_k \gamma_k &= M, \\
\sum_{kl} D_{kl} &= M(M-1).
\end{align}
Ordinarily, the trace of $D_{kl}$ would be $M^2$, but the choice to assign $D_{kk}=0$ reduces it by $M$. A sum rule for $P_{kl}$ can be deduced by computing the energy of the reduced BCS Hamiltonian in two ways: the energy can be evaluated as
\begin{align}
E_{BCS} = \sum_k \varepsilon_k \gamma_k - \frac{g}{2} \sum_{kl} P_{kl}
\end{align}
but it may be computed directly in terms of the EBV
\begin{align}
E_{BCS} = \frac{g}{2} M (M-N-1) +\frac{g}{2}\sum_k \varepsilon_k U_k.
\end{align}
Simply rearranging gives an expression for the trace
\begin{align} \label{eq:odlro}
\sum_{kl} P_{kl} = \sum_k \varepsilon_k \left( \frac{2}{g} \gamma_k - U_k \right) + M (N-M+1).
\end{align}
A violation of one of these conditions by $10^{-6}$ is judged to be unacceptable and causes the point to be rejected. In the optimized results, the conditions are respected nearly to machine-precision. Loss of precision appears when the single-particle energies $\{\varepsilon\}$ become too close to one another and could be foreseen by checking the condition number of the matrix $\bar{J}$ before numerical inversion. When $\bar{J}$ is ill-conditioned, the single-particle energies must be treated as explicitly degenerate, leading to a different construction we will report separately.

In the CI stage of the problem, we check conditions \eqref{eq:diag_consistency} and \eqref{eq:odlro} for \emph{each} computed RG state as well as off-diagonal conditions for each pair of states. In particular, for each pair of states the traces of the $\hat{n}_k$ correlations will vanish
\begin{align} \label{eq:off_diag_consistency}
\sum_k \gamma^{UV}_k &= 0 \\
\sum_{kl} D^{UV}_{kl} &= 0,
\end{align}
and a similar condition for the pair-transfer elements can be deduced
\begin{align} \label{eq:off_diag_odlro}
\sum_{kl} P^{UV}_{kl} = \frac{2}{g} \sum_k \varepsilon_k \gamma^{UV}_{k}
\end{align}
by projecting the RG state $\bra{\{u\}}$ against the Schr\"{o}dinger equation for the reduced BCS Hamiltonian
\begin{align}
\hat{H}_{BCS} \ket{\{v\}} = E_{BCS}[\{v\}] \ket{\{v\}}
\end{align}
acting on the RG state $\ket{\{v\}}$. In all computations, the largest violation of these consistency conditions observed is on the order of $10^{-10}$, and is usually on the order of $10^{-13}$.

In nearly every case, it is very difficult to visually discern the RG state energies from the OO-DOCI energies (not to mention RGCIS and RGCISD) so the RG energy curves will not be presented on top of the OO-DOCI results. We will instead plot their respective errors
\begin{subequations}
\begin{align}
\Delta_{RG} &= E_{RG} - E_{OO-DOCI} \\
\Delta_{RGCIS} &= E_{RGCIS} - E_{OO-DOCI} \\
\Delta_{RGCISD} &= E_{RGCISD} - E_{OO-DOCI}.
\end{align}
\end{subequations}

\subsection{H$_8$ chain}
We first looked at the linear H$_8$ chain to test the weak correlation functionals and RGCI on a system we have studied previously with RG states.\cite{johnson:2020,fecteau:2022,moisset:2022a} The optimal RG state is the N\'{e}el state 10101010. 
\begin{figure}[ht!] 
	\includegraphics[width=0.475\textwidth]{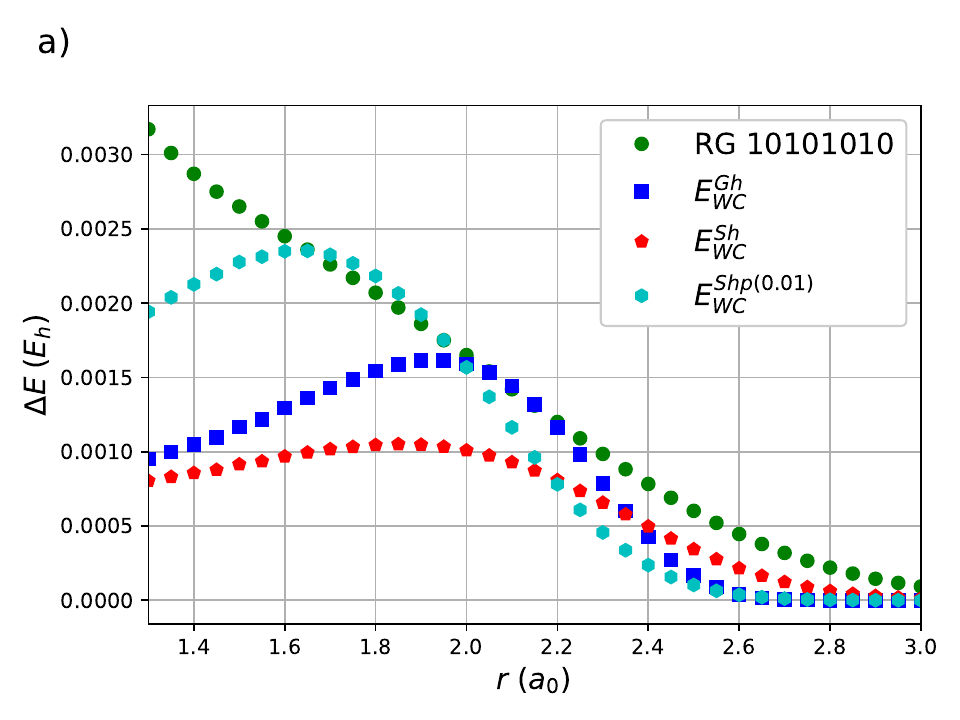} \hfill
	\includegraphics[width=0.475\textwidth]{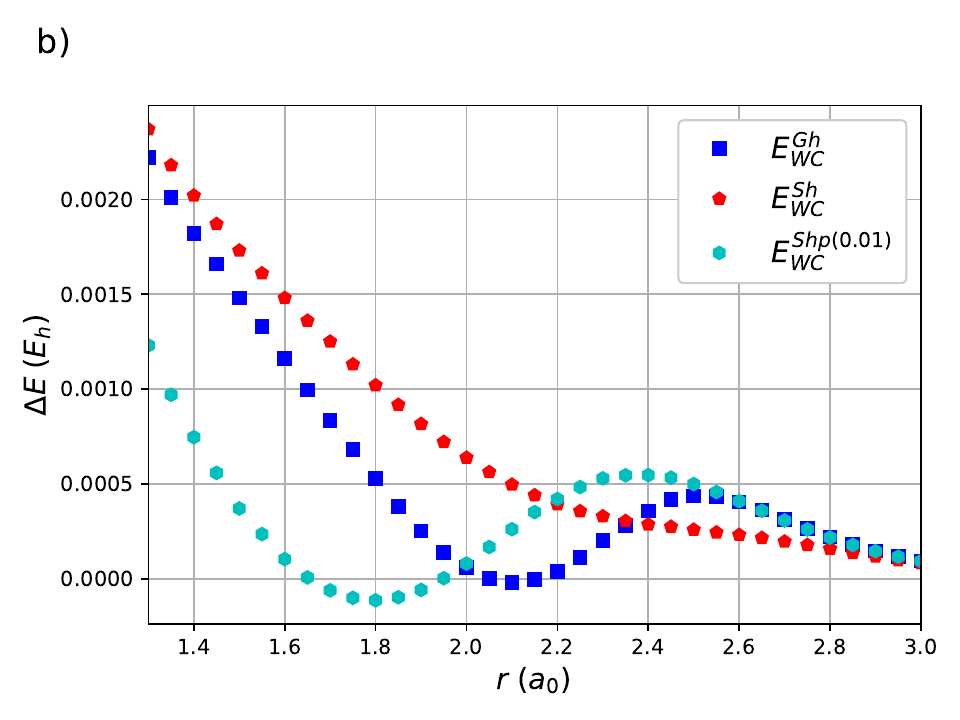}
	\caption{Weak correlation functionals for H$_8$ chain: (a) $|E_{WC}|$ compared with $\Delta_{RG}$. (b) Difference between $\Delta_{RG}$ and $|E_{WC}|$. 10101010 is the optimal RG state.}
	\label{fig:H8_wc}
\end{figure}
The best performing weak correlation functionals are $E^{Gh}_{WC}$ and $E^{Sh}_{WC}$, though $E^{Shp}_{WC}$ with a maximum at 0.01 is also reasonable. These results are shown in Figure \ref{fig:H8_wc}. Functionals of the type \eqref{eq:Ewc_piris} are always negative so their absolute values are plotted in Figure \ref{fig:H8_wc} to compare with $\Delta_{RG}$. The remaining error,
\begin{align}
\Delta_{RG+WC} = E_{RG} + E_{WC} - E_{OO-DOCI} = \Delta_{RG} - | E_{WC} | 
\end{align}
is plotted in Figure \ref{fig:H8_wc} (b). The other tested functionals either substantially over- or under-correlated (see supporting information). Of these three functionals, $E^{Sh}_{WC}$ appears to be the best behaved. All three under-correlate at short H--H distances. $E^{Shp(0.01)}_{WC}$ is the best at short H--H distances, but over-correlates at longer H--H separations, as does $E^{Gh}_{WC}$. $E^{Sh}_{WC}$ is the worst of the three at short H--H distances, but never over-correlates, has an error that decays monotonically, and is the best at longer H--H distances. All three functionals are reasonable given how they are essentially free to evaluate.

\begin{figure}[ht!] 
	\includegraphics[width=0.475\textwidth]{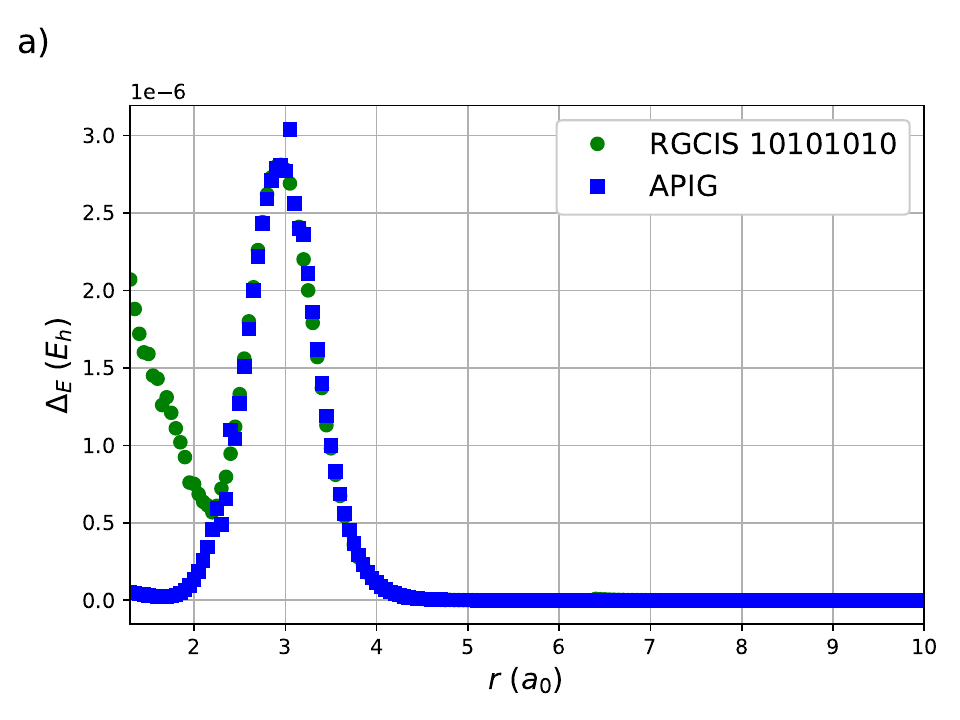} \hfill
	\includegraphics[width=0.475\textwidth]{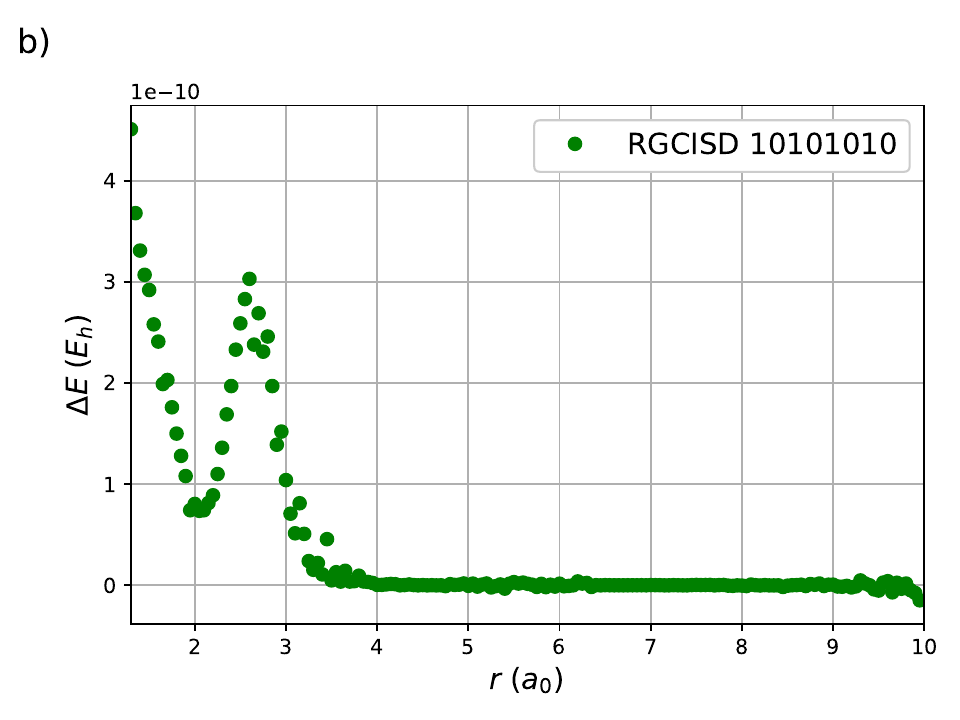}
	\caption{RGCIS and RGCISD treatment of H$_{8}$ chain: (a) $\Delta_{RGCIS}$ for the 10101010 RG state compared with similar results for APIG from ref.\cite{moisset:2022a} (b) $\Delta_{RGCISD}$ for the 10101010 RG state.   All results computed with the STO-6G basis set in the basis of OO-DOCI orbitals.}
	\label{fig:H8_ci}
\end{figure}
The RGCIS results in Figure \ref{fig:H8_ci} (a) seem to match the results obtained for APIG,\cite{moisset:2022a}
\begin{align}
\ket{\text{APIG}} = \prod^M_{a=1} \sum^N_{i=1} G_{ai} S^+_i \ket{\theta}
\end{align}
for which the geminal coefficients $G_{ai}$ are variationally optimized. For HF Slater determinants, Brillouin's theorem ensures that the optimal state does not couple with its CI singles. For RG states there is no Brillouin theorem, but the optimal state in terms of two-particle clusters would be APIG, which appears to be equivalent to RGCIS except at compressed geometries. The RGCISD results, in Figure \ref{fig:H8_ci} (b) are \emph{very} close to DOCI. Occasionally RGCISD is found to be very slightly below DOCI, which must be attributed to loss of precision on the order of $10^{-12}$.

\subsection{Paldus systems}
\begin{figure}[ht!] 
	\includegraphics[width=0.9\textwidth]{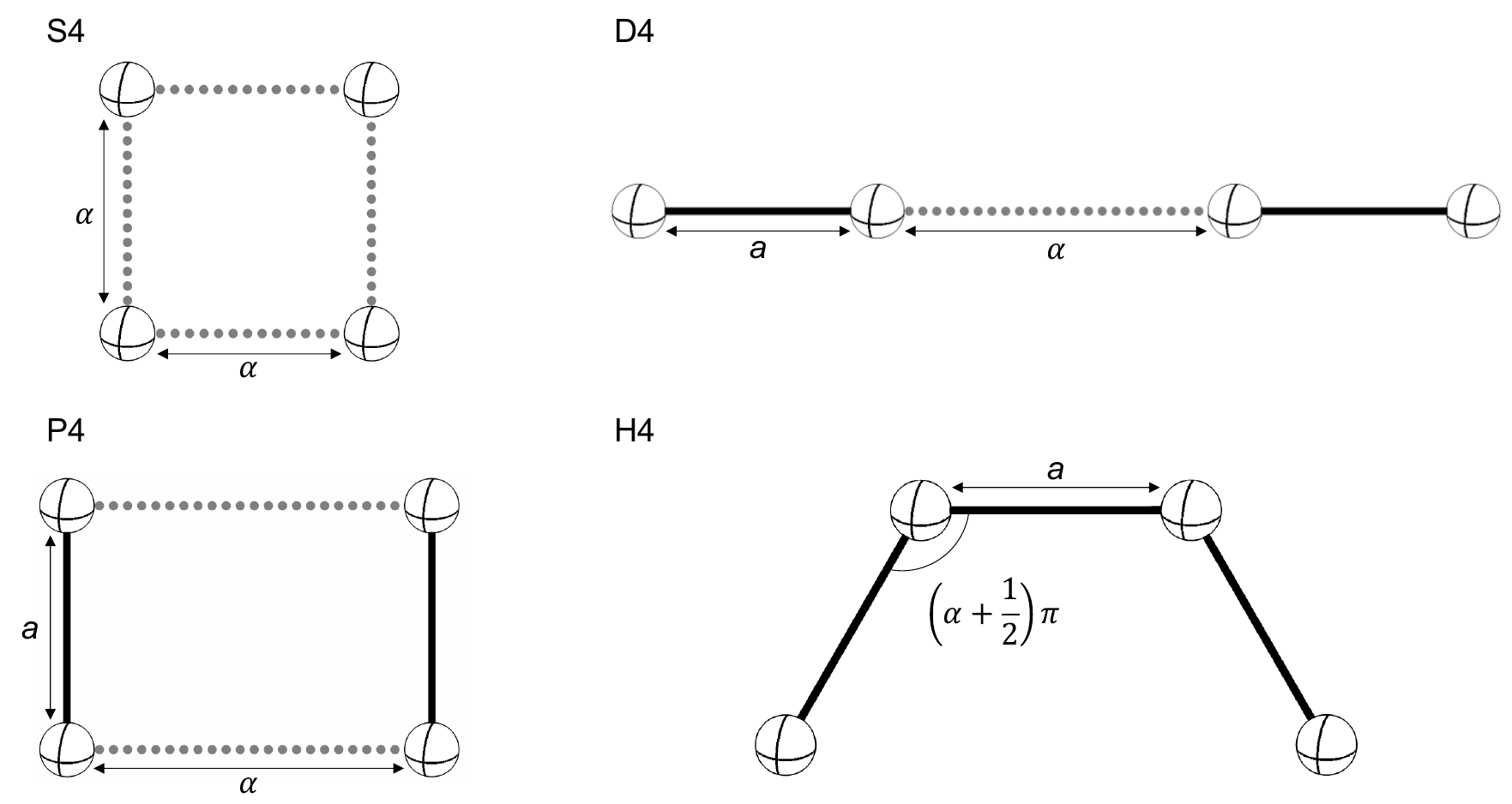}
	\caption{Paldus isomers of H$_{4}$. Solid lines have fixed length $a$ while dotted lines have variable length $\alpha$.} 
	\label{fig:H4_isomers}
\end{figure}

In ref.\cite{paldus:1993} Paldus and co-workers presented four isomers of H$_4$, see Figure \ref{fig:H4_isomers}, that are multi-reference in terms of Slater determinants. S4 is a square of hydrogen atoms whose side length $\alpha$ is increased to the dissociated limit. P4 and D4 consist of two H$_2$ subunits with a fixed ``bond-length'' $a$ and a variable distance between subunits, $\alpha$. As $a$ is increased, the HOMO/LUMO pairs of each H$_2$ subunit get closer in energy, which increases the multi-reference character of the problem. Thus, following ref.,\cite{paldus:1993} three values of $a$ are considered: $a=1.2\;a_0$, $a=1.6\;a_0$, and $a=2.0\;a_0$. The quoted FCI value for the H$_2$ bond is $1.667\;a_0$, so these three choices represent a shortened bond-length, a near-equilibrium bond-length, and a stretched bond-length. H4 represents a square that opens to a line with fixed bond-lengths $a$. Again, the three values $a=1.2\;a_0$, $a=1.6\;a_0$, and $a=2.0\;a_0$ are considered.

\begin{figure}[ht!] 
	\includegraphics[width=0.475\textwidth]{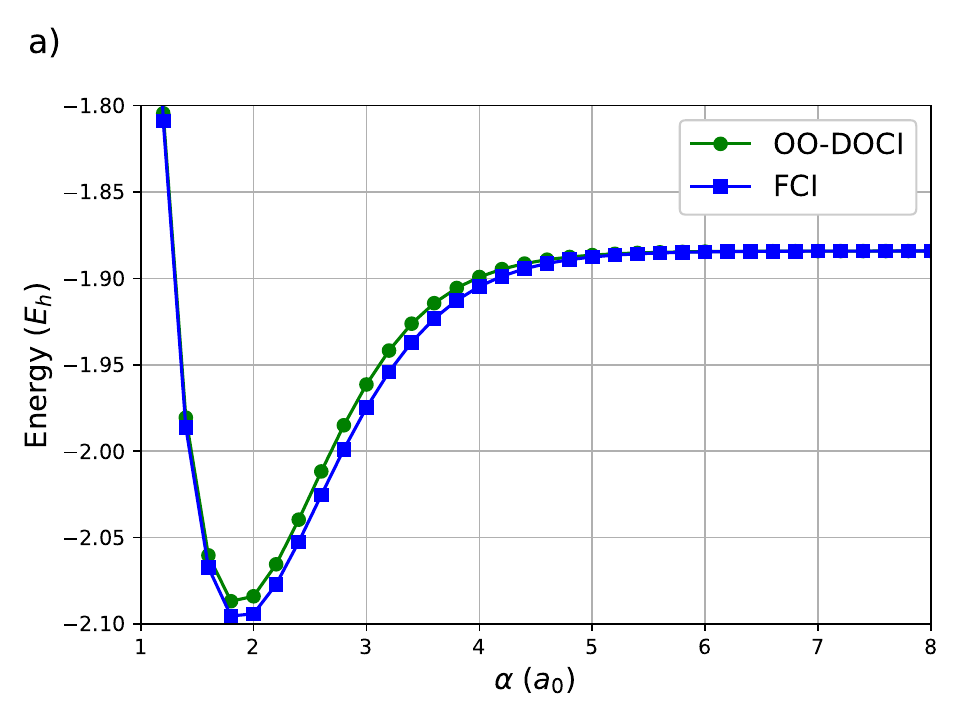} \hfill
	\includegraphics[width=0.475\textwidth]{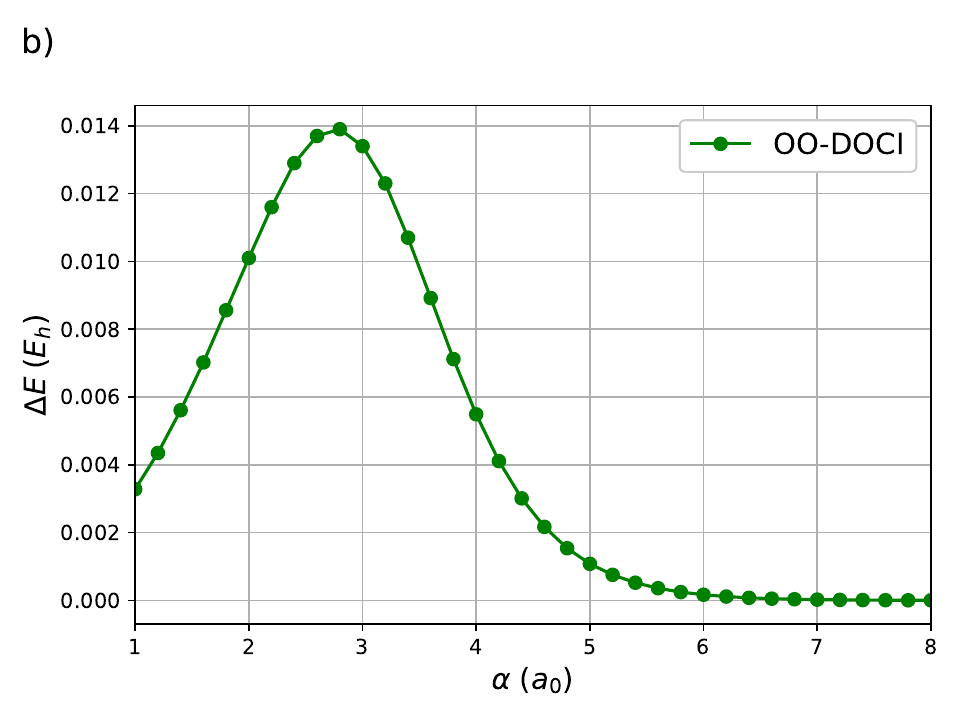}
	\caption{OO-DOCI treatment of Paldus S4: (a) FCI and OO-DOCI energies. (b) Error of OO-DOCI with respect to FCI. All results computed with the STO-6G basis set in the basis of OO-DOCI orbitals.}
	\label{fig:S4_DOCI_energies}
\end{figure}

\begin{figure}[ht!] 
	\includegraphics[width=0.475\textwidth]{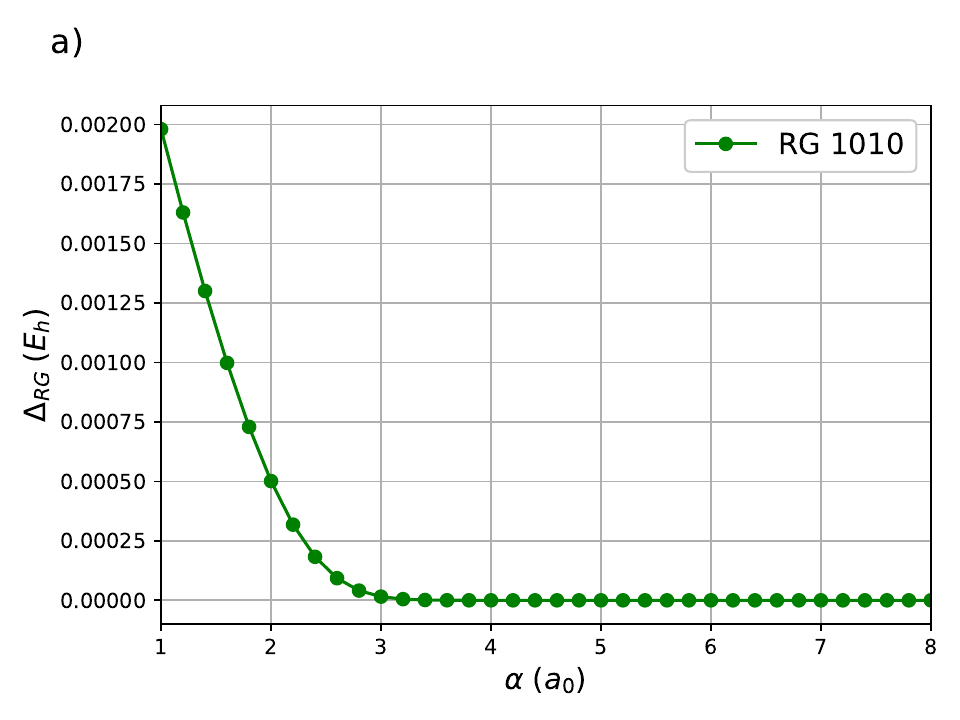} \hfill
	\includegraphics[width=0.475\textwidth]{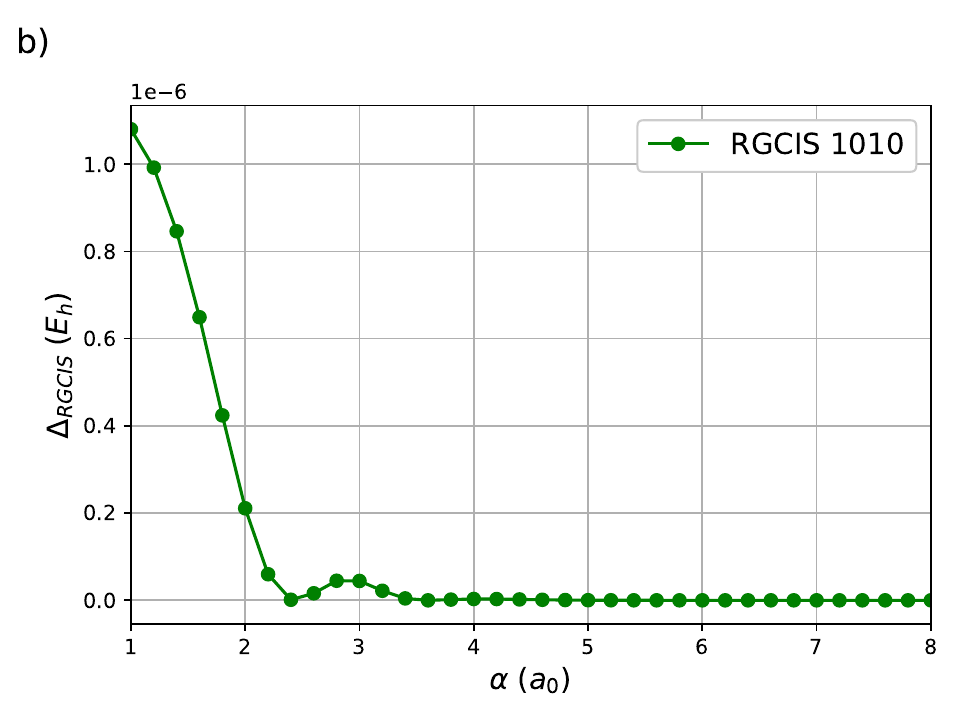}
	\caption{Variational RG and RGCI treatment of Paldus S4: (a) $\Delta_{RG}$ for the 1010 RG state. (b) $\Delta_{RGCIS}$ for the 1010 RG state. All results computed with the STO-6G basis set in the basis of OO-DOCI orbitals.}
	\label{fig:S4_RG_energies}
\end{figure}

Figure \ref{fig:S4_DOCI_energies} depicts FCI and OO-DOCI energies for S4 for a range of $\alpha = [1.0\;a_0,8.0\;a_0]$. OO-DOCI provides near-quantitative accuracy with respect to FCI-derived energies. The maximum error in the OO-DOCI energy is only $\approx 0.014$ E$_\text{h}$, around 2.8 $a_0$. 
Variationally optimized 1010 RG states, shown in Figure \ref{fig:S4_RG_energies}, are energetically similar to OO-DOCI, never displaying more than 2 $\times 10^{-3}$ E$_\text{h}$ error. Hence, the single RG state 1010 is similar to the OO-DOCI wavefunction throughout the potential energy curve. Figure \ref{fig:S4_RG_energies} (b) shows that RGCIS built from the 1010 RG state is near-exact, with respect to OO-DOCI; maximum errors are only on the order of 10$^{-6}$ E$_\text{h}$. Variational RG calculations were also performed for a 1100 RG state (see supporting information), but these states were found to be much too high in energy in the re-coupling region. Whereas for the 1010 RG state the $\{\varepsilon\}$ arrange themselves in a 2-2 pattern, for the 1100 RG state \emph{all} the $\{\varepsilon\}$ are near-degenerate, which substantially increases the computational cost of solving the EBV equations \eqref{eq:ebv_eq}. RGCISD results for the H$_4$ isomers are reported in the supporting information as the number of RG states included is the same as the number of Slater determinants in OO-DOCI. In these cases the RGCISD and OO-DOCI results agree on the order of $10^{-12}$. 

\begin{figure}[ht!] 
	\includegraphics[width=0.475\textwidth]{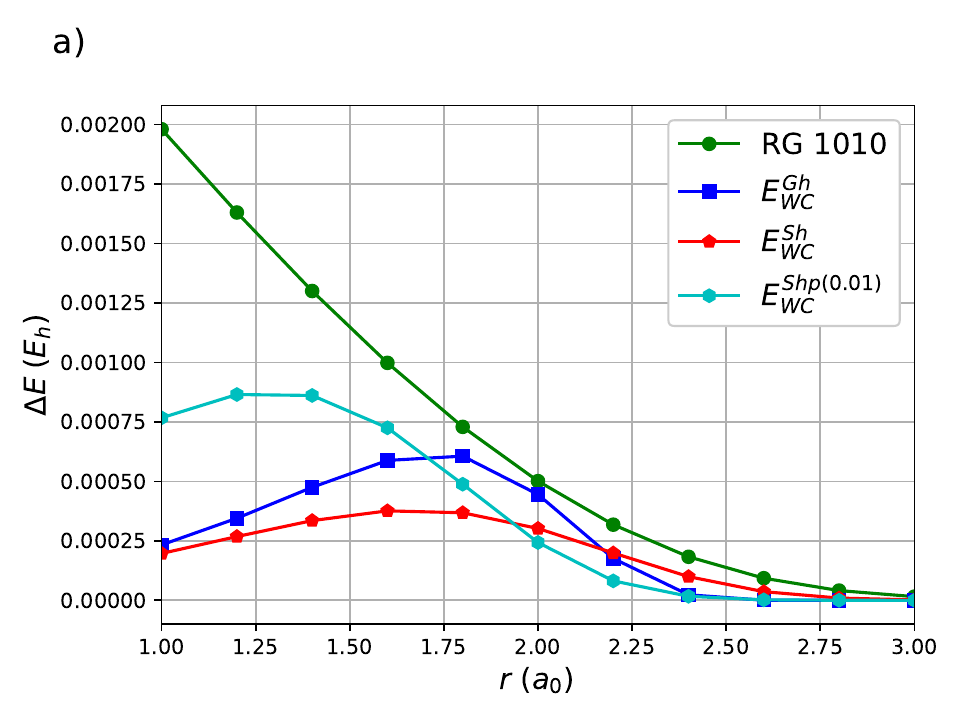} \hfill
	\includegraphics[width=0.475\textwidth]{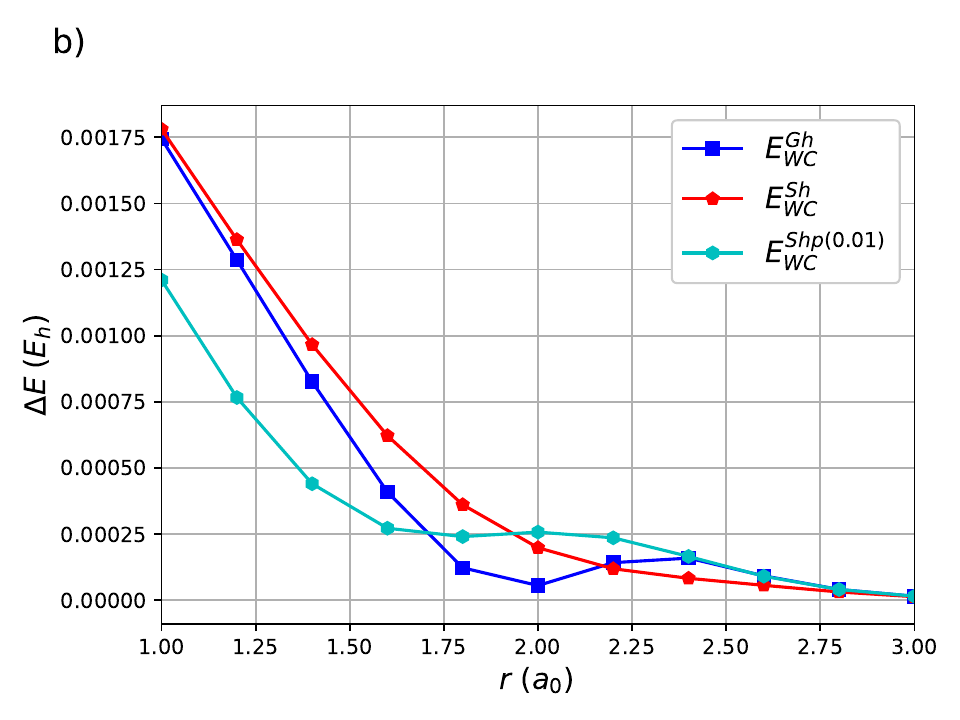}
	\caption{Weak correlation functionals for Paldus S4: (a) $|E_{WC}|$ compared with $\Delta_{RG}$. (b) Difference between $\Delta_{RG}$ and $|E_{WC}|$. 1010 is the optimal RG state. All results computed with the STO-6G basis set in the basis of OO-DOCI orbitals.}
	\label{fig:S4_wc}
\end{figure}

We also considered energy corrections from weak correlation functionals, applied to the 1010 RG reference, and the resulting data are shown in Figure \ref{fig:S4_wc}. In general, we observe the same pattern as for linear H$_8$: the error for $E^{Gh}_{WC}$ has some oscillatory behavior while the error for $E^{Sh}_{WC}$ decays monotonically and should therefore be preferred.

\begin{figure}[ht!] 
	\includegraphics[width=0.475\textwidth]{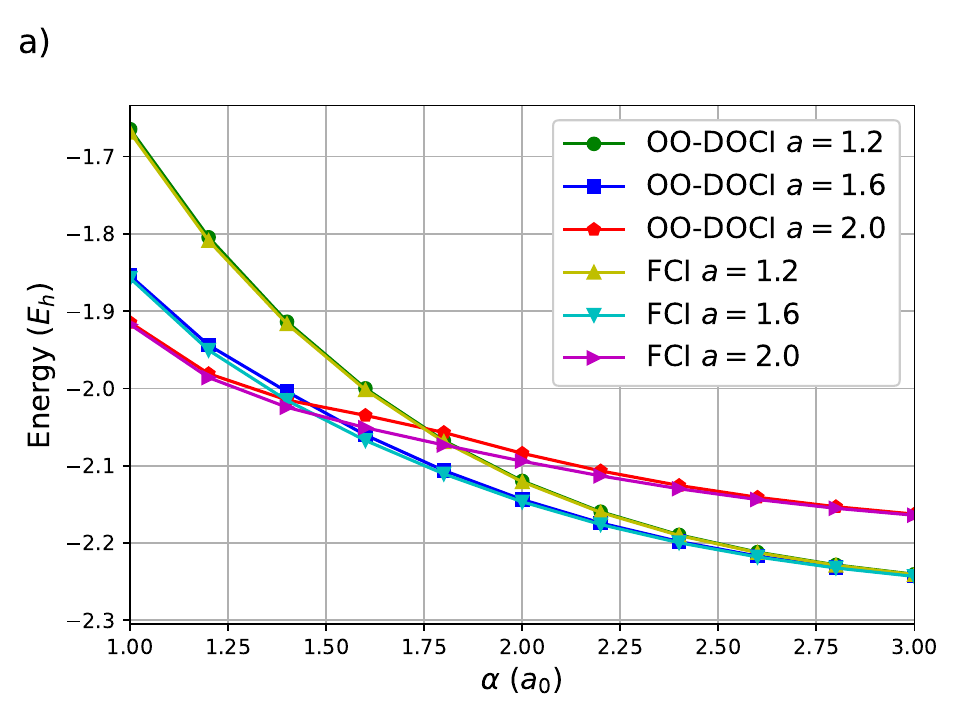} \hfill
	\includegraphics[width=0.475\textwidth]{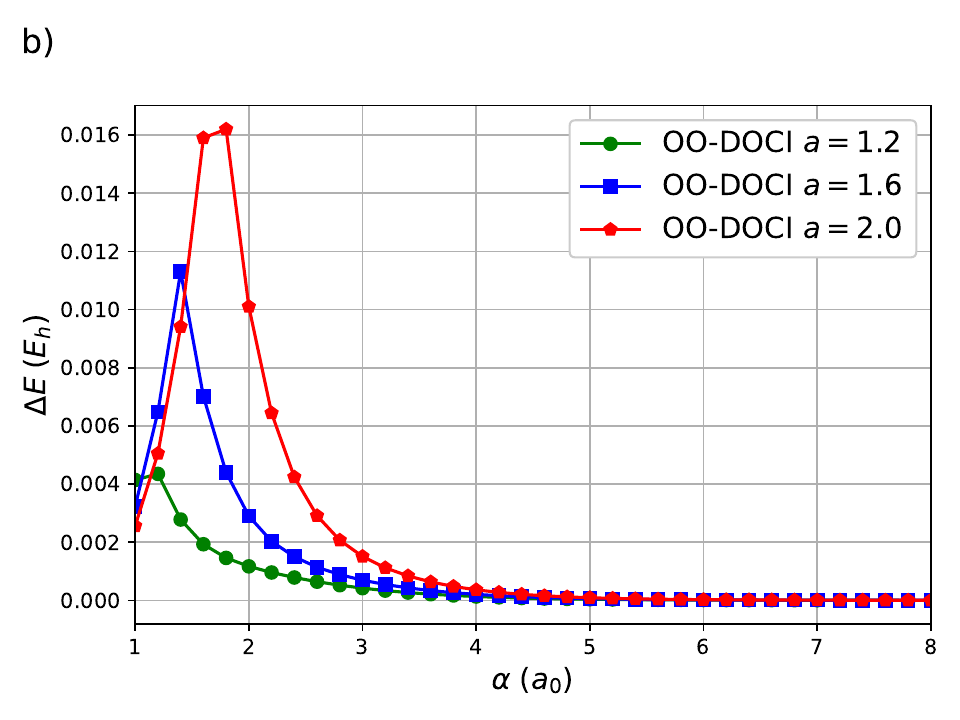}
	\caption{OO-DOCI treatment of Paldus P4 with $a=1.2\;a_0,1.6\;a_0,2.0\;a_0$: (a) FCI and OO-DOCI energies. Past $\alpha = 3.0\;a_0$ the curves become indiscernible. (b) Error of OO-DOCI with respect to FCI. All results computed with the STO-6G basis set in the basis of OO-DOCI orbitals.}
	\label{fig:P4_DOCI_energies}
\end{figure}

\begin{figure}[ht!] 
	\includegraphics[width=0.475\textwidth]{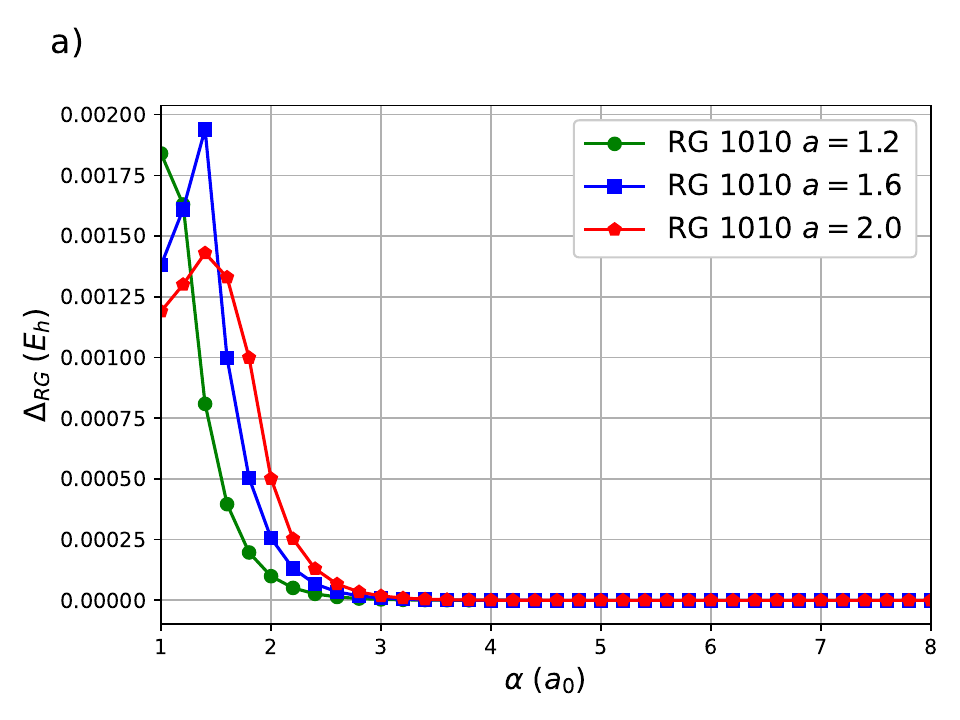} \hfill
	\includegraphics[width=0.475\textwidth]{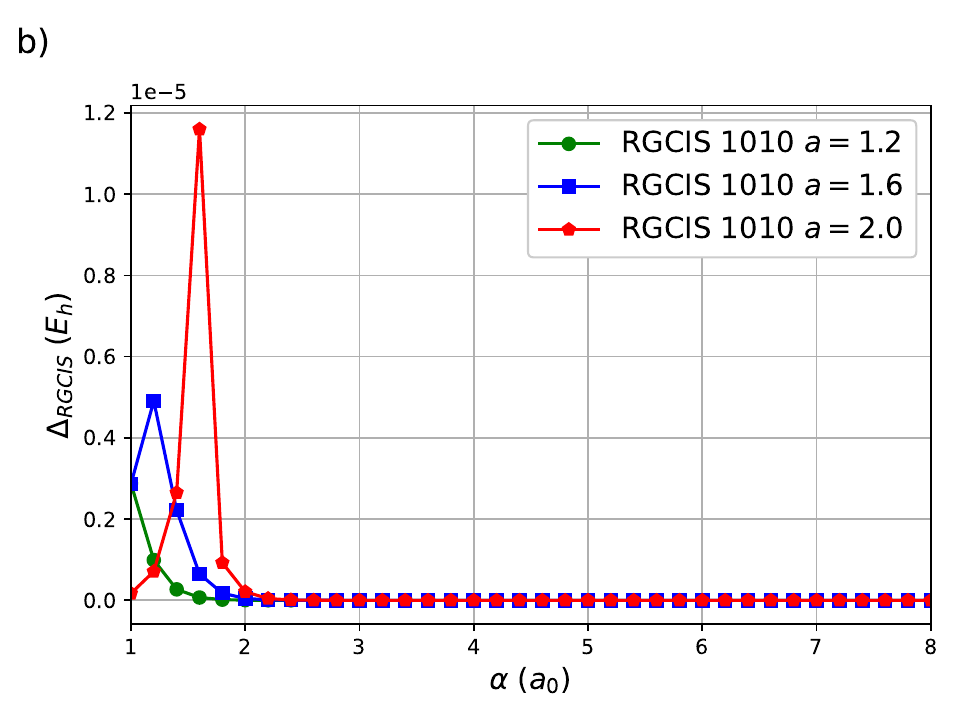}
	\caption{Variational RG and RGCI treatment of Paldus P4 with $a=1.2\;a_0,1.6\;a_0,2.0\;a_0$: (a) $\Delta_{RG}$ for the 1010 RG state. (b) $\Delta_{RGCIS}$ for the 1010 RG state. All results computed with the STO-6G basis set in the basis of OO-DOCI orbitals.}
	\label{fig:P4_RG_energies}
\end{figure}

FCI and OO-DOCI results were computed for P4 with $a=1.2\;a_0,1.6\;a_0,2.0\;a_0$ for a range of $\alpha = [1.0\;a_0,8.0\;a_0]$ and the results are presented in Figure \ref{fig:P4_DOCI_energies}. Generally OO-DOCI agrees with FCI, though there are visible gaps between the respective energies for both $a=1.6\;a_0$ and $a=2.0\;a_0$. This gap arises since the long and short sides of the rectangle switch, and hence the dominant Slater determinant the FCI expansion changes. Valence bonds form along the short sides of the rectangle, and for the square geometry two valence bond descriptions are degenerate.
Figure \ref{fig:P4_RG_energies} depicts deviations between energies from RG (1010) and RGCIS (built from 1010) states and the those from OO-DOCI. The correct RG state is 1010 for all values of $a$ and $\alpha$. The 1100 state was also variationally optimized, but it was found to have much too large an energy (see supporting information). RGCIS is near-exact, with errors with respect to OO-DOCI never exceeding $1.2 \times 10^{-5}$ E$_{\text{h}}$.
The weak correlation functionals do not provide any noticeable improvement upon the 1010 RG state (see supporting information). All of them under-correlate at small $\alpha$ and over-correlate at large $\alpha$.

\begin{figure}[ht!] 
	\includegraphics[width=0.475\textwidth]{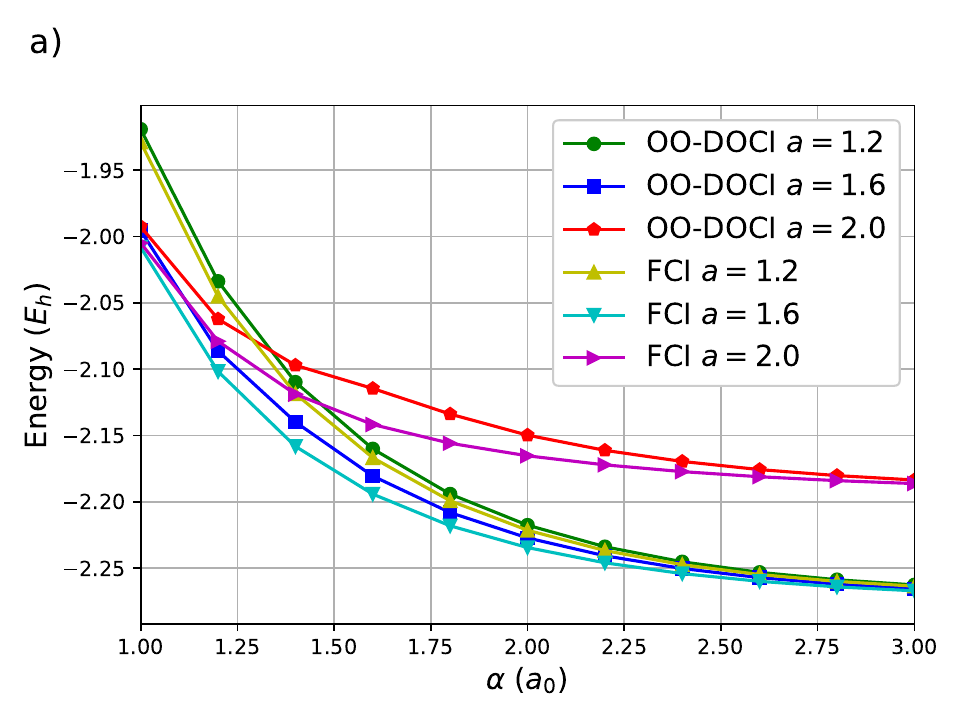} \hfill
	\includegraphics[width=0.475\textwidth]{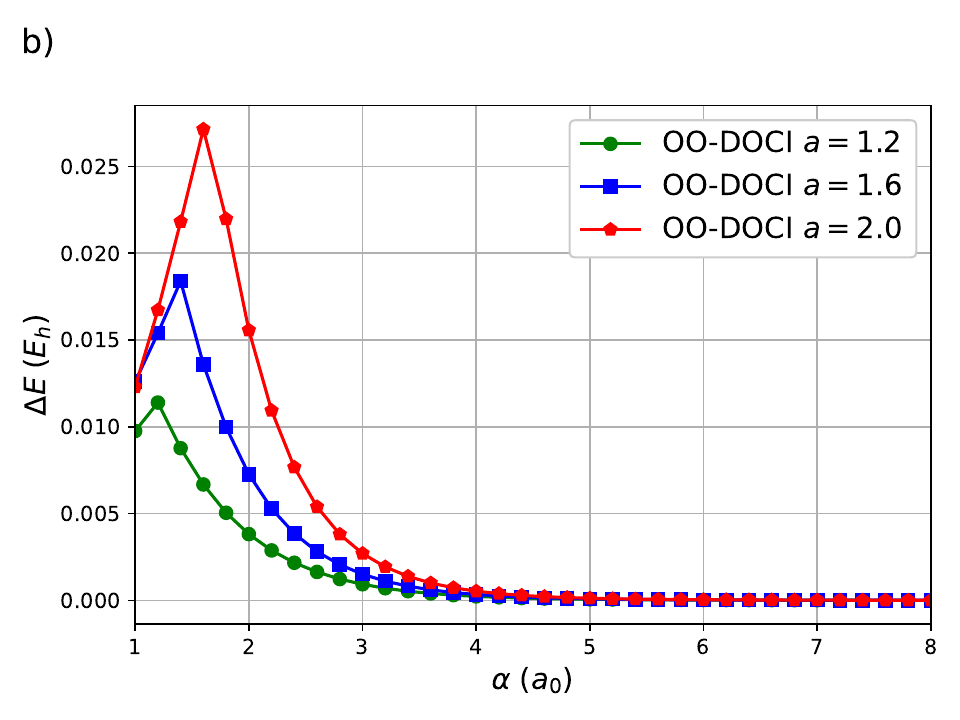}
	\caption{OO-DOCI treatment of Paldus D4 with $a=1.2\;a_0,1.6\;a_0,2.0\;a_0$: (a) FCI and OO-DOCI energies. (b) Error of OO-DOCI with respect to FCI. All results computed with the STO-6G basis set in the basis of OO-DOCI orbitals.}
	\label{fig:D4_DOCI_energies}
\end{figure}

FCI and OO-DOCI results were computed for D4 with $a=1.2\;a_0,1.6\;a_0,2.0\;a_0$ for a range of $\alpha = [1.0\;a_0,8.0\;a_0]$ and the results are presented in Figure \ref{fig:D4_DOCI_energies}. The agreement between OO-DOCI and FCI is comparable to the case of P4, with maximum deviations between the methods being roughly twice as large ($\approx$ 0.03 E$_{\text{h}}$ at 1.6 $a_0$)
This gap occurs for a similar reason as for P4: when $\alpha < a$, the bonding pattern should be centred between the second and third hydrogen atoms as they are closest. Contributions from the seniority two and four sectors are required for quantitative agreement with FCI.

Variational RG calculations capture the transition in the bonding pattern explicitly. When $\alpha < a$, the correct RG state is 1100 with a set of $\{\varepsilon\}$ arranged in a 1-2-1 energetic pattern: 1 small $\varepsilon$, 2 near-degenerate $\varepsilon$, and 1 large $\varepsilon$. This arrangement corresponds to one doubly-occupied orbital, two partially-occupied valence orbitals, and one empty virtual orbital. Once $\alpha \geq a$, the correct RG state is 1010 with a set of $\{\varepsilon\}$ in two sets of near-degenerate pairs. With the correct RG reference, the RGCIS results never differ from OO-DOCI by more than $2\times 10^{-6}E_{\text{h}}$.
\begin{figure}[ht!] 
	\includegraphics[width=0.475\textwidth]{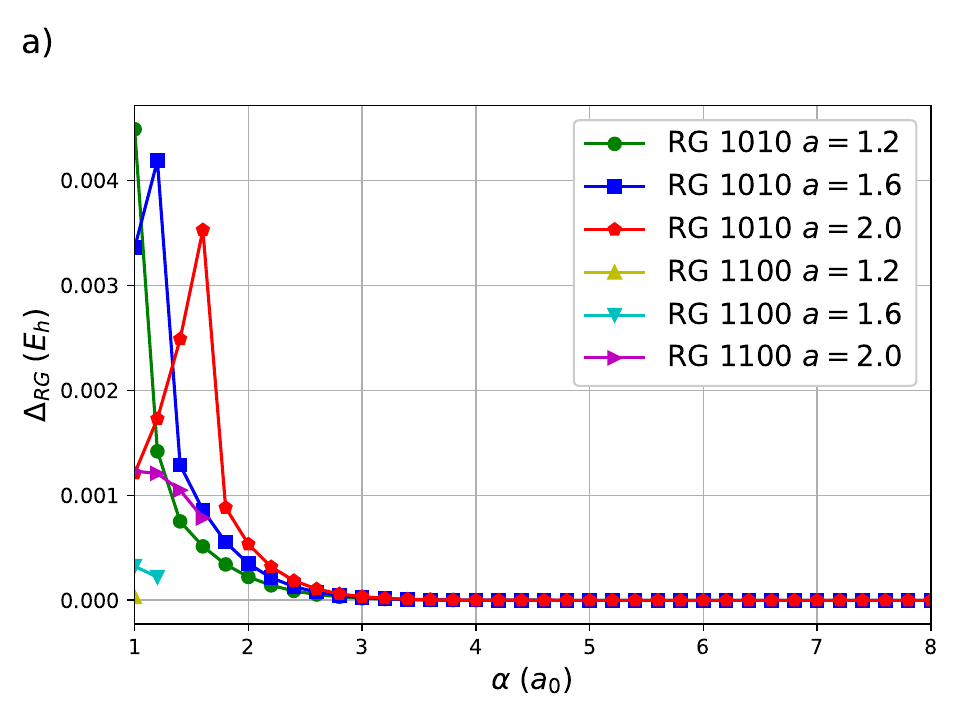} \hfill
	\includegraphics[width=0.475\textwidth]{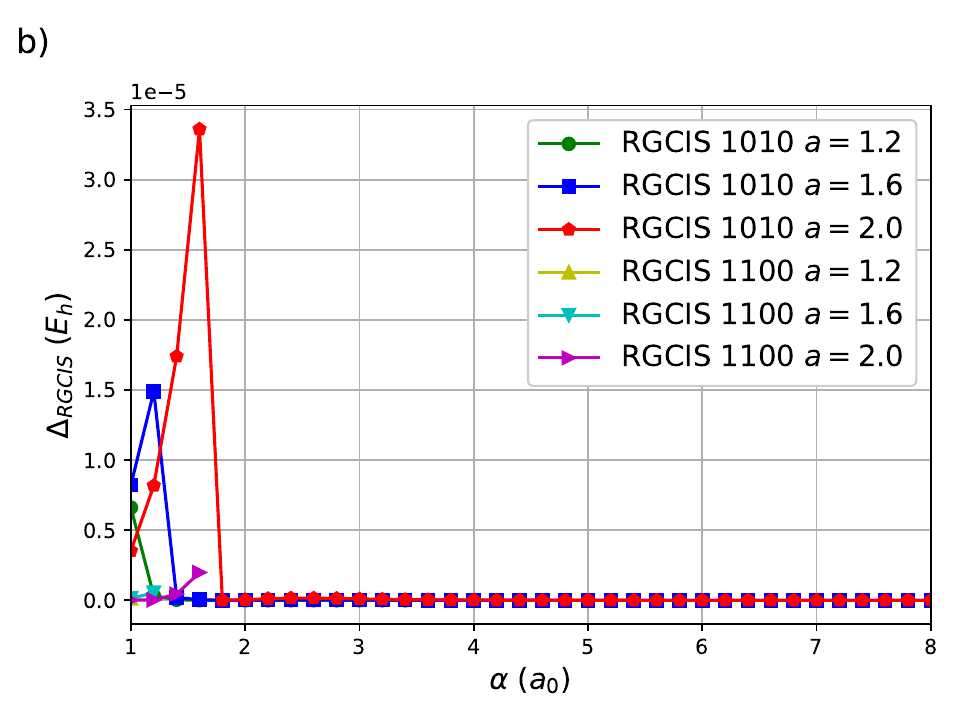}
	\caption{Variational RG and RGCI treatment of Paldus D4 with $a=1.2\;a_0,1.6\;a_0,2.0\;a_0$: (a) $\Delta_{RG}$ for the 1010 and 1100 RG states. (b) $\Delta_{RGCIS}$ for the 1010 and 1100 RG states. 1100 is only optimal at short distances and quickly jumps off the scale once $\alpha > a$. All results computed with the STO-6G basis set in the basis of OO-DOCI orbitals.}
	\label{fig:D4_RG_energies}
\end{figure}
As was the case for P4, the weak-correlation functionals do not provide any useful improvement (see supporting information). At short distances they all under-correlate substantially and at long distances they all over-correlate substantially.

\begin{figure}[ht!] 
	\includegraphics[width=0.475\textwidth]{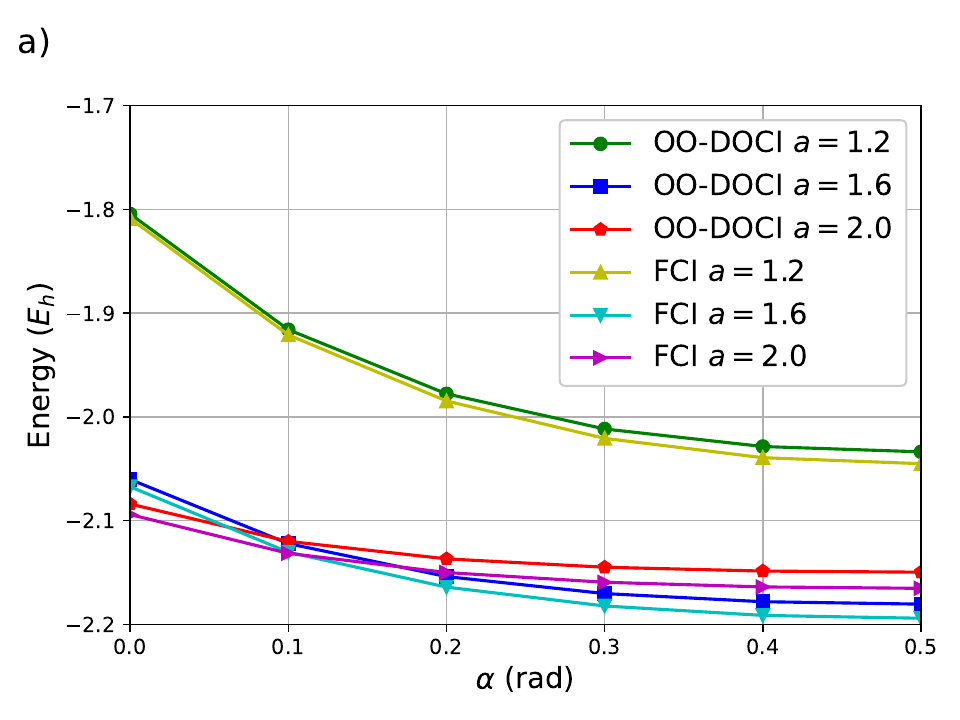} \hfill
	\includegraphics[width=0.475\textwidth]{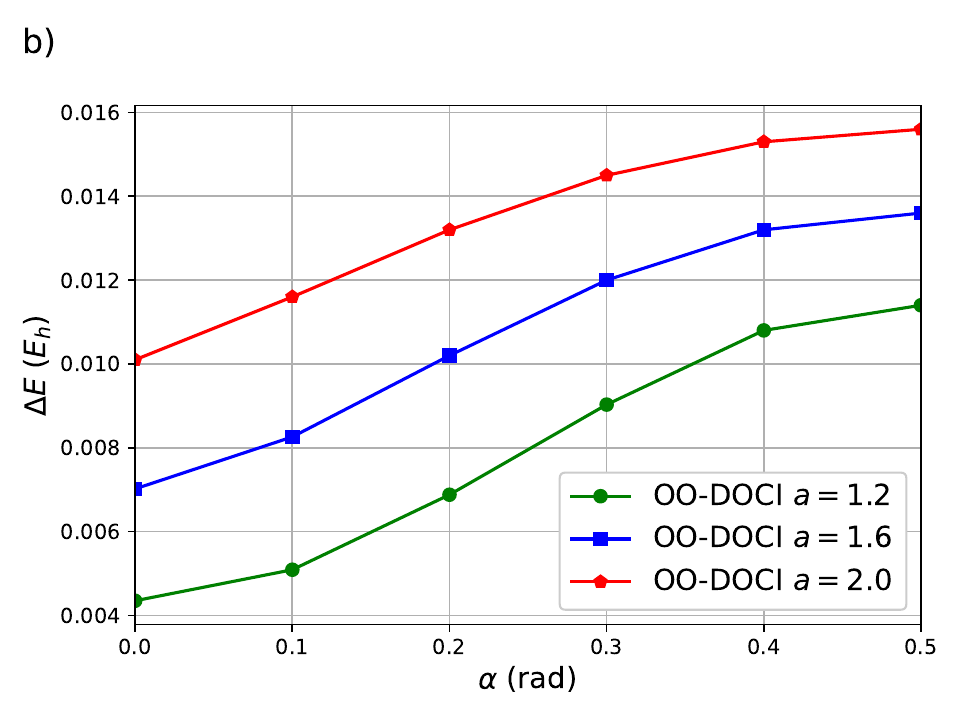}
	\caption{OO-DOCI treatment of Paldus H4 with $a=1.2\;a_0,1.6\;a_0,2.0\;a_0$: (a) FCI and OO-DOCI energies. (b) Error of OO-DOCI with respect to FCI. All results computed with the STO-6G basis set in the basis of OO-DOCI orbitals.}
	\label{fig:H4_DOCI_energies}
\end{figure}

Figure \ref{fig:H4_DOCI_energies} depicts FCI and OO-DOCI potential energy curves for the H4 model with $a=1.2\;a_0,1.6\;a_0,2.0\;a_0$. The OO-DOCI and FCI curves are qualitatively similar though the agreement is not quantitative, which indicates the need for weak correlation contributions from higher seniority sectors.

\begin{figure}[ht!] 
	\includegraphics[width=0.475\textwidth]{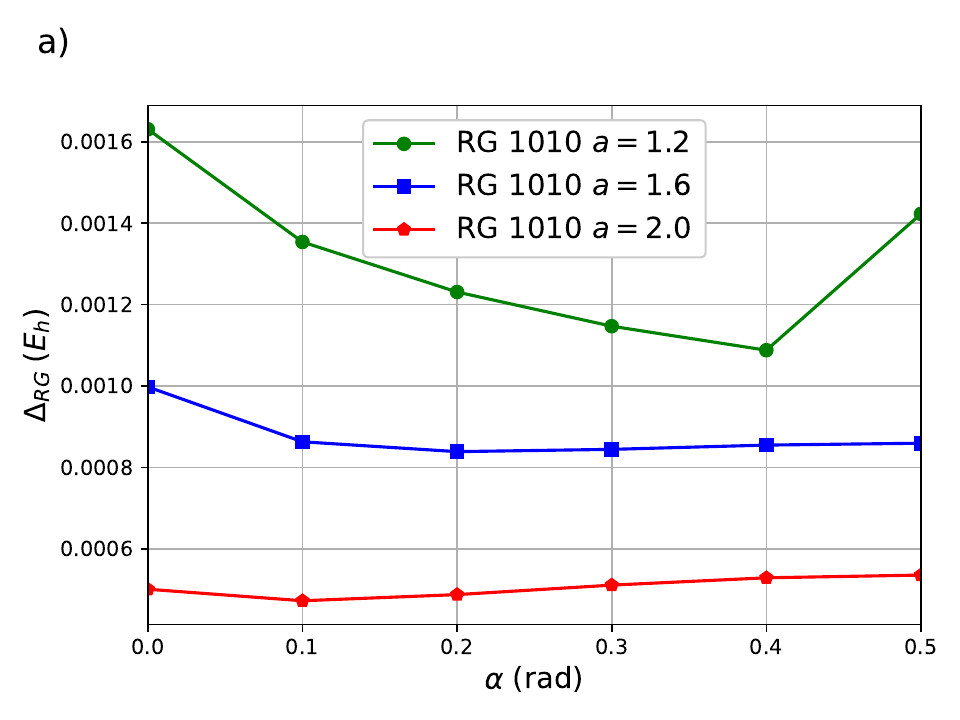} \hfill
	\includegraphics[width=0.475\textwidth]{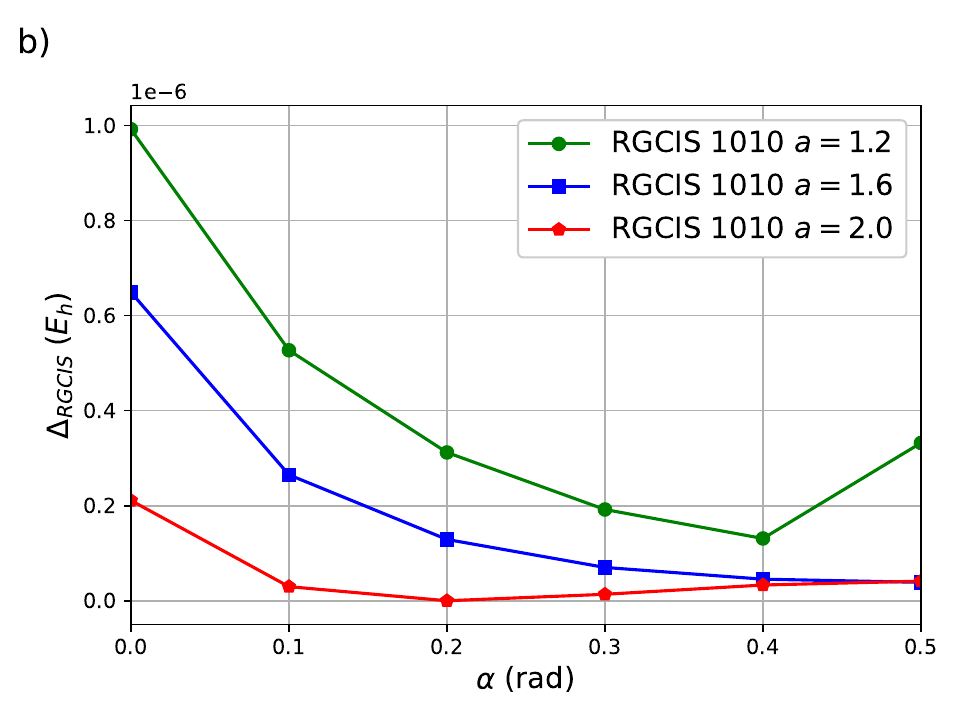}
	\caption{Variational RG and RGCI treatment of H4 with $a=1.2\;a_0,1.6\;a_0,2.0\;a_0$: (a) $\Delta_{RG}$ for the 1010 RG state. (b) $\Delta_{RGCIS}$ for the 1010 RG state. All results computed with the STO-6G basis set in the basis of OO-DOCI orbitals.}
	\label{fig:H4_RG_energies}
\end{figure}

\begin{figure}[ht!] 
	\includegraphics[width=0.475\textwidth]{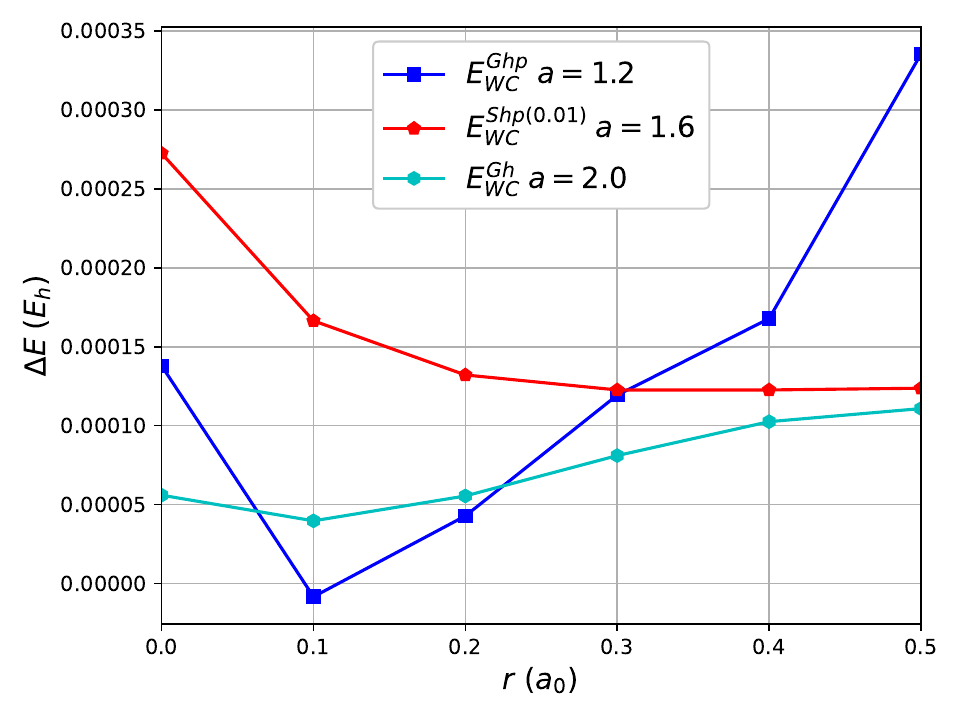}
	\caption{Weak correlation functionals for Paldus H4 with $a=1.2\;a_0,1.6\;a_0,2.0\;a_0$: Difference between $\Delta_{RG}$ and $|E_{WC}|$. 1010 is the optimal RG state. All results computed with the STO-6G basis set in the basis of OO-DOCI orbitals.}
	\label{fig:H4_wc}
\end{figure}

Variational RG calculations, presented in Figure \ref{fig:H4_RG_energies}, show that a 1010 RG state recovers OO-DOCI energies to within roughly 1.6 $\times 10^{-3}$ E$_{\text{h}}$ throughout the curve, and several orders of magnitude of improvement can be obtained with RGCIS.
In this case, the relative performance of the weak correlation functionals is less clear: for each choice of $a$, a different functional was found to be optimal (Figure \ref{fig:H4_wc}). At $a=1.2\;a_0$ $E^{Ghp}_{WC}$ performs the best, at $a=1.6\;a_0$ $E^{Shp}_{WC}$ with a maximum at 0.01 is best, while at $a=2.0\;a_0$ the best performing functional is $E^{Gh}_{WC}$. Results for all eight tested functionals are included in the supporting information. 

\subsection{H$_{10}$ isomers}
\begin{figure}[ht!] 
	\includegraphics[width=0.9\textwidth]{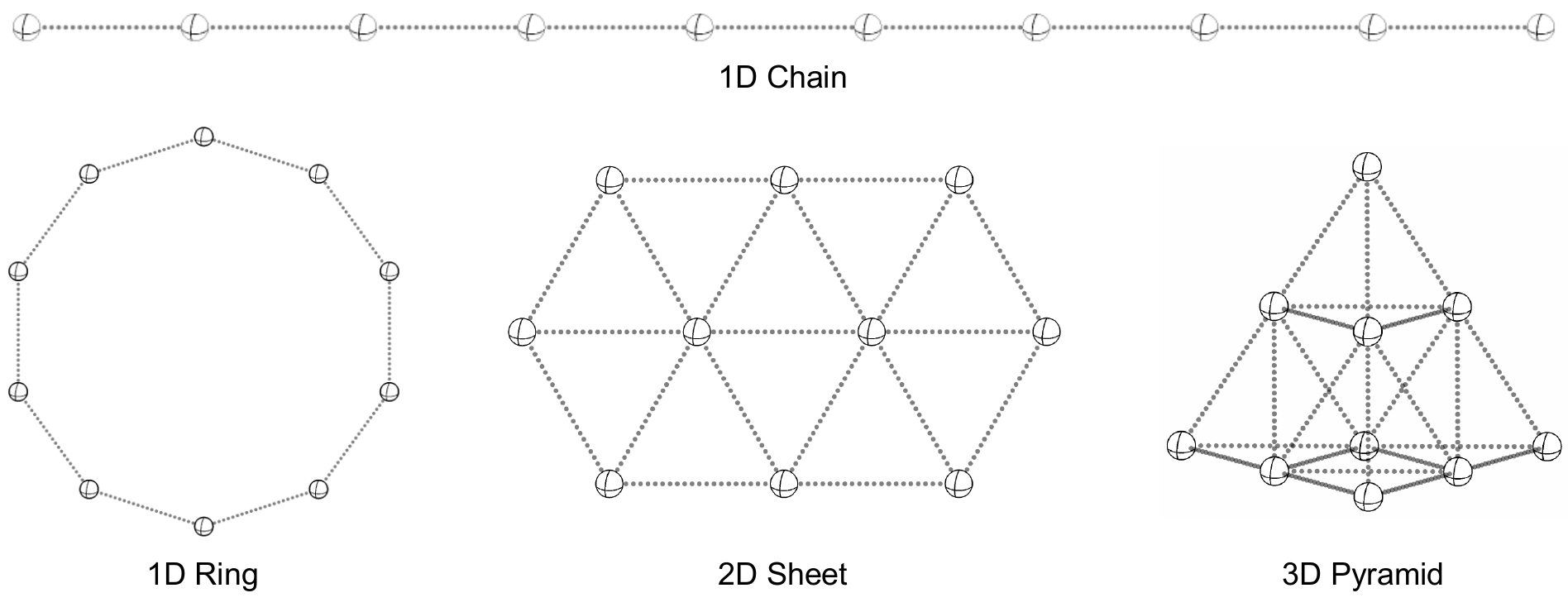}
	\caption{Isomers of H$_{10}$ considered in this work. The dashed lines are of equivalent distance.}
	\label{fig:H10_isomers}
\end{figure}

Recently, Stair and Evangelista presented four isomers of H$_{10}$, shown in Figure \ref{fig:H10_isomers}, to assess not only the accuracy of some common quantum chemistry approaches, but also their ability to provide compact representations of the electronic structure of these complex systems. Each isomer can be thought of as a proxy for a finite-size Hubbard model: the single variable is the inter-atomic distance which modulates on-site repulsion. The chain and the ring are similar to 1-dimensional Hubbard models without and with periodic boundary conditions, respectively. In the sheet system, the hydrogen atoms are arranged in a 3-4-3 pattern, with all of the nearest-neighbours being equidistant. The pyramid is a tetrahedron with four hydrogen atoms at the vertices and the six remaining hydrogen atoms at the midpoint of each of the edges.

\begin{figure}[ht!] 
	\includegraphics[width=0.475\textwidth]{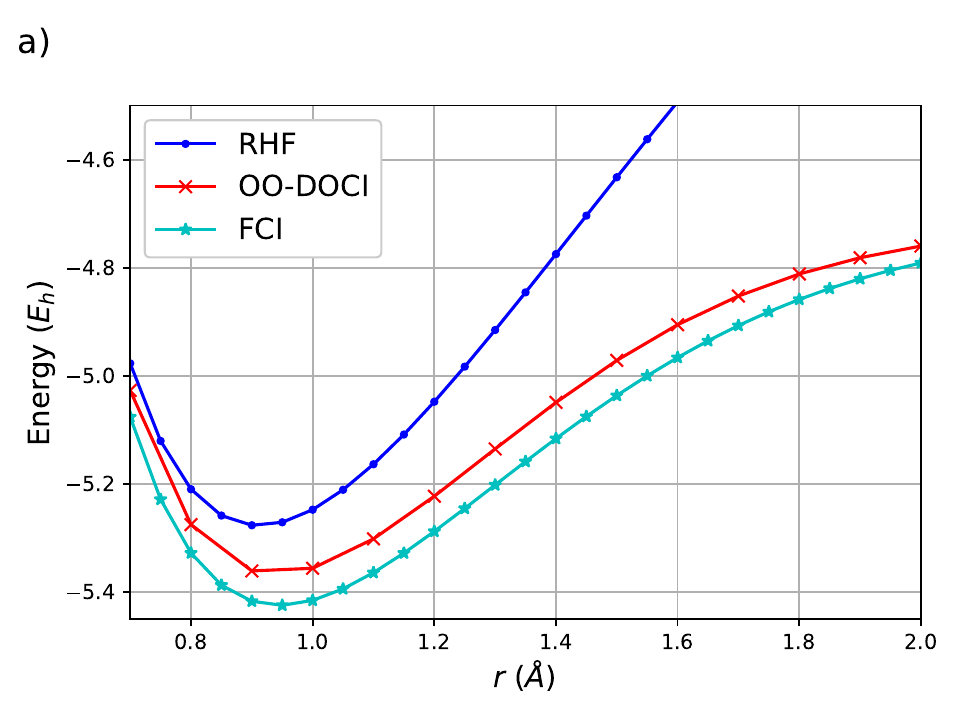} \hfill
	\includegraphics[width=0.475\textwidth]{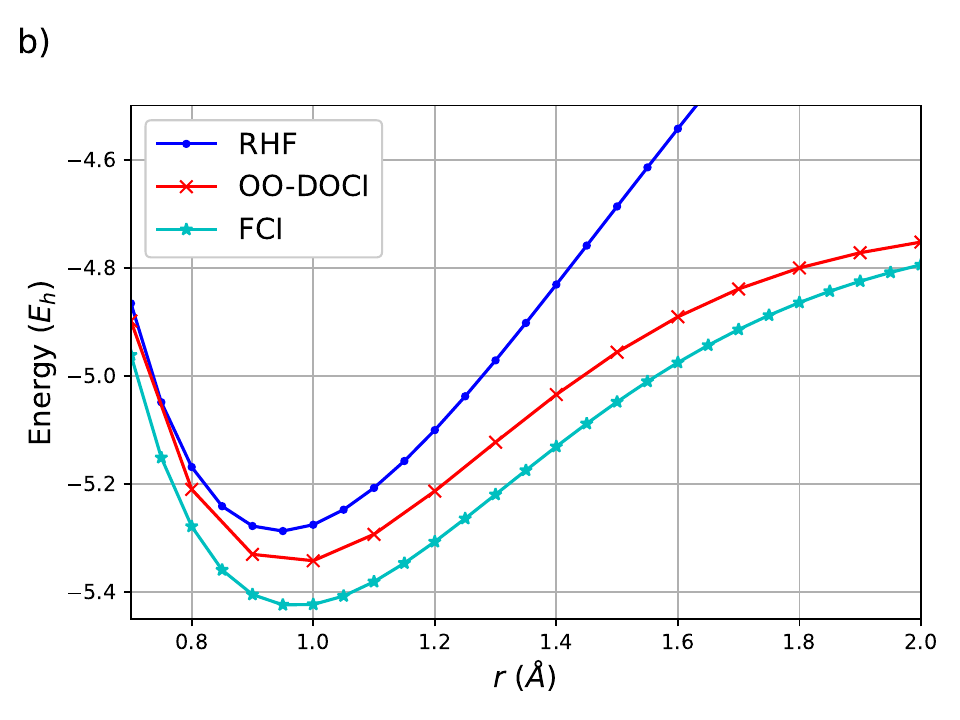} \\
	\includegraphics[width=0.475\textwidth]{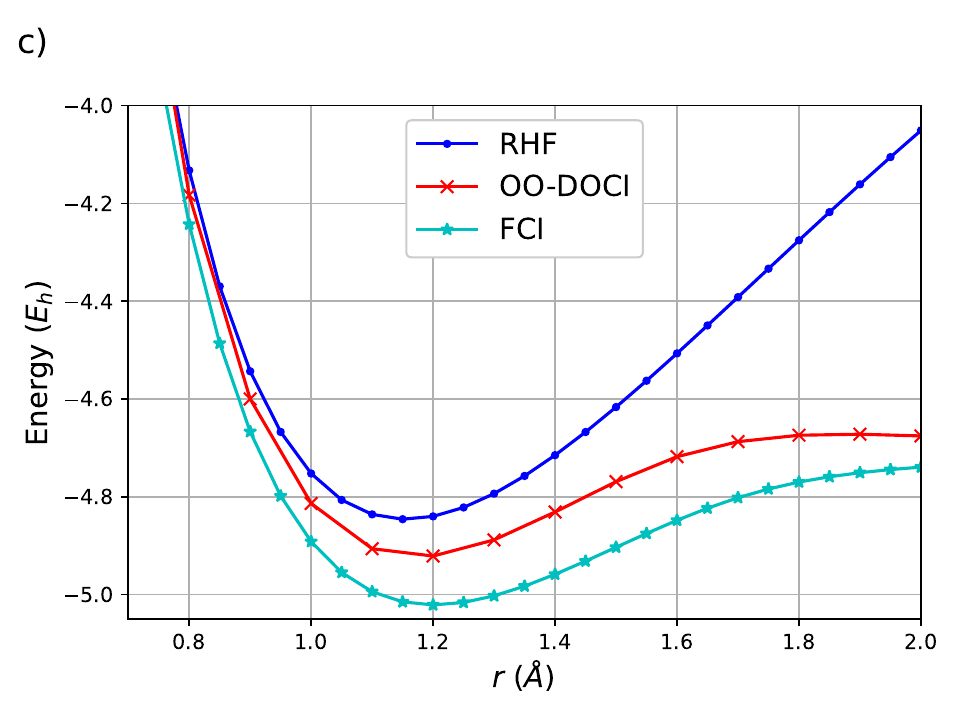} \hfill
	\includegraphics[width=0.475\textwidth]{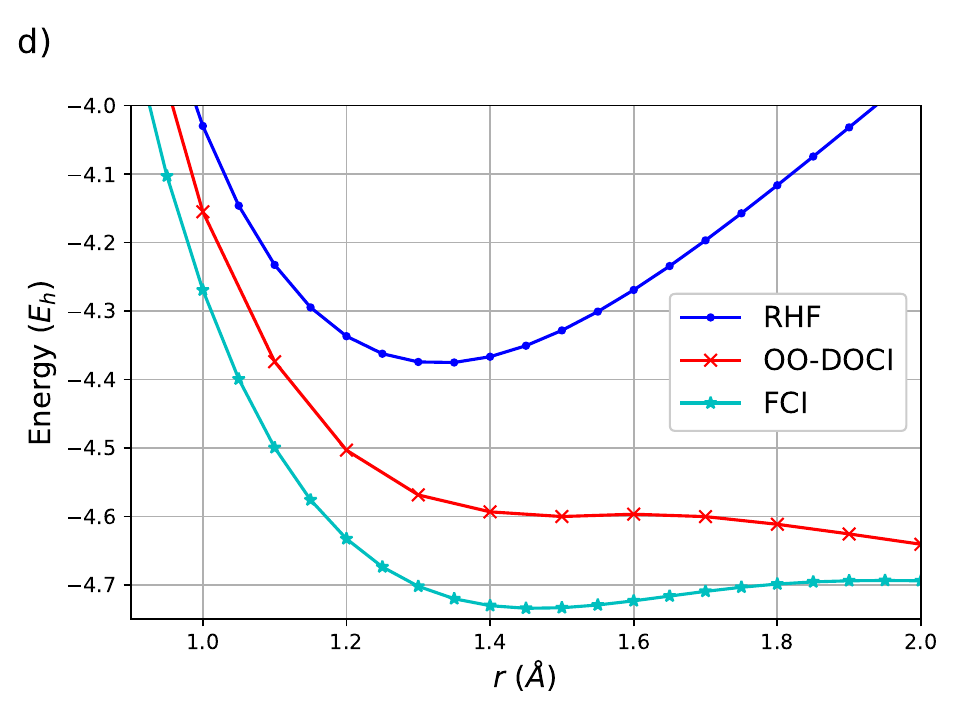}
	\caption{OO-DOCI treatment of H$_{10}$ isomers: RHF, OO-DOCI and FCI energies for (a) 1-dimensional chain. (b) 1-dimensional ring. (c) 2-dimensional sheet. (d) 3-dimensional pyramid. OO-DOCI results computed with the STO-6G basis set in the basis of OO-DOCI orbitals. RHF and FCI results are from ref.\cite{stair:2020}}
	\label{fig:H10_DOCI_energies}
\end{figure}

\begin{figure}[ht!] 
	\includegraphics[width=0.475\textwidth]{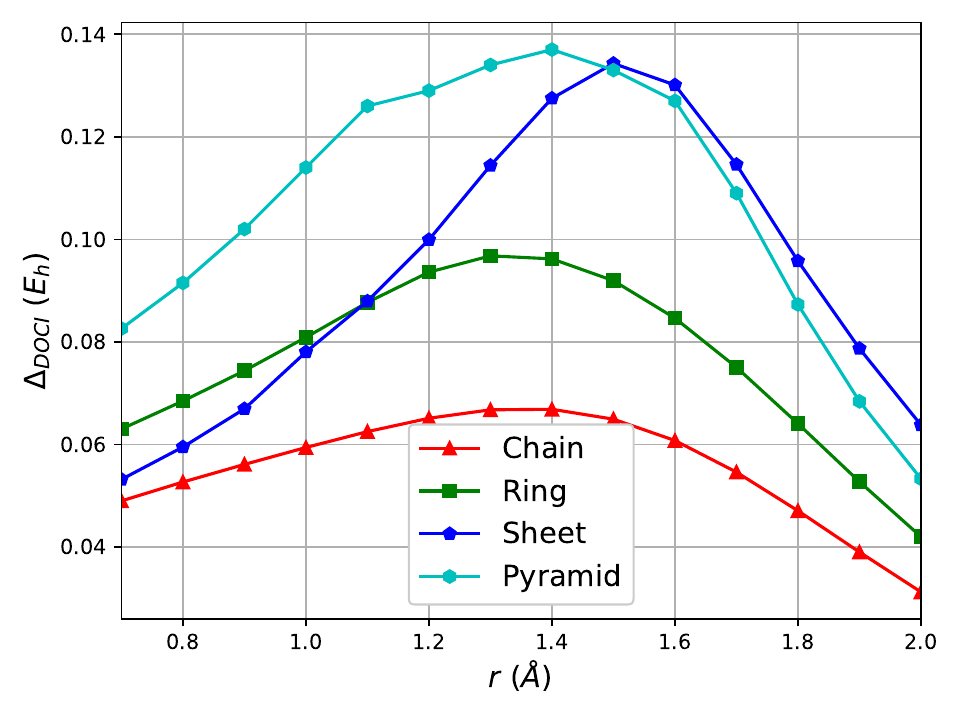}
	\caption{OO-DOCI error with respect to FCI for H$_{10}$ isomers. All results computed with the STO-6G basis set in the basis of OO-DOCI orbitals.}
	\label{fig:H10_DOCI_FCI}
\end{figure}

Figure \ref{fig:H10_DOCI_energies} depicts potential energy curves for the four H$_{10}$ isomers computed at the OO-DOCI / STO-6G level of theory. FCI results computed the same basis, which were taken from Ref.,\cite{stair:2020} are also provided. As expected, OO-DOCI does a reasonable job of reproducing the overall shapes of the FCI curves for both of the 1D structures. As can be seen in Fig. \ref{fig:H10_DOCI_FCI}, the maximum deviations between OO-DOCI and FCI are $\approx 0.67$ E$_{\text{h}}$ and $\approx 0.97$ E$_{\text{h}}$ for the 1D chain and ring structures respectively. For the 2D and 3D structures, the largest deviations between OO-DOCI and FCI energies are somewhat larger; moreover, for these systems, we also begin to see some qualitative differences in the shapes of the respective curves. For the 2D sheet, the curvature at intermediate H--H distances is different than that from FCI. For the 3D pyramid, OO-DOCI predicts only a shallow minimum at roughly the correct H--H distance, after which it exhibits a small hump before it appears to approach the correct dissociation limit. Even so, when considering additional correlation treatments, OO-DOCI should be a much better starting point than restricted Hartree-Fock (RHF). We now show that OO-DOCI can itself be very well approximated by a single RG state in all cases.

\subsubsection{1D structures}
\begin{figure}[ht!] 
	\includegraphics[width=0.475\textwidth]{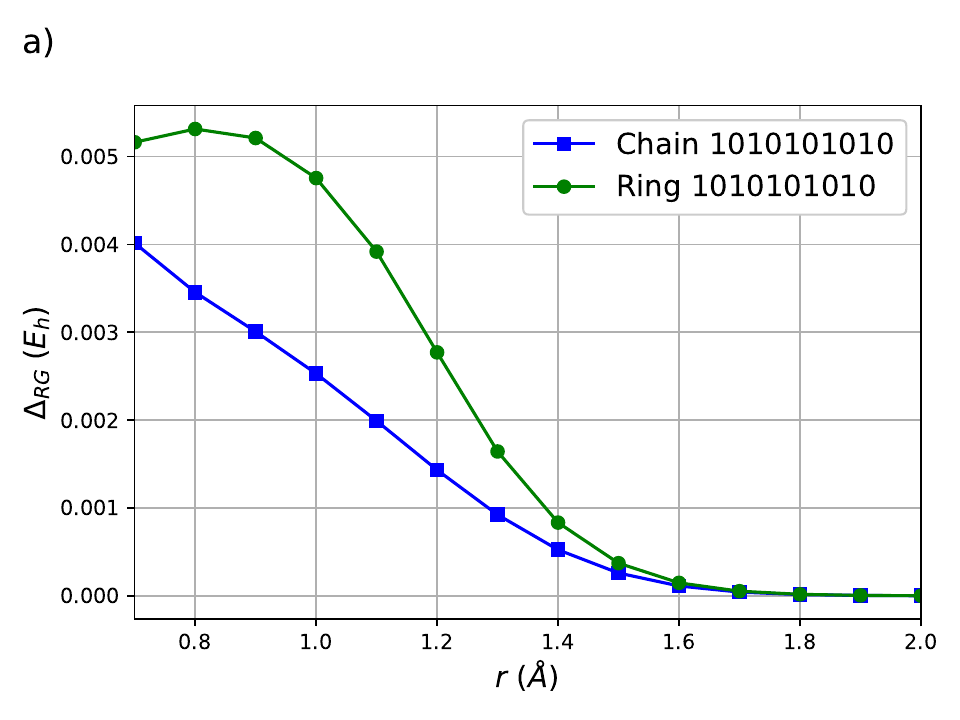} \hfill
	\includegraphics[width=0.475\textwidth]{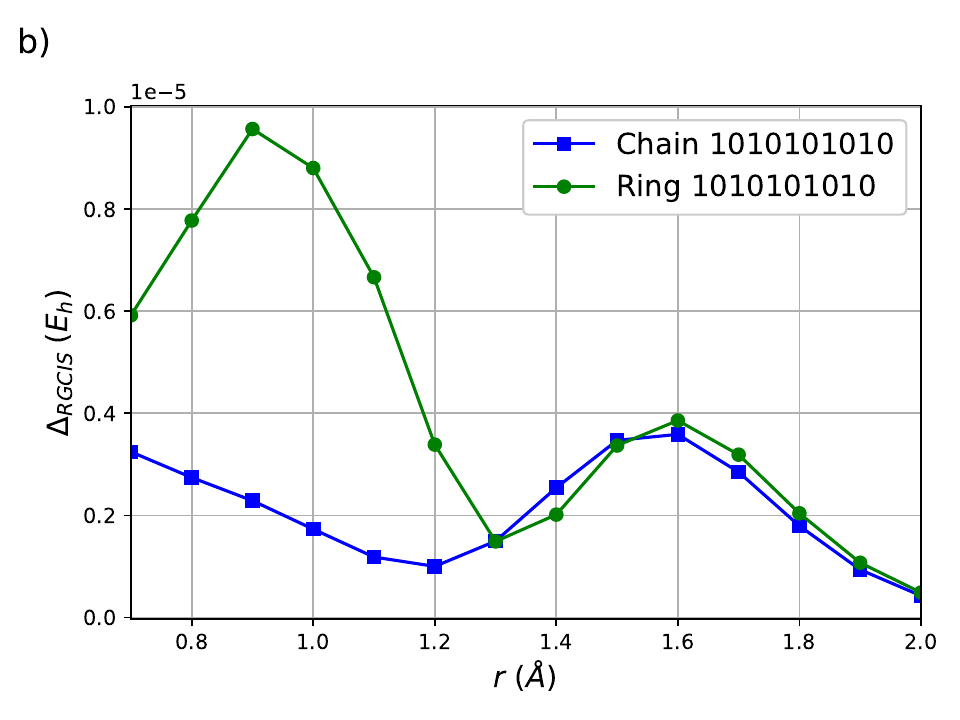}
	\caption{Variational RG and RGCIS treatment of H$_{10}$ chain and ring: (a) $\Delta_{RG}$ for the 1010101010 RG state. (b) $\Delta_{RGCIS}$ for the 1010101010 RG state. All results computed with the STO-6G basis set in the basis of OO-DOCI orbitals.}
	\label{fig:H10_1d_rgcis}
\end{figure}

The linear chain is exactly the same as those we have already studied, except that the chain is one GVB pair longer. Hence, the optimal RG state is the N\'{e}el RG state (1010101010). The error in the RG energy with respect to that from OO-DOCI is shown in Figure \ref{fig:H10_1d_rgcis} (a). The maximum deviation from the OO-DOCI energy ($\approx$4 $\times 10^{-3}$ E$_{\text{h}}$) occurs at the most compressed geometry considered, and the RG state effectively recovers the OO-DOCI energy at stretched geometries. As seen in Fig. \ref{fig:H10_1d_rgcis} (b), the additional correlation provided by RGCIS brings the error with respect to OO-DOCI down several orders of magnitude in the equilibrium region. The errors in the RGCIS energy are all less than 4 $\times 10^{-6}$ E$_{\text{h}}$ throughout the entire curve. To the precision we can trust our results, RGCISD is essentially indiscernible from OO-DOCI (see supporting information).

The ring is a finite size ``periodic'' 1D structure. We expect that the relative performance of RG and OO-DOCI should be of similar quality to the case of the chain structure, and, indeeed, it is: the optimal RG state is the N\'{e}el RG state and the error in its energy with respect to OO-DOCI, shown in Figure \ref{fig:H10_1d_rgcis} (a), is only slightly larger than for the chain at compressed geometries and quite similar at stretched geometries. Again, the additional correlation afforded by RGCIS brings the error down by several orders of magnitude (Fig. \ref{fig:H10_1d_rgcis} (b)).

\begin{figure}[ht!] 
	\includegraphics[width=0.475\textwidth]{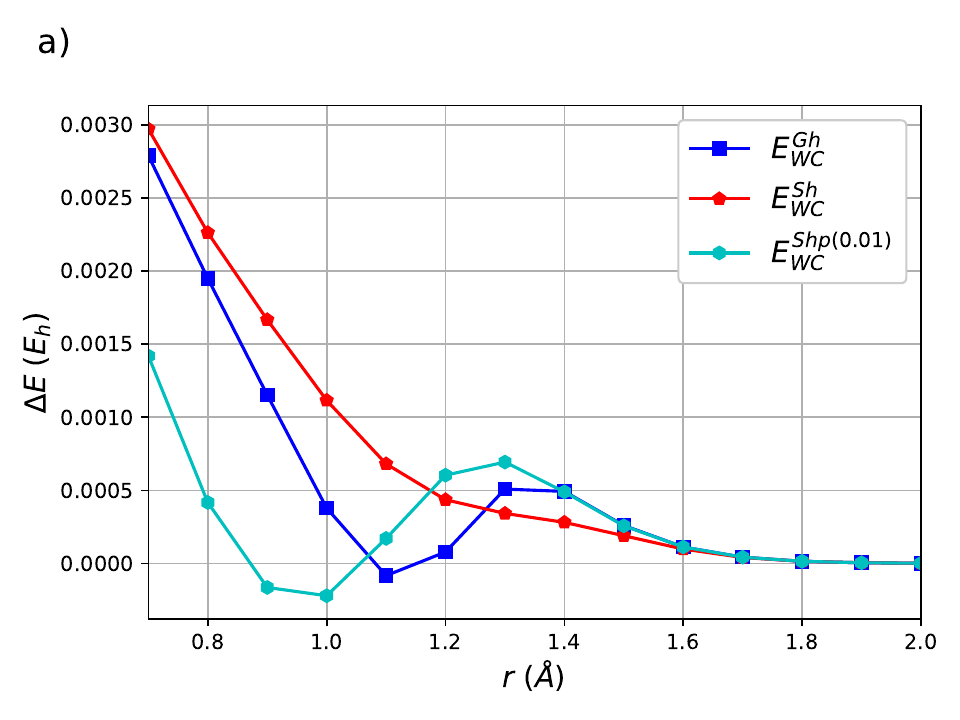} \hfill
	\includegraphics[width=0.475\textwidth]{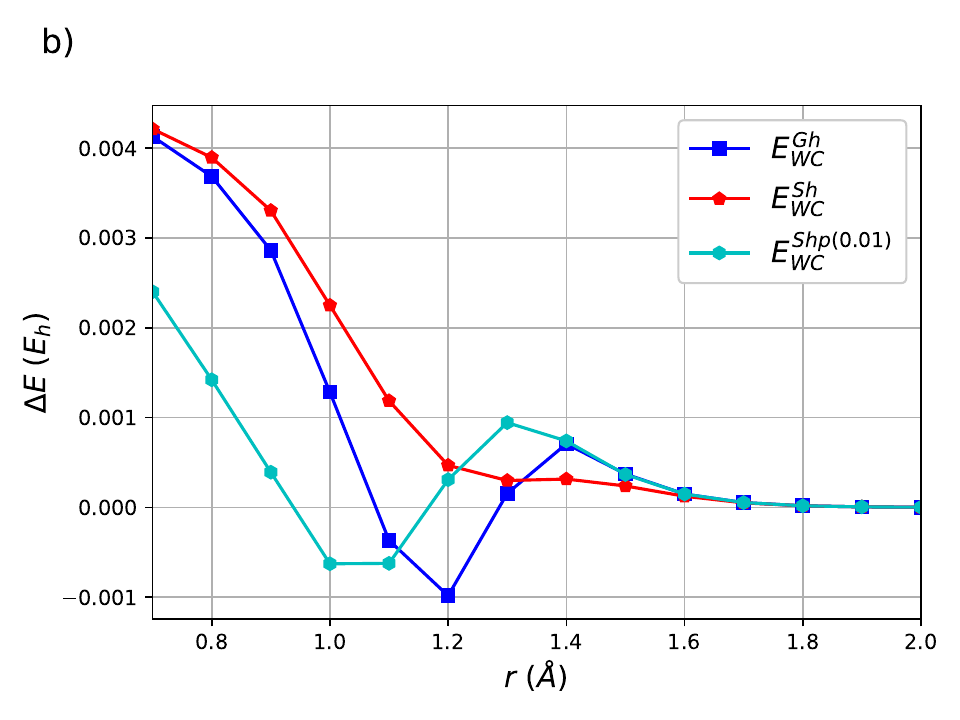}	
	\caption{Weak correlation functionals for 1D H$_{10}$ isomers: (a) Difference between $\Delta_{RG}$ and $|E_{WC}|$ for 1D chain. (b) Difference between $\Delta_{RG}$ and $|E_{WC}|$ for 1D ring. 1010101010 is the optimal RG state. All results computed with the STO-6G basis set in the basis of OO-DOCI orbitals.}
	\label{fig:H10_1d_wc}
\end{figure}

Corrections from the weak correlation functionals, shown in Figure \ref{fig:H10_1d_wc}, follow the same pattern as for linear H$_8$: $E^{Gh}_{WC}$ and $E^{Sh}_{WC}$ are both reasonable, though $E^{Sh}_{WC}$ should be preferred as it never over-correlates and its error with respect to OO-DOCI decays monotonically.

\subsubsection{Sheet and pyramid}
\begin{figure}[ht!] 
	\includegraphics[width=0.475\textwidth]{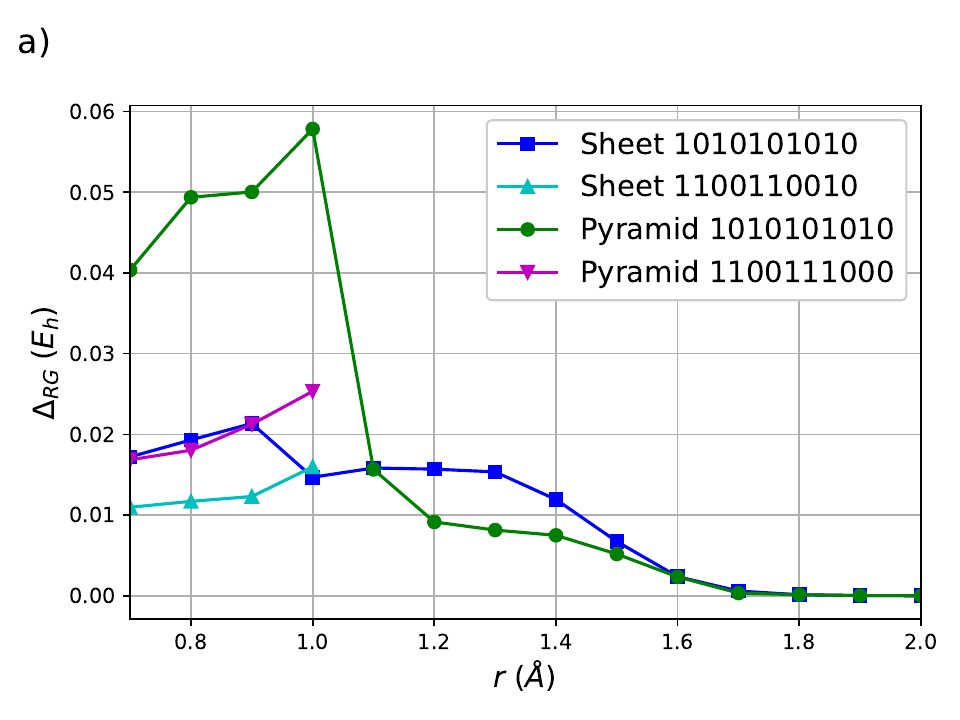} \hfill
	\includegraphics[width=0.475\textwidth]{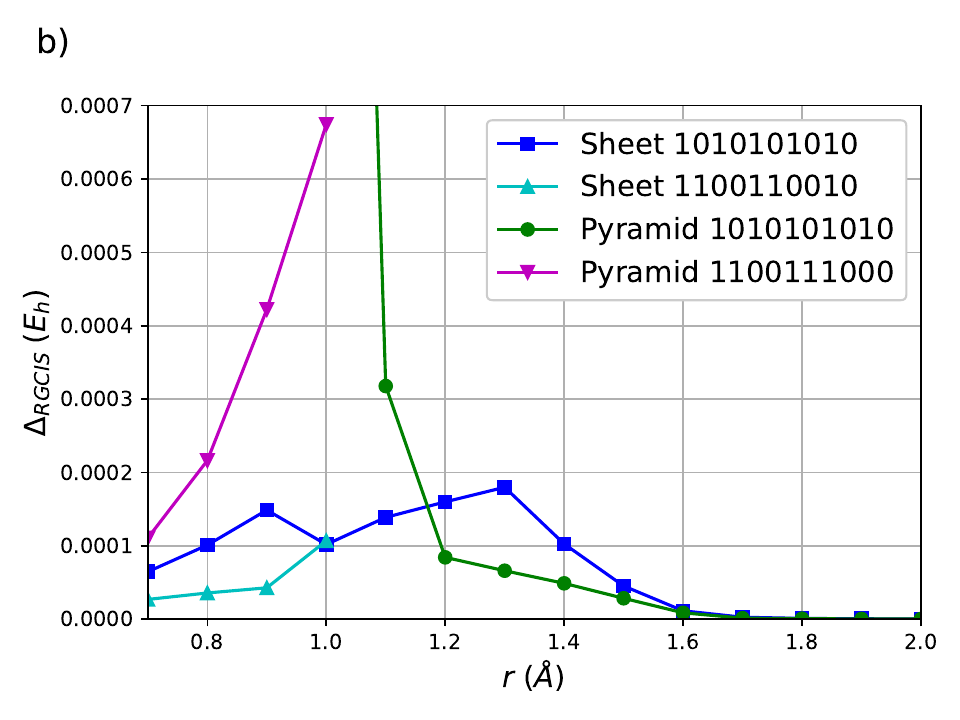}
	\caption{Variational RG and RGCIS treatment of H$_{10}$ sheet and pyramid: (a) $\Delta_{RG}$ for the 1010101010, 1100110010 (sheet) and 1100111000 (pyramid) RG states. (b) $\Delta_{RGCIS}$ for the 1010101010, 1100110010 (sheet) and 1100111000 (pyramid) RG states. Past 1.1, the curves for 1100110010 (sheet) and 1100111000 (pyramid) jump off the graph. All results computed with the STO-6G basis set in the basis of OO-DOCI orbitals.}
	\label{fig:H10_nd_rgcis}
\end{figure}

The sheet is the first case where there is an appreciable qualitative difference between OO-DOCI and FCI. We also found that the optimal variational RG states \emph{change} in this case. As can be seen in Fig. \ref{fig:H10_nd_rgcis} (a), the N\'{e}el RG state is optimal everywhere except at compressed geometries. 
For H--H distances less than or equal to 1.0 \AA, we found that the optimal RG state is instead 1100110010, for which the single-particle energies $\{\varepsilon\}$ arrange themselves in a 4-4-2 pattern: there are 4 orbitals that contain 2 pairs, another 4 orbitals that contain another 2 pairs, and 2 orbitals that contain 1 pair. This structure is perhaps more clear from the corresponding geminal coefficient matrix
\begin{align}
G_{ai} = \frac{1}{u_a - \varepsilon_i} \approx
\begin{pmatrix}
* & * & * & * & 0 & 0 & 0 & 0 & 0 & 0 \\
* & * & * & * & 0 & 0 & 0 & 0 & 0 & 0 \\
0 & 0 & 0 & 0 & * & * & * & * & 0 & 0 \\
0 & 0 & 0 & 0 & * & * & * & * & 0 & 0 \\
0 & 0 & 0 & 0 & 0 & 0 & 0 & 0 & * & * 
\end{pmatrix}
\end{align}
which is block diagonal. Again, the elements in the off-diagonal blocks are not numerically zero, but they are much smaller than those in the diagonal blocks. Remarkably, this state does not seem to be describable as GVB/APSG, for which the matrix $G_{ai}$ is \eqref{eq:gvb_matrix}, nor AP1roG/pCCD, for which $G_{ai}$ is
\begin{align}
G_{ai} = 
\begin{pmatrix}
1 & 0 & 0 & 0 & 0 & * & * & * & * & * \\
0 & 1 & 0 & 0 & 0 & * & * & * & * & * \\
0 & 0 & 1 & 0 & 0 & * & * & * & * & * \\
0 & 0 & 0 & 1 & 0 & * & * & * & * & * \\
0 & 0 & 0 & 0 & 1 & * & * & * & * & * 
\end{pmatrix}.
\end{align}
For GVB/APSG and AP1roG/pCCD, the elements marked ``0'' in $G_{ai}$ are identically zero. 

As for the 1D chain and ring structures, RGCIS improves upon the single, optimal RG state, decreasing the energy error by several orders of magnitude (Fig. \ref{fig:H10_nd_rgcis} (b)). However, in this case, the RGCIS error is somewhat larger at compressed geometries. The maximum error is $\approx 2 \times 10^{-4}$ E$_{\text{h}}$, at an H--H distance of 1.3 \AA, but this error decreases rapidly (as does that for the single RG state) at larger H--H distances. RGCISD, with the correct RG reference, has a maximum error of $\approx 1 \times 10^{-6}$ E$_{\text{h}}$ at 1.3 \AA\; (see supporting information).

Variational RG minimization for the 3D pyramid, shown in Figure \ref{fig:H10_nd_rgcis} (a), again found the N\'{e}el RG state to be optimal, except at short H--H distances where the optimal RG state found was 1100111000. In this case, the single-particle energies $\{\varepsilon\}$ arrange themselves in a 4-6 pattern: there are 4 sites that contain 2 pairs and 6 sites that contain the remaining 3. The corresponding geminal coefficient matrix has the form
\begin{align}
G_{ai} = \frac{1}{u_a - \varepsilon_i} \approx
\begin{pmatrix}
* & * & * & * & 0 & 0 & 0 & 0 & 0 & 0 \\
* & * & * & * & 0 & 0 & 0 & 0 & 0 & 0 \\
0 & 0 & 0 & 0 & * & * & * & * & * & * \\
0 & 0 & 0 & 0 & * & * & * & * & * & * \\
0 & 0 & 0 & 0 & * & * & * & * & * & * 
\end{pmatrix}.
\end{align}
Again, the elements labelled ``0'' are not numerically zero, but they are much smaller than those labelled ``*''. This state also does not appear to be describable as GVB/APSG nor AP1roG/pCCD. As seen in the other systems, the energies obtained from RGCIS improve significantly upon those from a single RG state, although the RGCIS error is somewhat larger for the pyramid than for the sheet at short H--H distances. Nonetheless, the RGCIS error never exceeds $7 \times 10^{-4}$ E$_{\text{h}}$, and the error decreases rapidly (as does that for the single RG state) at larger distances. RGCISD, with the correct RG reference, has a maximum error of $\approx 3 \times 10^{-6}$ E$_{\text{h}}$ at 1.0 \AA (see supporting information).

\begin{figure}[ht!] 
	\includegraphics[width=0.475\textwidth]{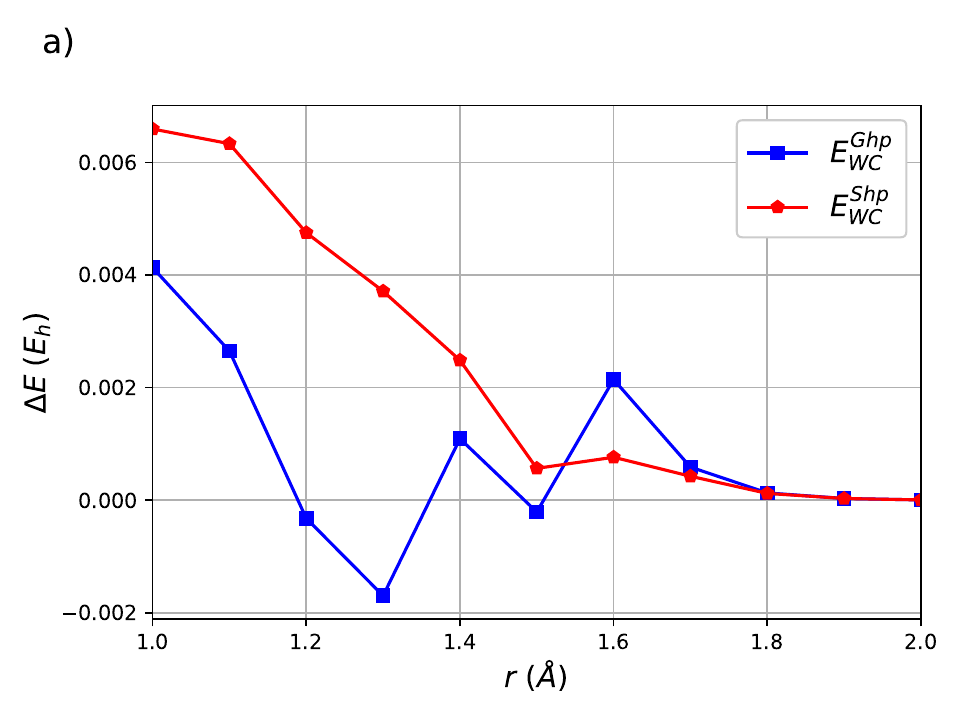} \hfill
	\includegraphics[width=0.475\textwidth]{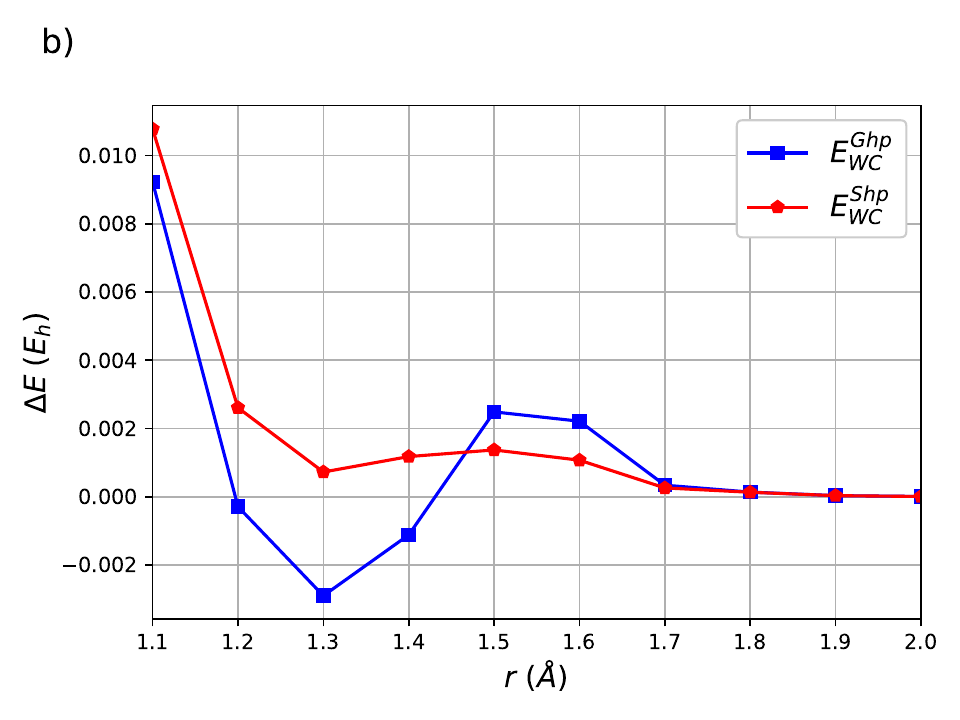}
	\caption{Weak correlation functionals for H$_{10}$ sheet and pyramid: (a) Differences between $\Delta_{RG}$ and $|E_{WC}|$ for 2D sheet. (b) Differences between $\Delta_{RG}$ and $|E_{WC}|$ for 3D pyramid. Results are only computed for the region where 1010101010 is the optimal RG state. All results computed with the STO-6G basis set in the basis of OO-DOCI orbitals.}
	\label{fig:H10_nd_wc}
\end{figure}

The best performing weak correlation functionals, shown in Figure \ref{fig:H10_nd_wc}, are different than for the 1D structures. $E^{Ghp}_{WC}$ and $E^{Shp}_{WC}$ are now the best for both the 2D sheet and the 3D pyramid. $E^{Shp}_{WC}$ should be preferred as its error with respect to OO-DOCI is much better behaved.

A discussion of the correct choice of RG state is warranted. Given that there are $\binom{10}{5}$ possible RG states for 5 pairs in 10 orbitals, how can we be sure we have the correct one? With a few observations we can reduce this number to one that is completely manageable. First, we have observed that the single-particle energies $\{\varepsilon\}$ tend to separate into partitions of $2k$ elements for $k$ pairs. Second, within a partition we always want the 1s listed first, as otherwise the pairs will be placed principally in higher energy orbitals which leads to \emph{much} higher energies. Third, the order of the partitions in a bitstring does not matter, e.g. the bitstrings 1100111000 and 1110001100 refer to different RG states for a given set of $\{\varepsilon\}$ and $g$, but when optimized will find equivalent solutions. In this contribution we found results for 1100111000 first. Thus, the only RG states that must be considered are in one-to-one correspondence with the partitions of the number of pairs:
\begin{align}
(5) &\mapsto 1111100000 \nonumber \\
(4,1) &\mapsto 11110000\;10 \nonumber \\
(3,2) &\mapsto 111000\;1100 \nonumber \\
(3,1,1) &\mapsto 111000\;10\;10 \nonumber \\
(2,2,1) &\mapsto 1100\;1100\;10 \nonumber \\
(2,1,1,1) &\mapsto 1100\;10\;10\;10 \nonumber \\
(1,1,1,1,1) &\mapsto 10\;10\;10\;10\;10.
\end{align}
Finally, symmetry arguments can reduce this list to a few possible candidates. In the pyramid, the four vertices are equivalent and the six edges are equivalent, so it is not surprising that 1100111000 is the optimal RG state at short distances. In the sheet, the two inner hydrogen atoms are equivalent and the four corners are equivalent. It is not obvious whether the other four hydrogen atoms are equivalent or not, and, thus it is necessary to try all bitstrings containing partitions (2,1). These rules are meant to rationalize our results. In this study many more possible RG states were tested, but none fell outside of these observations. These arguments notwithstanding, our experience is that, at large H--H separations distances, the N\'{e}el RG state is \emph{always} optimal.

\section{Conclusion}
We have computed OO-DOCI wavefunctions and RG states for hydrogen clusters containing four, eight, and ten hydrogen atoms. With the exception of the pyramid structure for H$_{10}$, OO-DOCI captures the correct qualitative behavior of the energy as a function of H--H distance in all cases. OO-DOCI itself has been shown to be well approximated by a single RG state, while RGCIS significantly improves this approximation. RGCISD states are effectively indiscernible from OO-DOCI states, in terms of the energy. Hence, when the electronic structure of these systems is represented in terms of RG states, rather than Slater determinants, these systems all appear to be effectively single-reference. Given the good performance of RGCIS and RGCISD, it is worth pursuing improved expressions of the TDM elements, as the expressions reported herein are expensive to evaluate but are certainly improvable.  

As an alternative to RGCIS or RGCISD, we explored the utility of weak correlation functionals within the seniority-zero channel, which are computationally inexpensive to evaluate. For 1D structures we found that the best performance was obtained for the $E^{Sh}_{WC}$ functional, while for 2D and 3D structures we best-performing functional was $E^{Shp}_{WC}$.  In general, the Slater type functionals we considered are to be preferred over Gaussian type as their error with respect to OO-DOCI is better behaved. The exception is the Paldus ring-opening system (H4) for which $E^{Ghp}_{WC}$ was the best for $a=1.2a_0$ while $E^{Gh}_{WC}$ was the best at longer bond lengths.

Regardless of the route chosen to recover the full energy within the seniority-zero sector, quantitative agreement with FCI will ultimately require contributions from the seniority-two and seniority-four channels. Such contributions can be included post-hoc in a number of ways.\cite{kobayashi:2010,limacher:2014b,boguslawski:2015,johnson:2017,pernal:2018b} 

Finally, we reiterate that the optimal RG state for the H$_{10}$ isomers is the N\'{e}el state except for the 2D sheet and the 3D pyramid at short distances. There, the single particle energies $\{\varepsilon\}$ arrange into larger clusters that accommodate more than one pair. This behavior is not allowed by the geminal coefficients of GVB/APSG, nor AP1roG/pCCD. 

\begin{acknowledgement}

The authors thank Mario Piris and Alexandre Faribault for helpful discussions. P.A.J. was supported by NSERC and the Digital Research Alliance of Canada. A.E.D. acknowledges support from the National Science Foundation under Grant No. OAC-2103705.

\end{acknowledgement}

\bibliography{h10}

\end{document}


\pagebreak
\begin{figure}[ht!] 
	\includegraphics[width=0.475\textwidth]{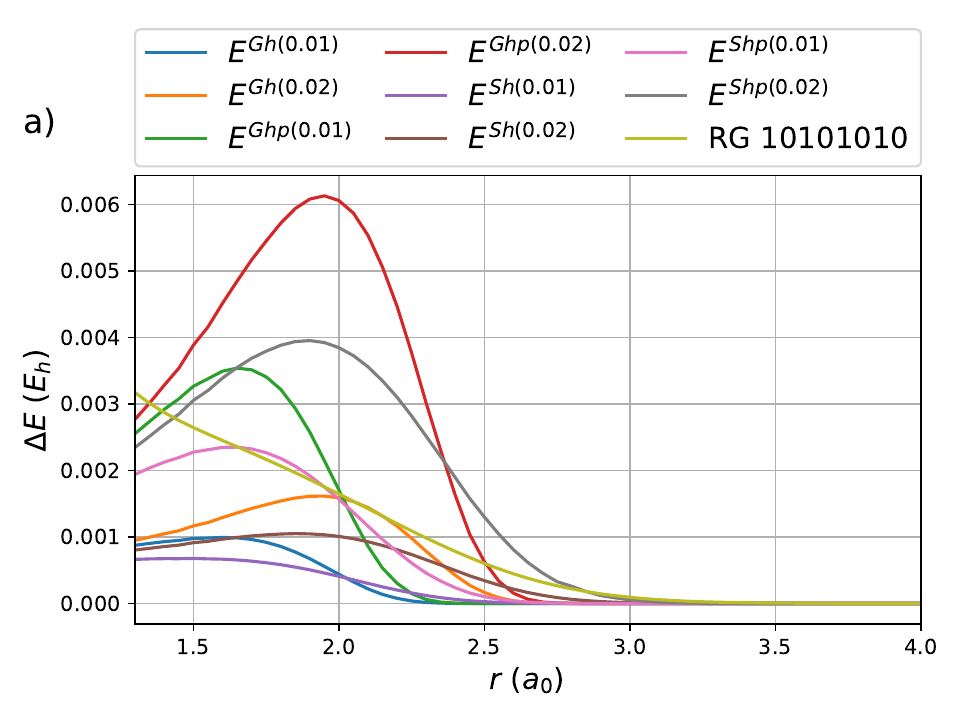} \hfill
	\includegraphics[width=0.475\textwidth]{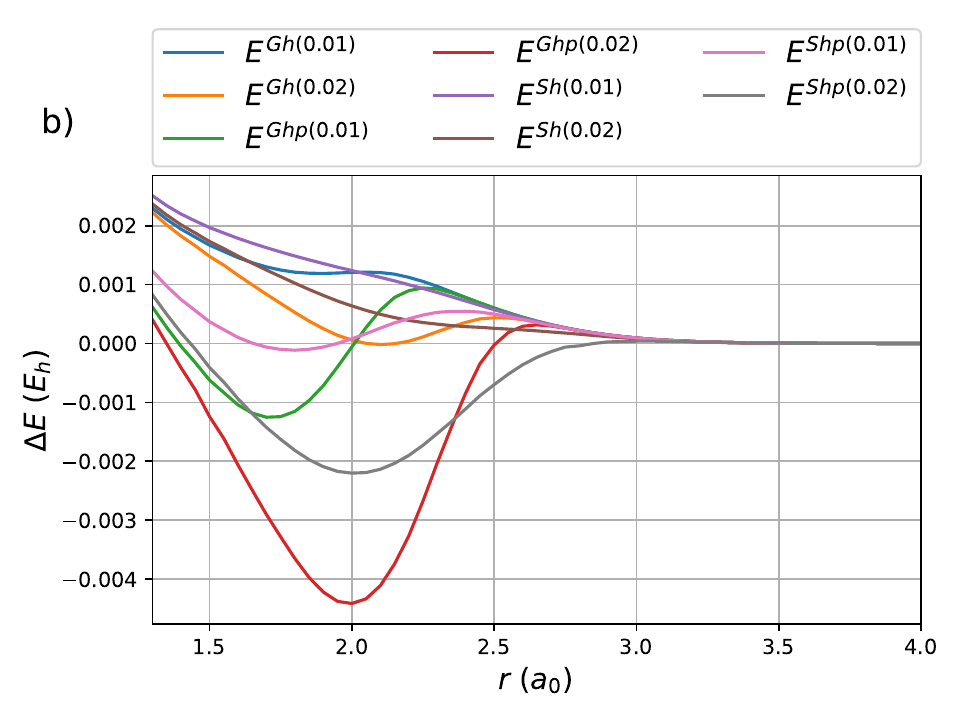} 
	\caption{Weak correlation functionals for H$_8$ chain: (a) $|E_{WC}|$ compared with $\Delta_{RG}$. (b) Difference between $\Delta_{RG}$ and $|E_{WC}|$. Results computed with the STO-6G basis in the OO-DOCI orbitals on top of the 10101010 RG state.} 
	\label{fig:H8_wc}
\end{figure}
\begin{figure}[ht!] 
	\includegraphics[width=0.475\textwidth]{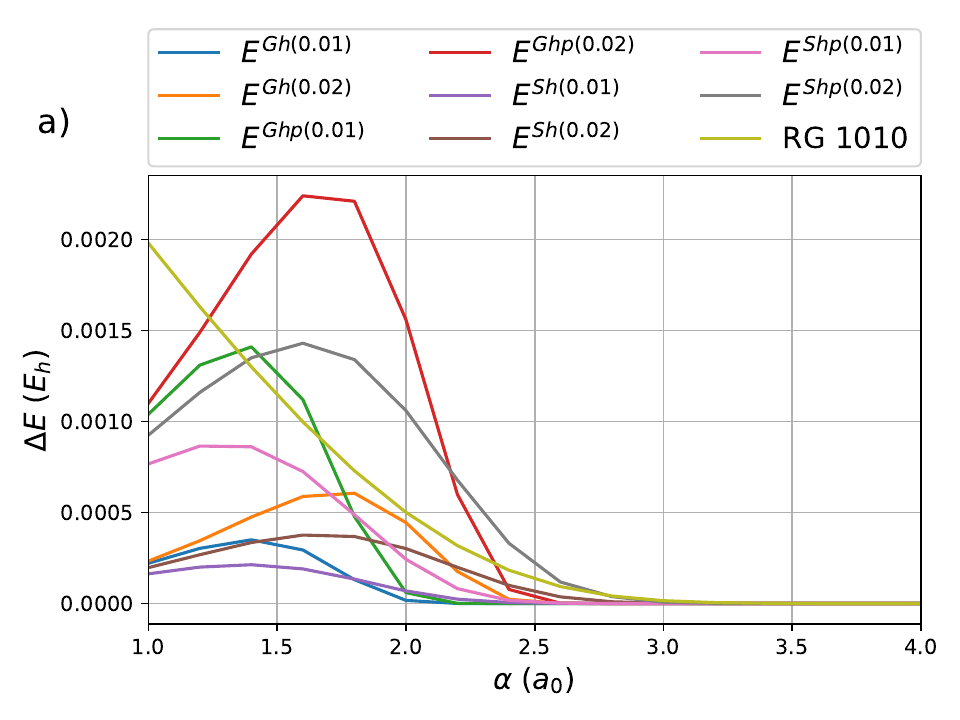} \hfill
	\includegraphics[width=0.475\textwidth]{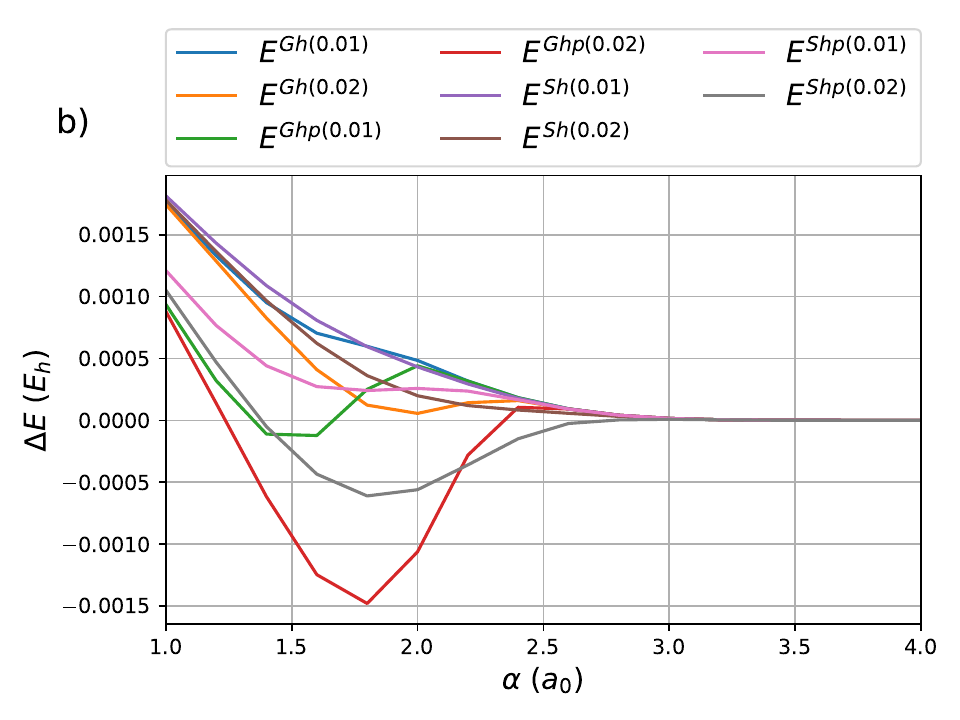} 
	\caption{Weak correlation functionals for Paldus S4: (a) $|E_{WC}|$ compared with $\Delta_{RG}$. (b) Difference between $\Delta_{RG}$ and $|E_{WC}|$. Results computed with the STO-6G basis in the OO-DOCI orbitals on top of the 1010 RG state.} 
	\label{fig:S4_wc}
\end{figure}
\begin{figure}[ht!] 
	\includegraphics[width=0.475\textwidth]{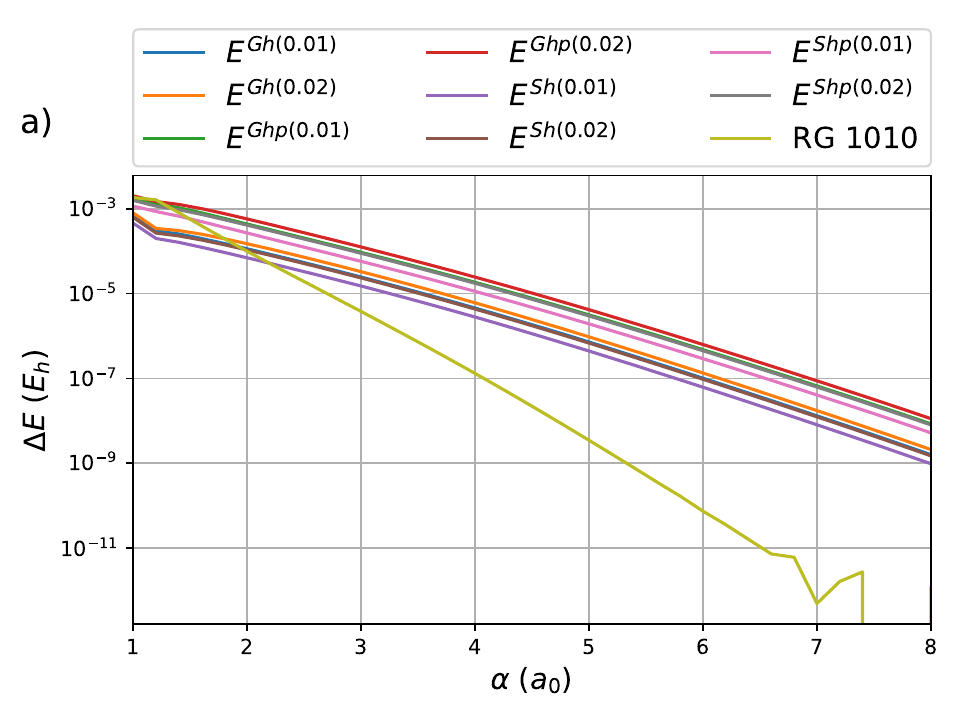} \hfill
	\includegraphics[width=0.475\textwidth]{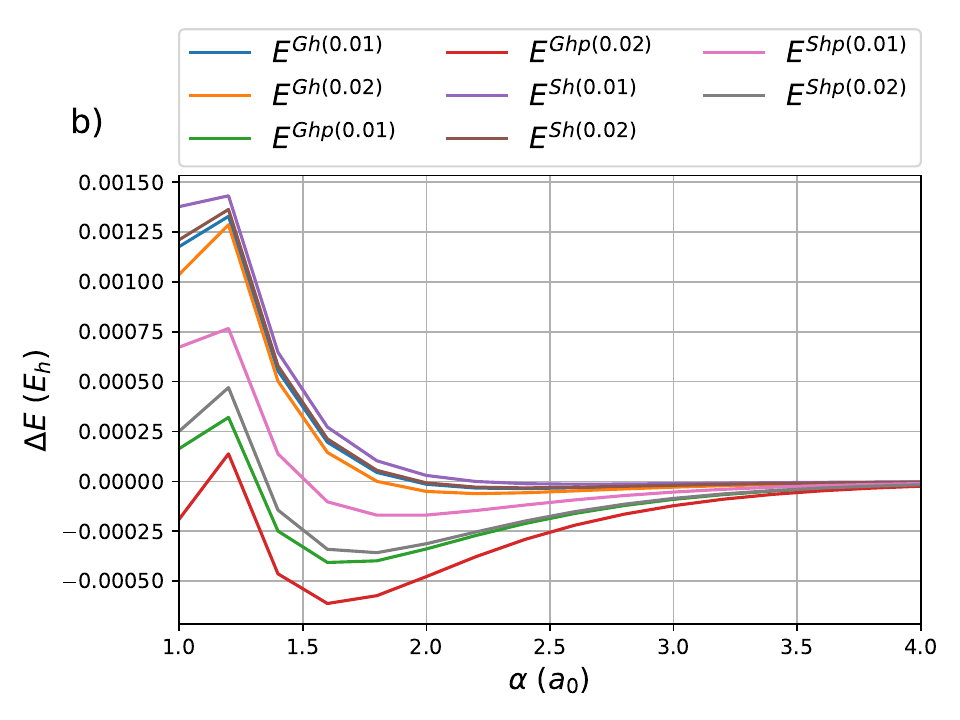} \\
	\includegraphics[width=0.475\textwidth]{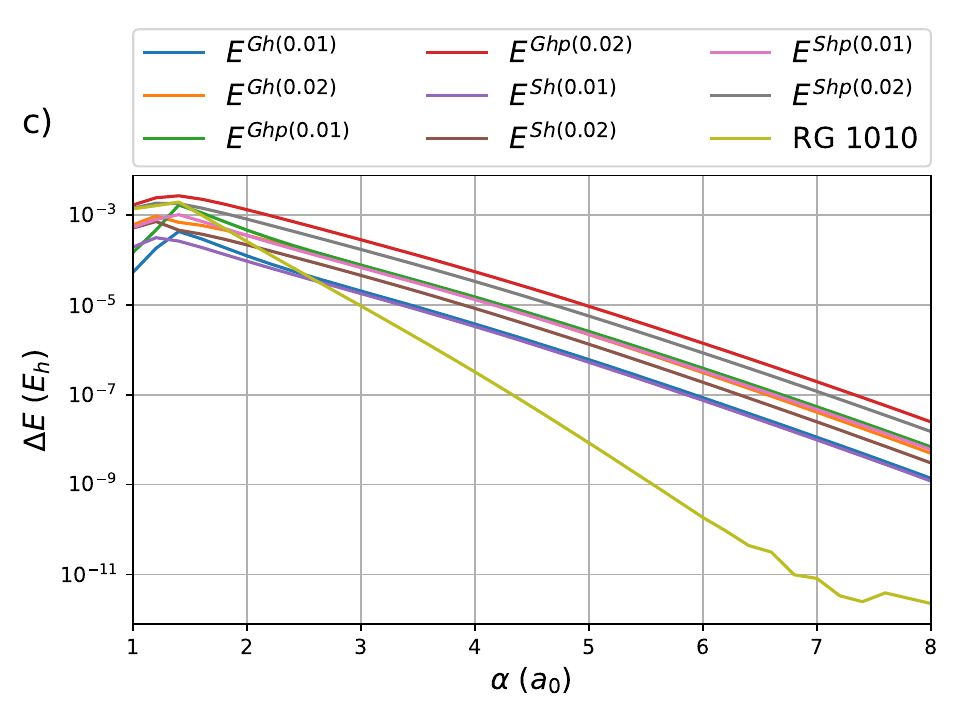} \hfill
	\includegraphics[width=0.475\textwidth]{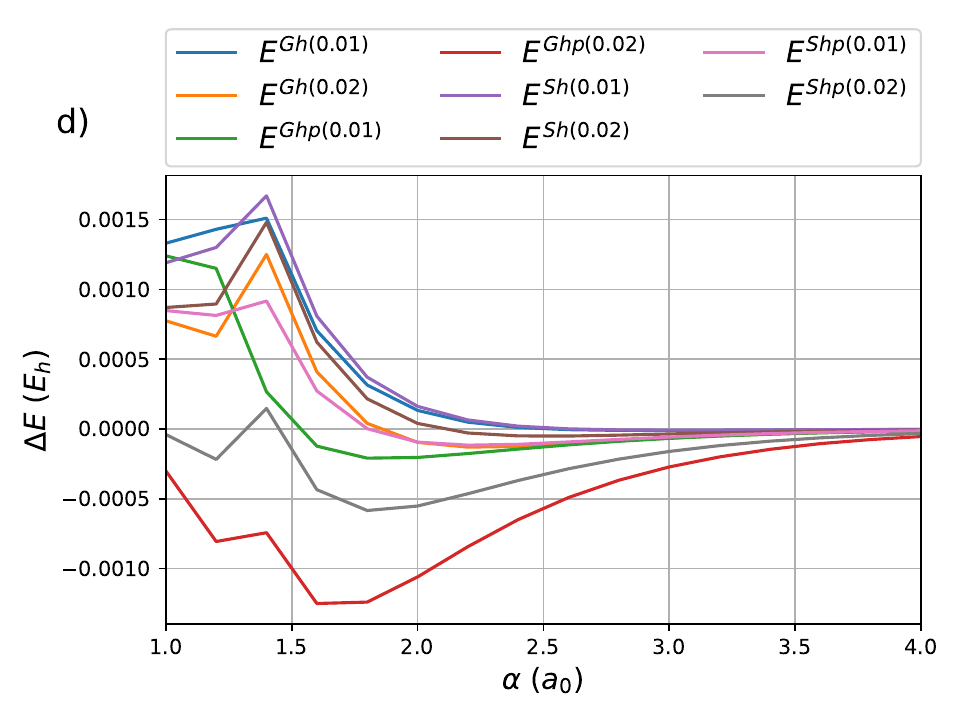} \\
	\includegraphics[width=0.475\textwidth]{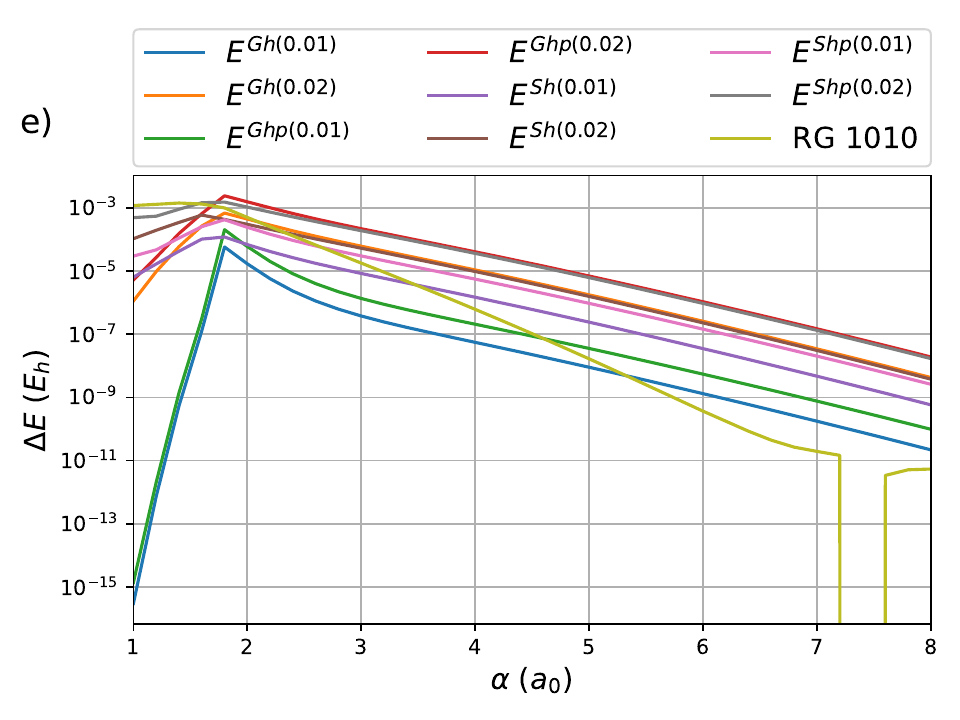} \hfill
	\includegraphics[width=0.475\textwidth]{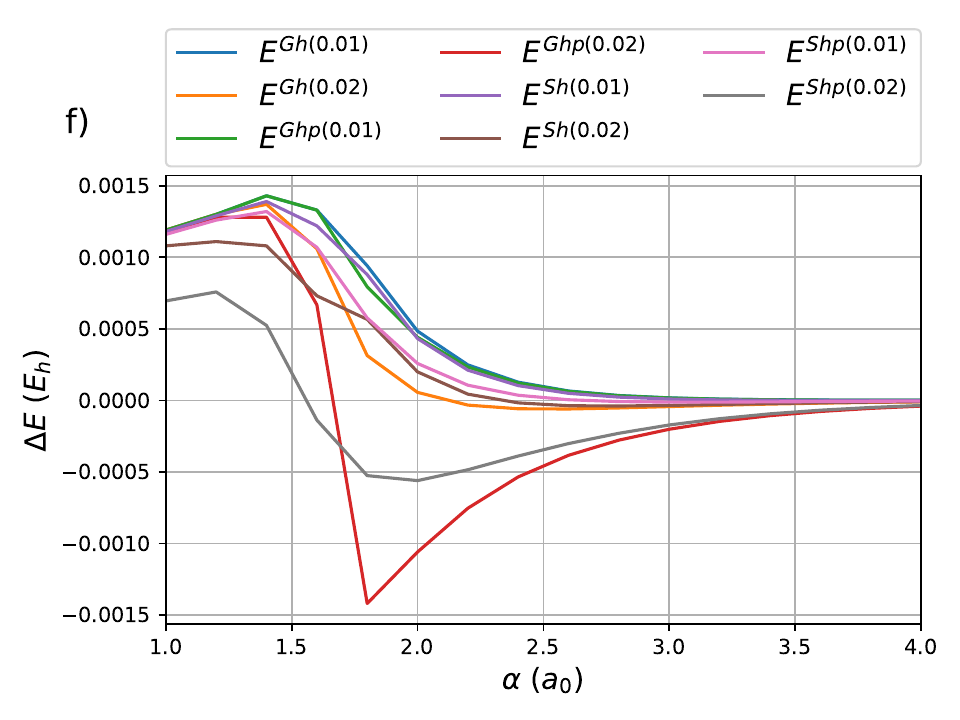}
	\caption{Weak correlation functionals for Paldus P4: (a,c,e) $|E_{WC}|$ compared with $\Delta_{RG}$ for (a) $a=1.2\;a_0$, (c) $a=1.6\;a_0$, and (e) $a=2.0\;a_0$. (b,d,f) Difference between $\Delta_{RG}$ and $|E_{WC}|$ for (b) $a=1.2\;a_0$, (d) $a=1.6\;a_0$, and (f) $a=2.0\;a_0$. Results computed with the STO-6G basis in the OO-DOCI orbitals on top of the 1010 RG state.} 
	\label{fig:P4_wc}
\end{figure}
\begin{figure}[ht!] 
	\includegraphics[width=0.475\textwidth]{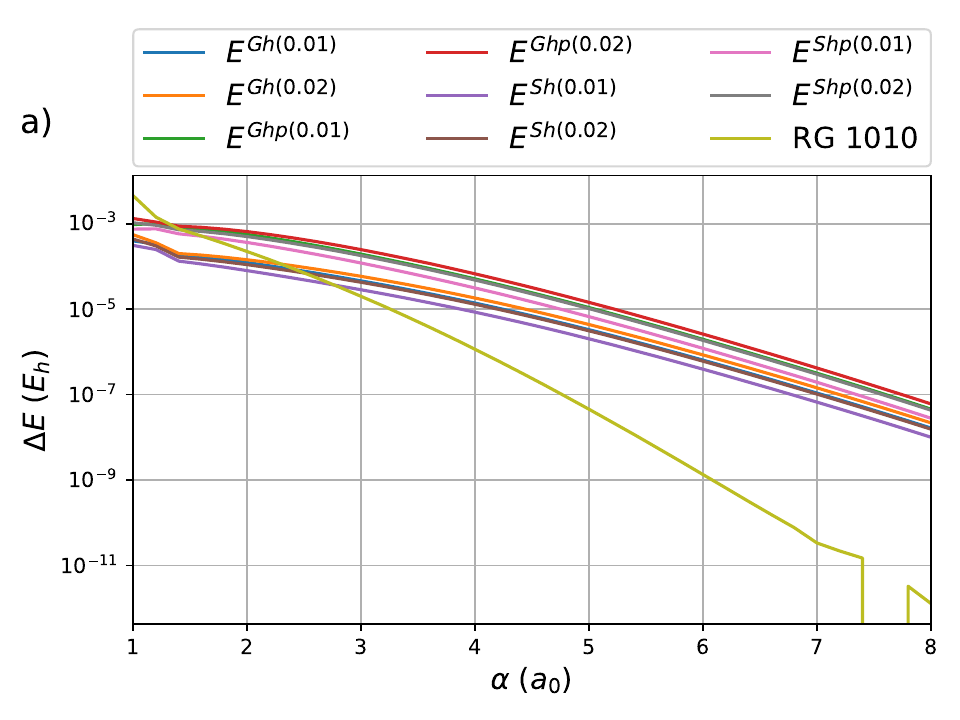} \hfill
	\includegraphics[width=0.475\textwidth]{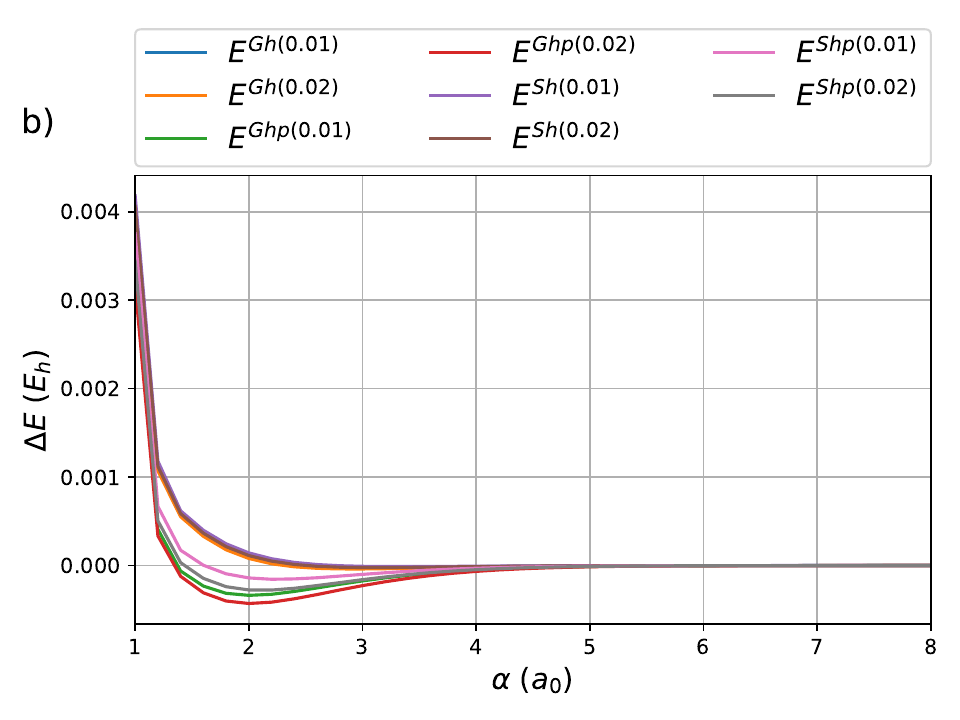} \\
	\includegraphics[width=0.475\textwidth]{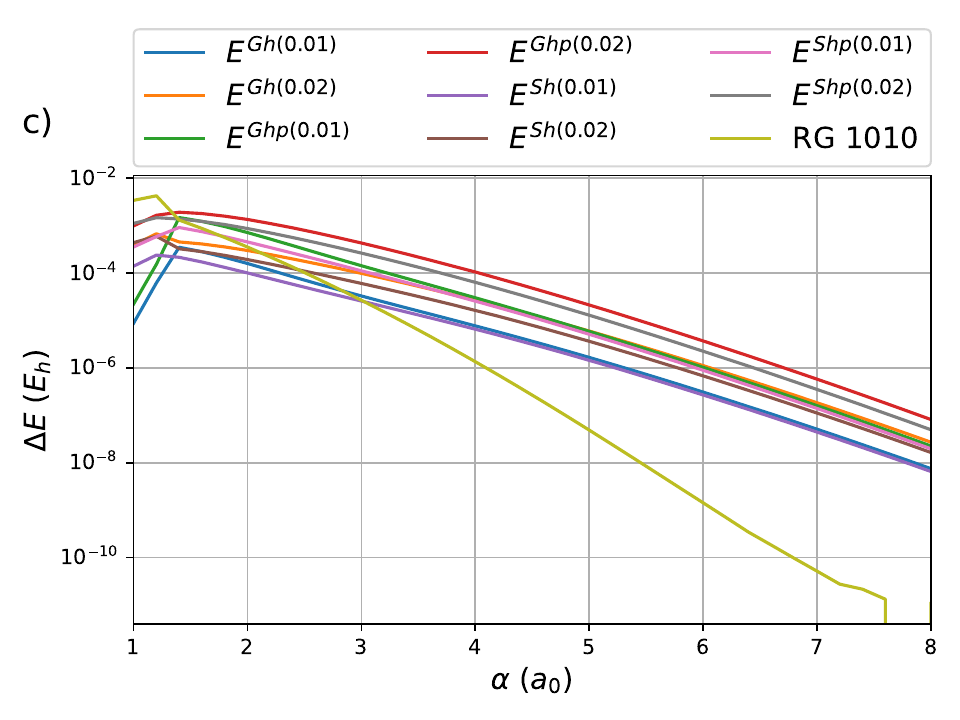} \hfill
	\includegraphics[width=0.475\textwidth]{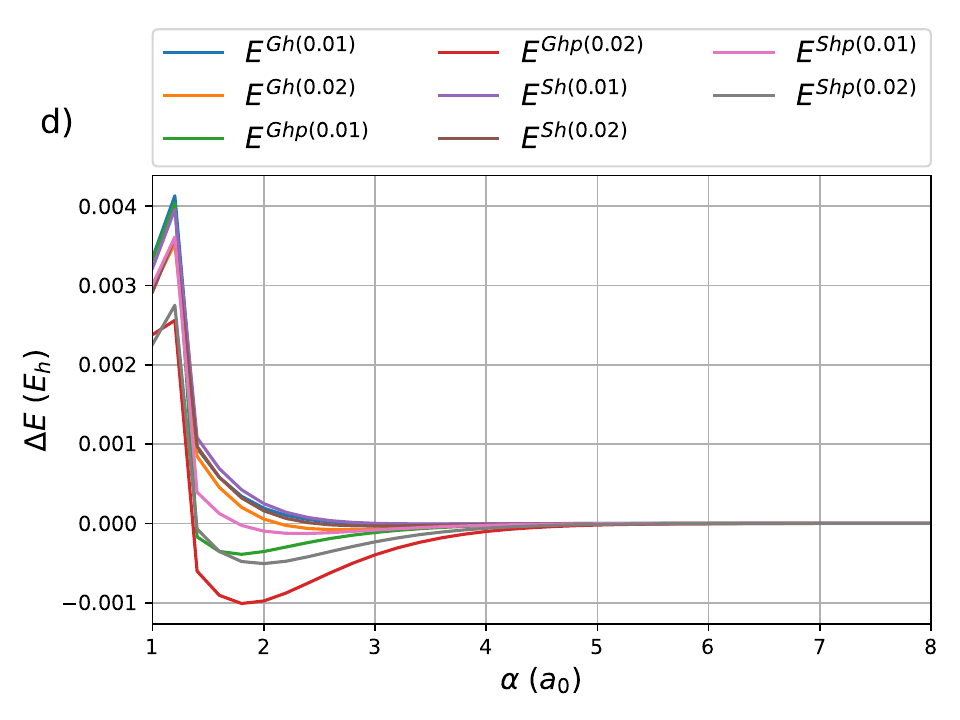} \\
	\includegraphics[width=0.475\textwidth]{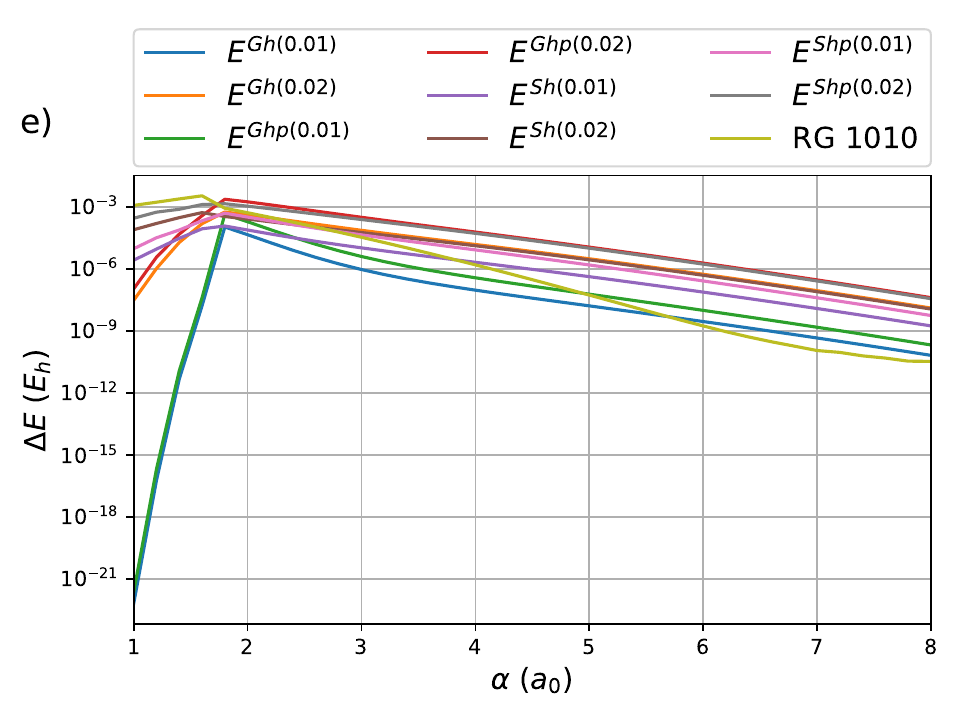} \hfill
	\includegraphics[width=0.475\textwidth]{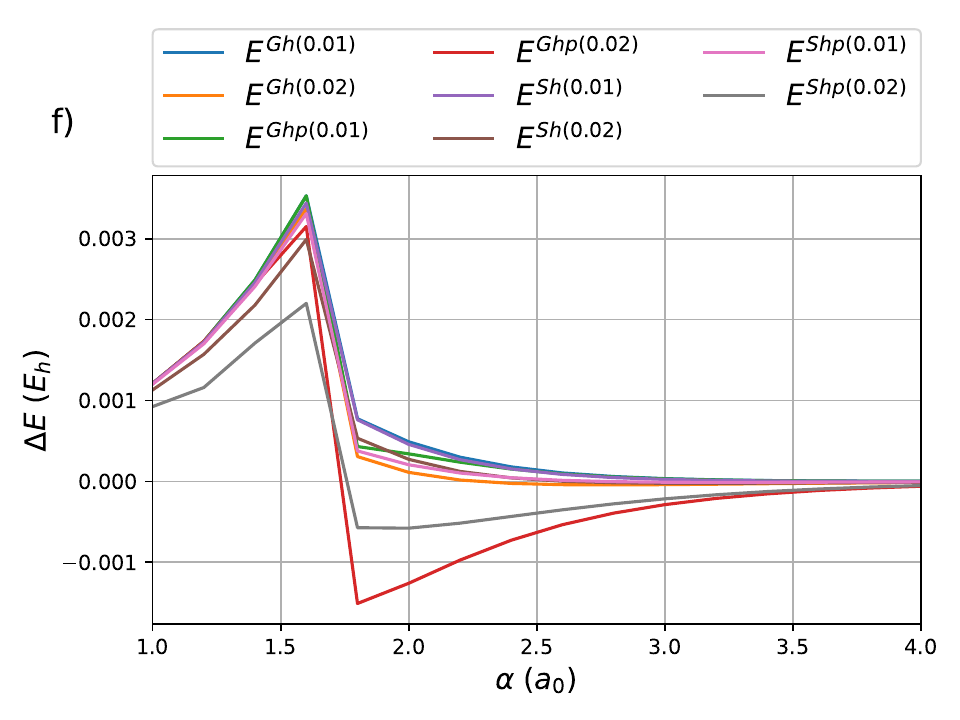}
	\caption{Weak correlation functionals for Paldus D4: (a,c,e) $|E_{WC}|$ compared with $\Delta_{RG}$ for (a) $a=1.2\;a_0$, (c) $a=1.6\;a_0$, and (e) $a=2.0\;a_0$. (b,d,f) Difference between $\Delta_{RG}$ and $|E_{WC}|$ for (b) $a=1.2\;a_0$, (d) $a=1.6\;a_0$, and (f) $a=2.0\;a_0$. Results computed with the STO-6G basis in the OO-DOCI orbitals on top of the 1010 RG state.} 
	\label{fig:D4_wc}
\end{figure}
\begin{figure}[ht!] 
	\includegraphics[width=0.475\textwidth]{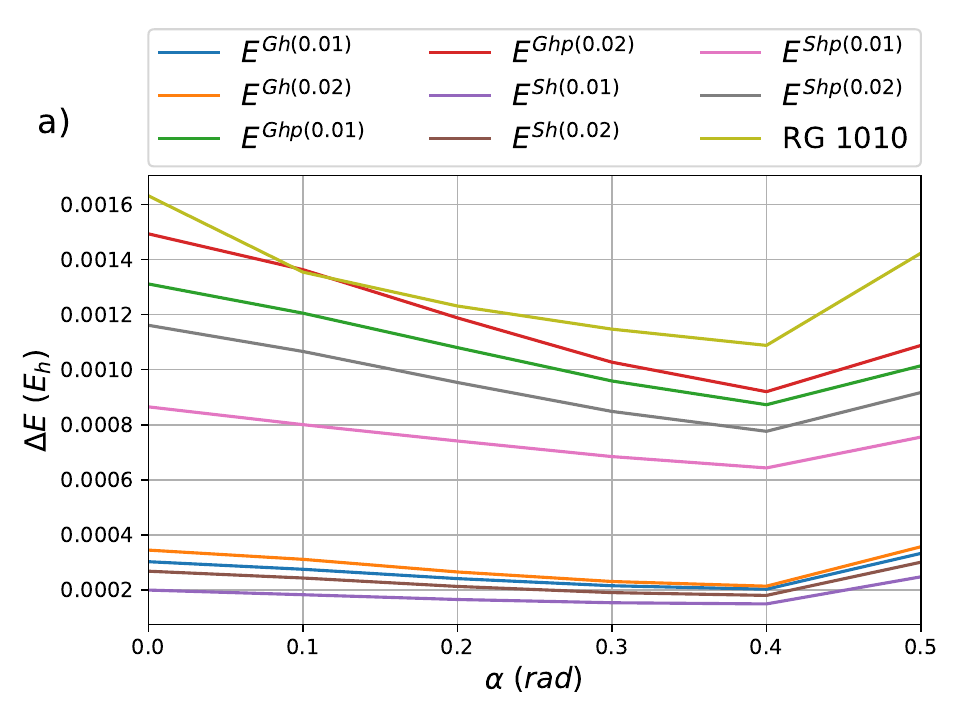} \hfill
	\includegraphics[width=0.475\textwidth]{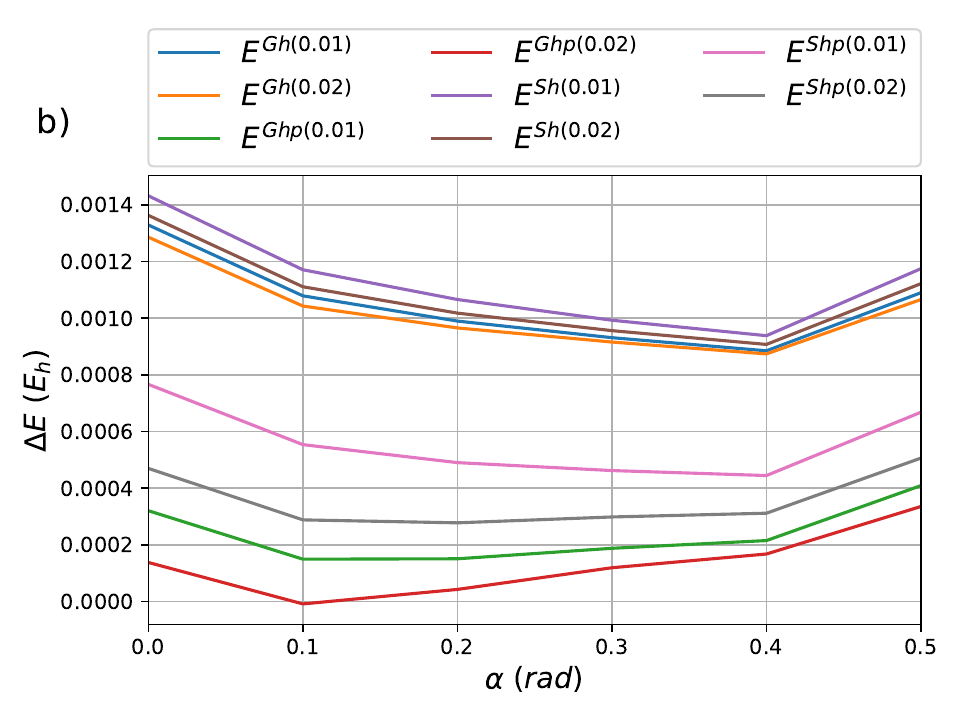} \\
	\includegraphics[width=0.475\textwidth]{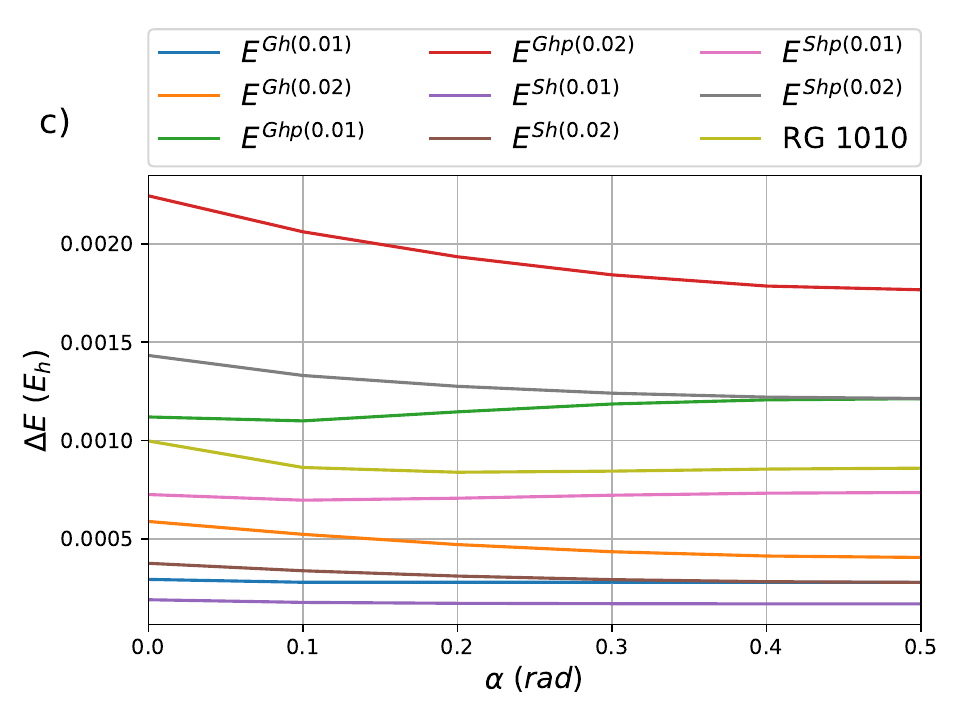} \hfill
	\includegraphics[width=0.475\textwidth]{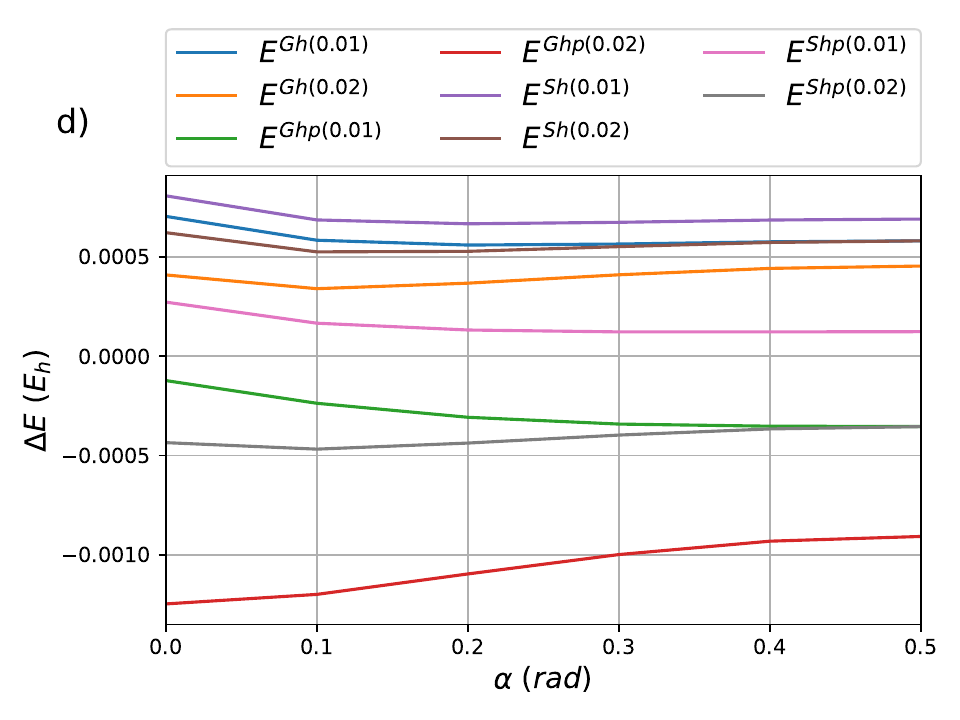} \\
	\includegraphics[width=0.475\textwidth]{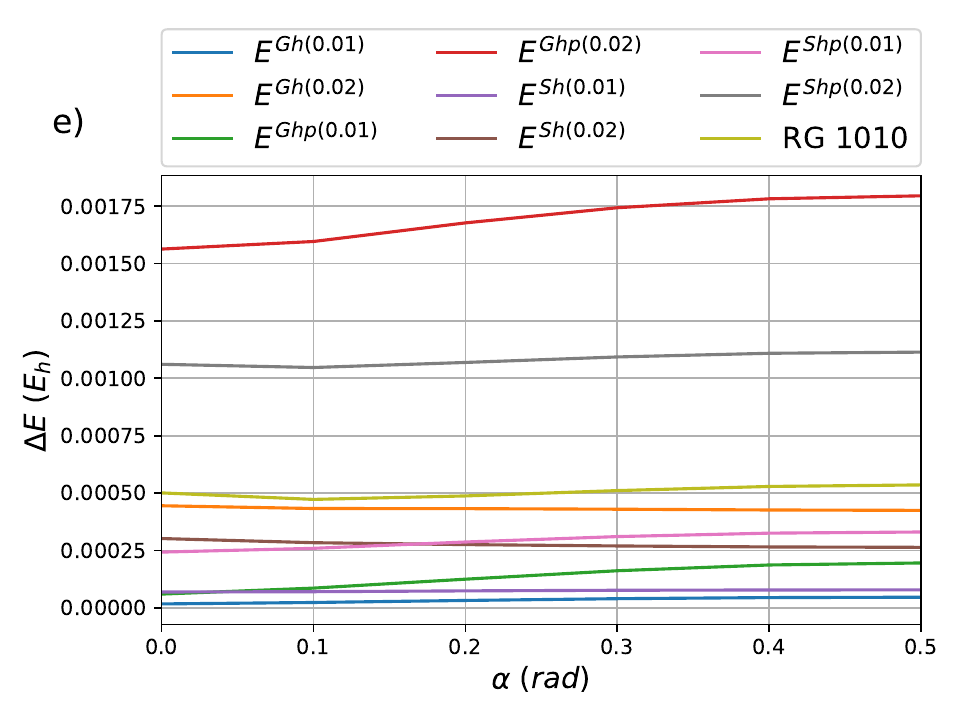} \hfill
	\includegraphics[width=0.475\textwidth]{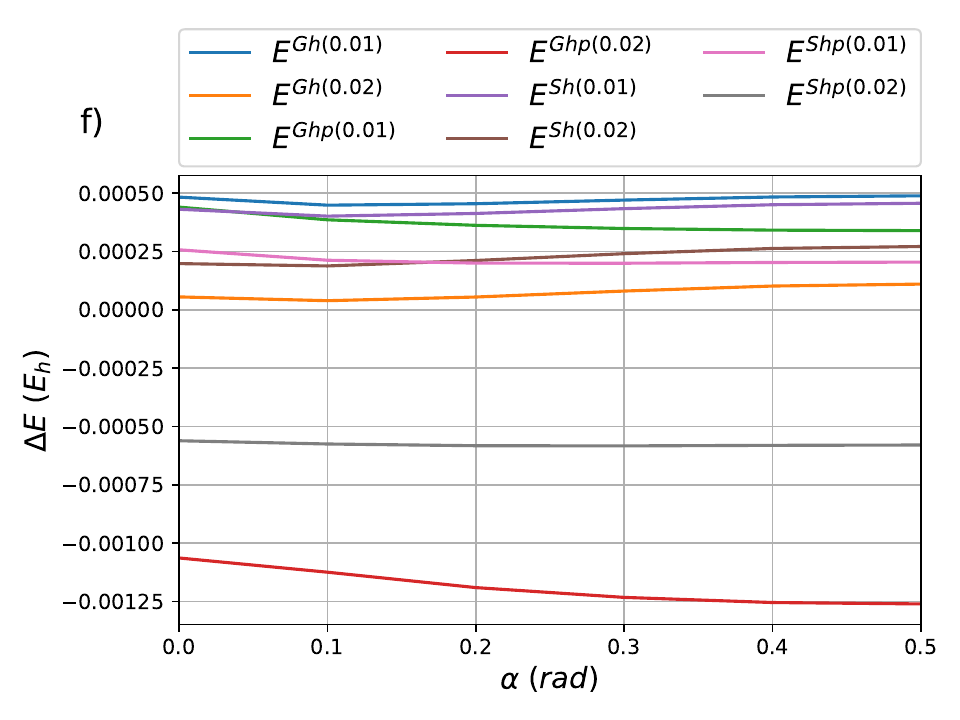}
	\caption{Weak correlation functionals for Paldus H4: (a,c,e) $|E_{WC}|$ compared with $\Delta_{RG}$ for (a) $a=1.2\;a_0$, (c) $a=1.6\;a_0$, and (e) $a=2.0\;a_0$. (b,d,f) Difference between $\Delta_{RG}$ and $|E_{WC}|$ for (b) $a=1.2\;a_0$, (d) $a=1.6\;a_0$, and (f) $a=2.0\;a_0$. Results computed with the STO-6G basis in the OO-DOCI orbitals on top of the 1010 RG state.} 
	\label{fig:H4_wc}
\end{figure}
\begin{figure}[ht!] 
	\includegraphics[width=0.475\textwidth]{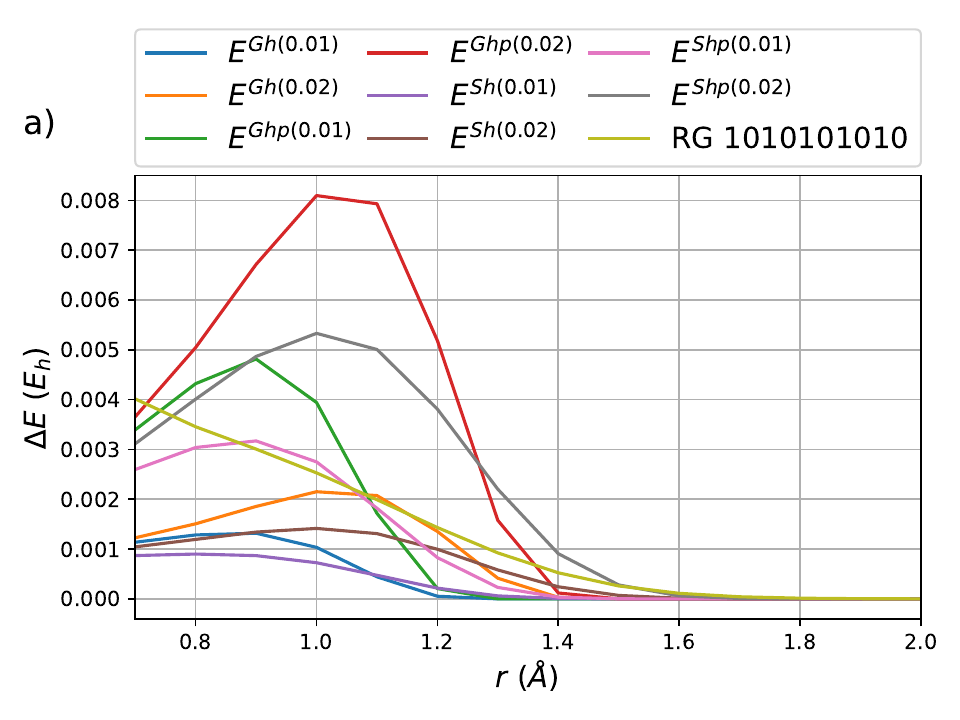} \hfill
	\includegraphics[width=0.475\textwidth]{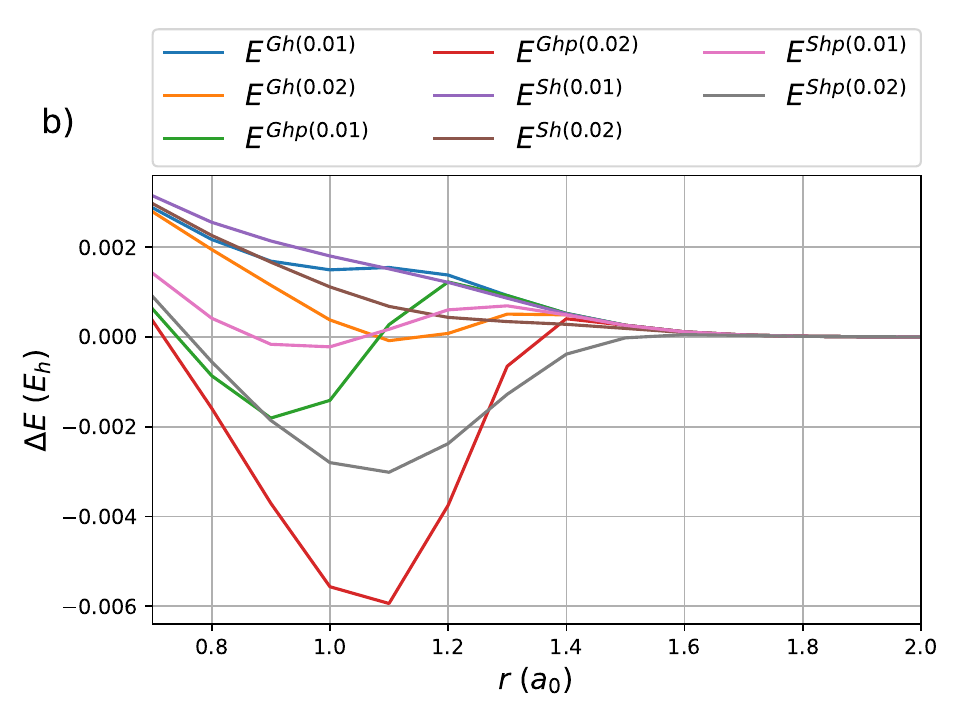} 
	\caption{Weak correlation functionals for H$_{10}$ chain: (a) $|E_{WC}|$ compared with $\Delta_{RG}$. (b) Difference between $\Delta_{RG}$ and $|E_{WC}|$. Results computed with the STO-6G basis in the OO-DOCI orbitals on top of the 1010101010 RG state.} 
	\label{fig:H10_chain_wc}
\end{figure}
\begin{figure}[ht!] 
	\includegraphics[width=0.475\textwidth]{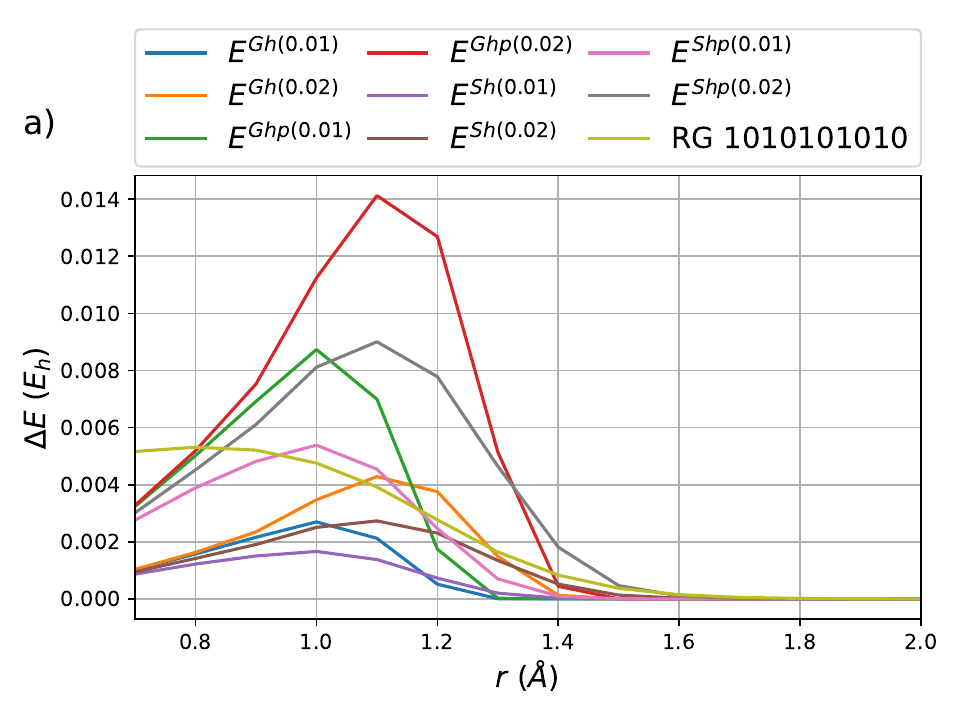} \hfill
	\includegraphics[width=0.475\textwidth]{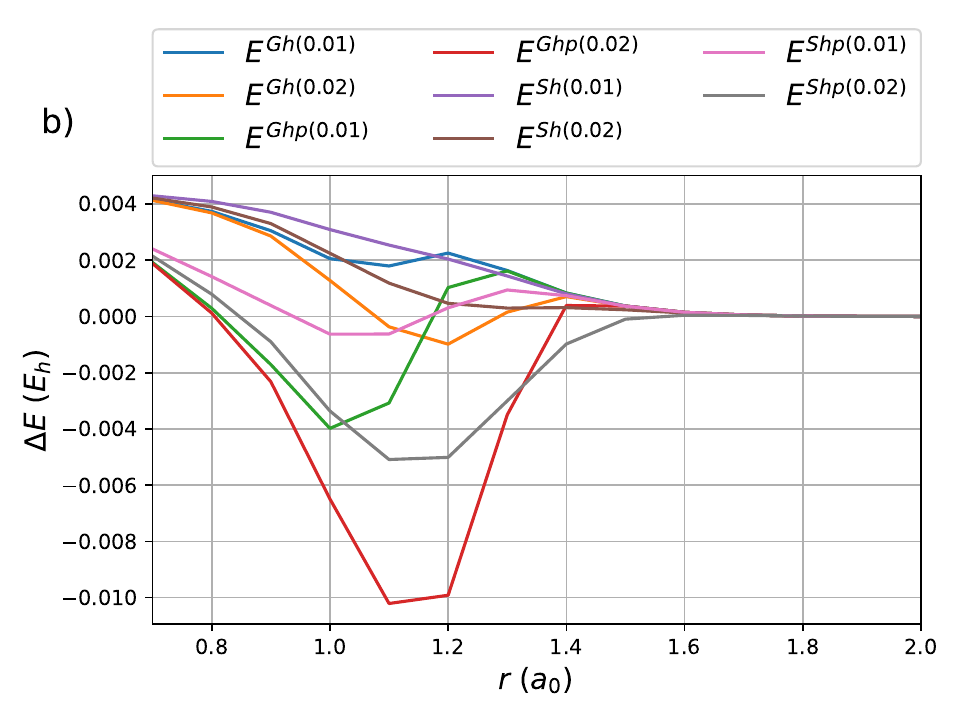} 
	\caption{Weak correlation functionals for H$_{10}$ ring: (a) $|E_{WC}|$ compared with $\Delta_{RG}$. (b) Difference between $\Delta_{RG}$ and $|E_{WC}|$. Results computed with the STO-6G basis in the OO-DOCI orbitals on top of the 1010101010 RG state.} 
	\label{fig:H10_ring_wc}
\end{figure}
\begin{figure}[ht!] 
	\includegraphics[width=0.475\textwidth]{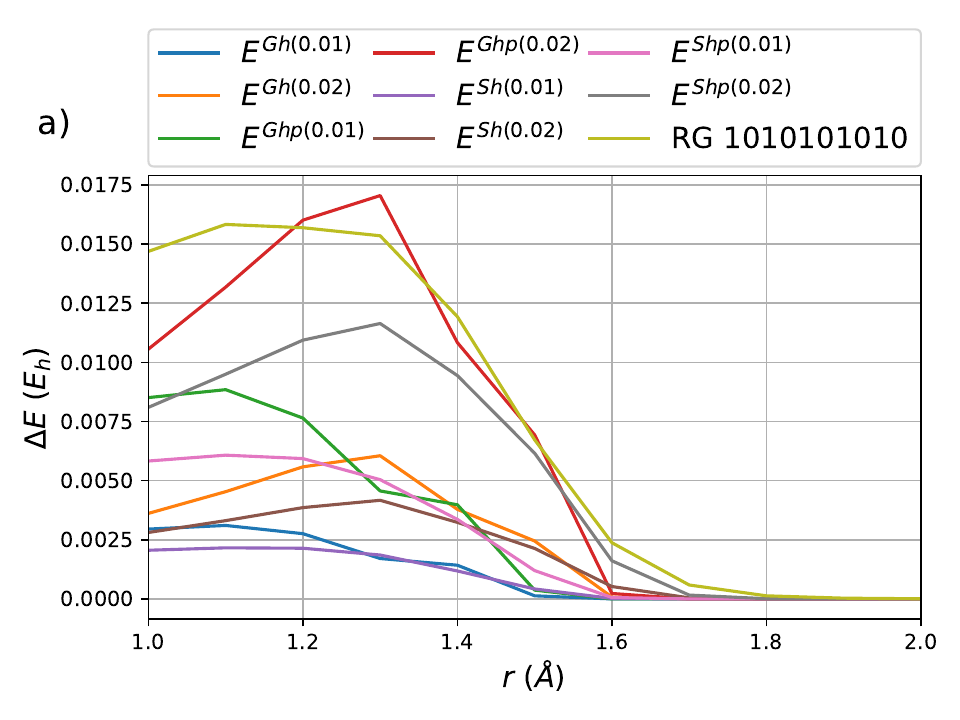} \hfill
	\includegraphics[width=0.475\textwidth]{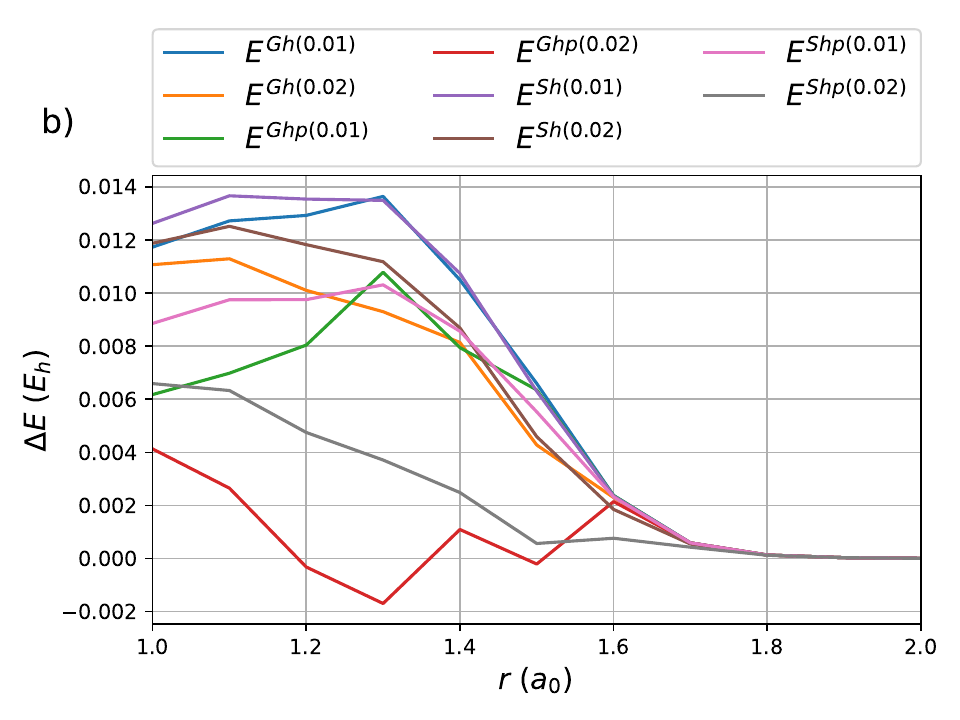} 
	\caption{Weak correlation functionals for H$_{10}$ sheet: (a) $|E_{WC}|$ compared with $\Delta_{RG}$. (b) Difference between $\Delta_{RG}$ and $|E_{WC}|$. Results computed with the STO-6G basis in the OO-DOCI orbitals on top of the 1010101010 RG state.} 
	\label{fig:H10_sheet_wc}
\end{figure}
\begin{figure}[ht!] 
	\includegraphics[width=0.475\textwidth]{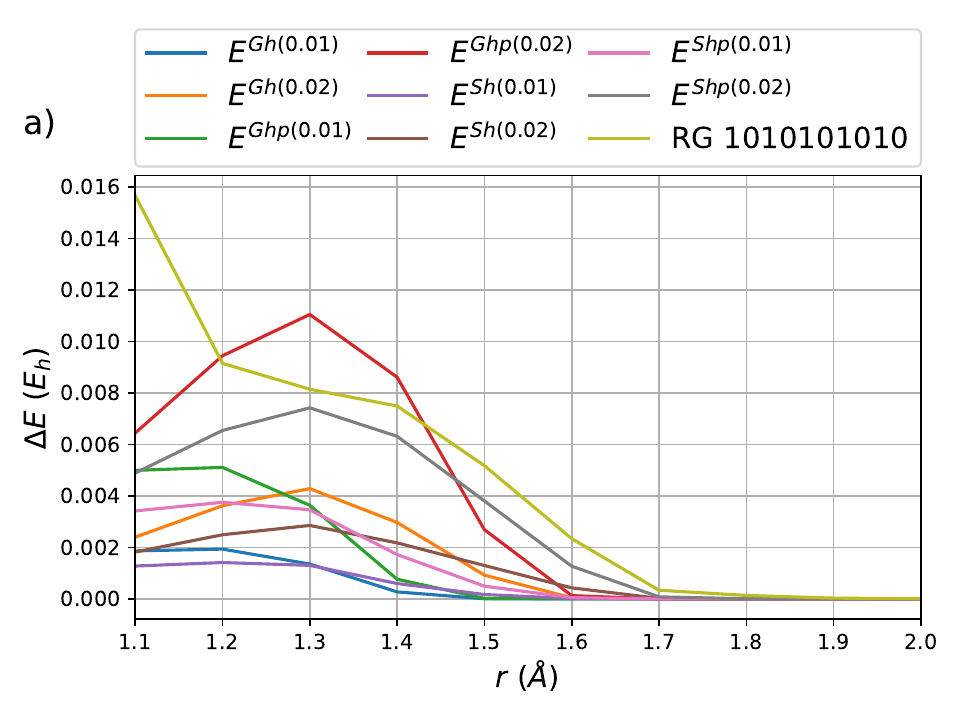} \hfill
	\includegraphics[width=0.475\textwidth]{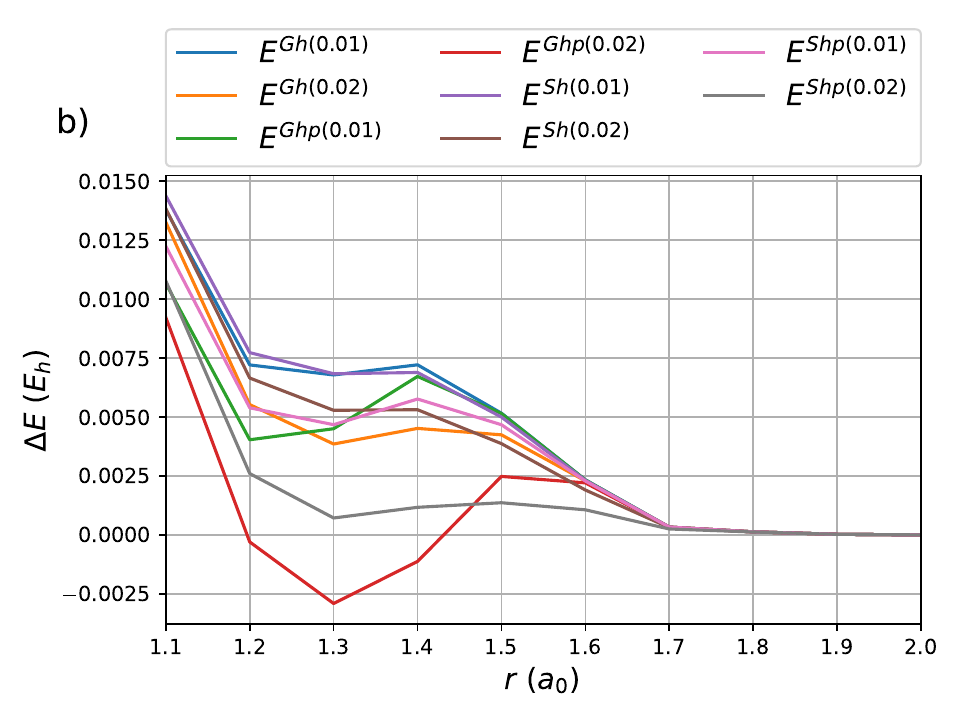} 
	\caption{Weak correlation functionals for H$_{10}$ pyramid: (a) $|E_{WC}|$ compared with $\Delta_{RG}$. (b) Difference between $\Delta_{RG}$ and $|E_{WC}|$. Results computed with the STO-6G basis in the OO-DOCI orbitals on top of the 1010101010 RG state.} 
	\label{fig:H10_pyramid_wc}
\end{figure}

\begin{figure}[ht!] 
	\includegraphics[width=0.475\textwidth]{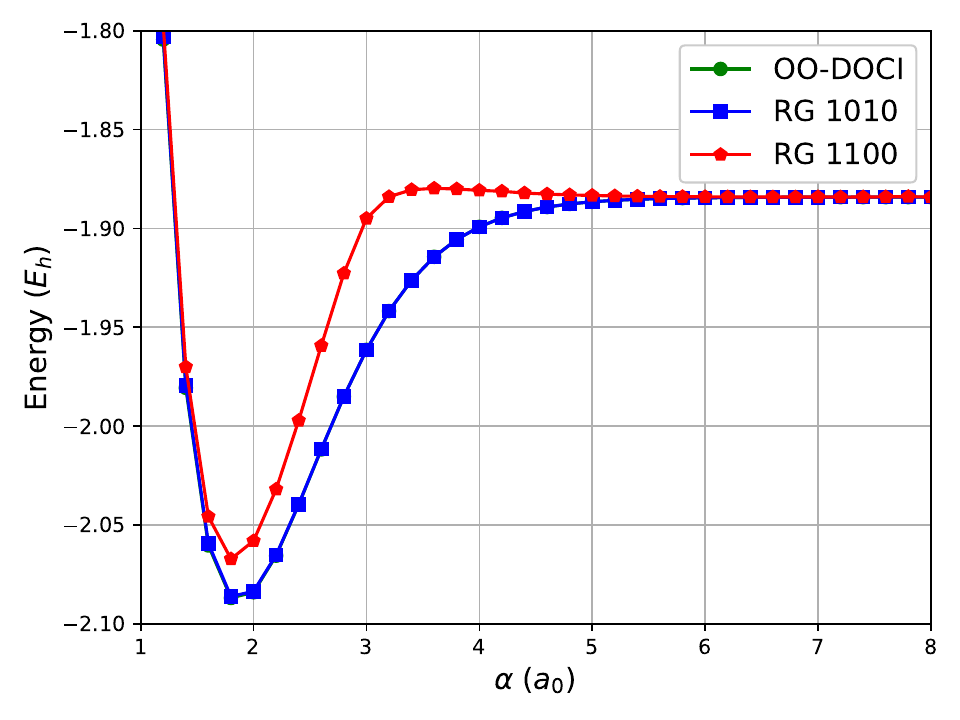} \hfill
	\caption{Variational RG treatment of Paldus S4. OO-DOCI and variational 1010 results are superimposed. Results computed with the STO-6G basis in the OO-DOCI orbitals.} 
	\label{fig:S4_1100_raw}
\end{figure}
\begin{figure}[ht!] 
	\includegraphics[width=0.475\textwidth]{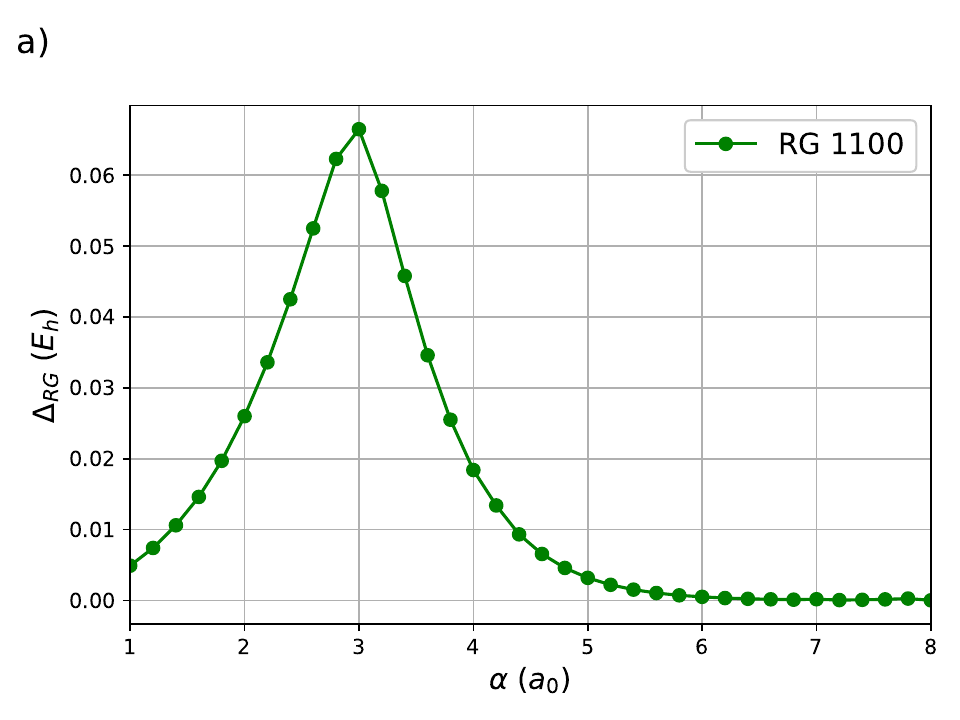} \hfill
	\includegraphics[width=0.475\textwidth]{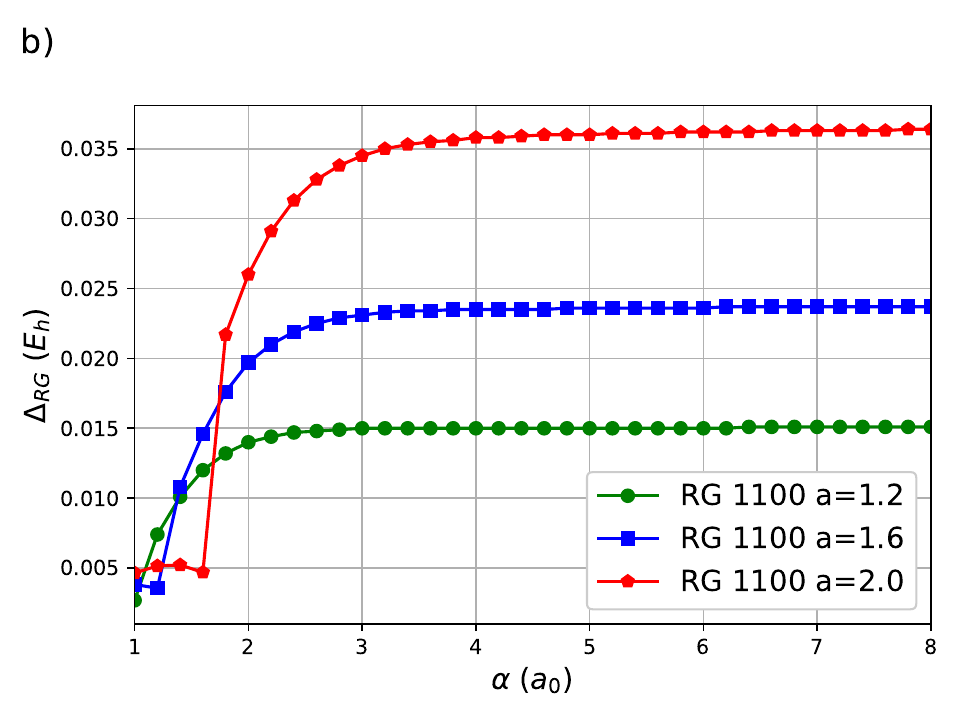} \\
	\includegraphics[width=0.475\textwidth]{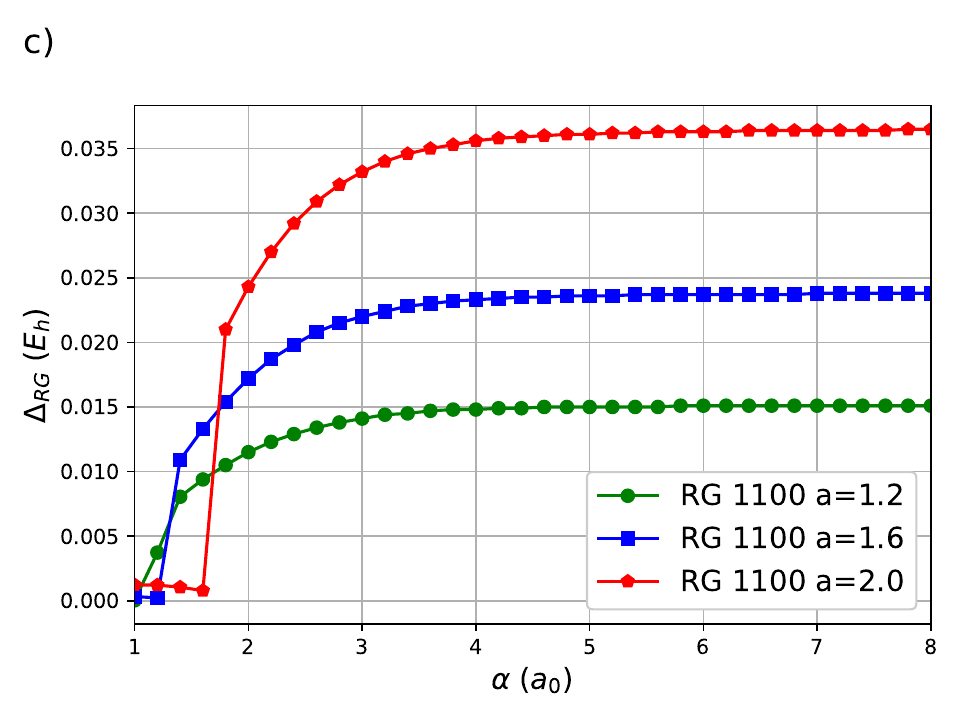} \hfill
	\includegraphics[width=0.475\textwidth]{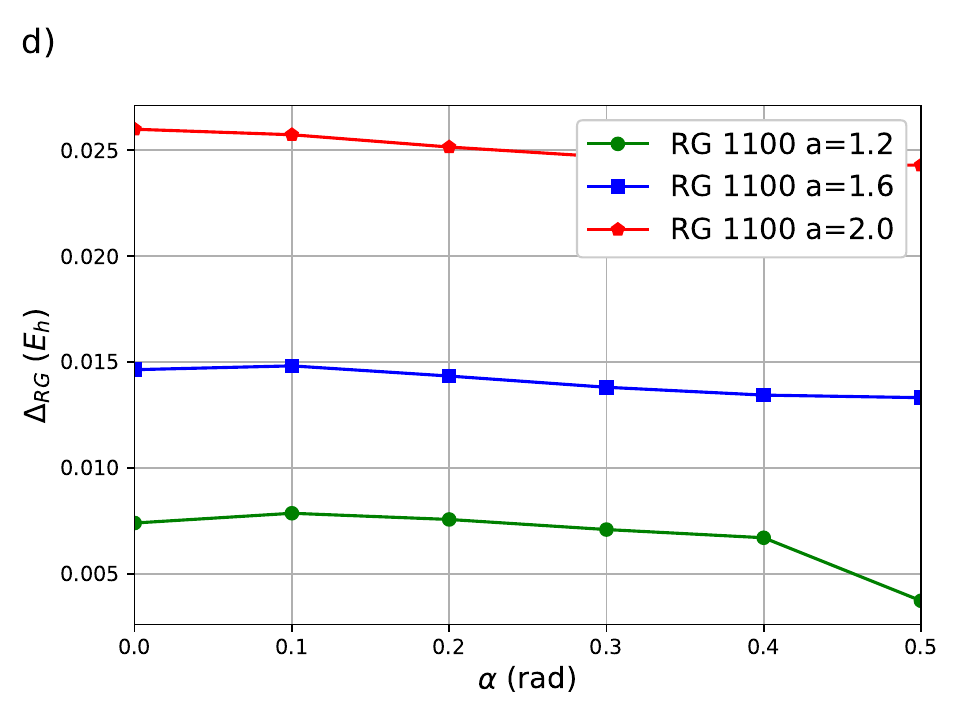} 
	\caption{Variational RG treatment of Paldus systems with a 1100 RG state: $\Delta_{RG}$ for (a) S4, (b) P4, (c) D4, (d) H4. Results computed with the STO-6G basis in the OO-DOCI orbitals.} 
	\label{fig:H4_1100_diff}
\end{figure}

\begin{figure}[ht!] 
	\includegraphics[width=0.475\textwidth]{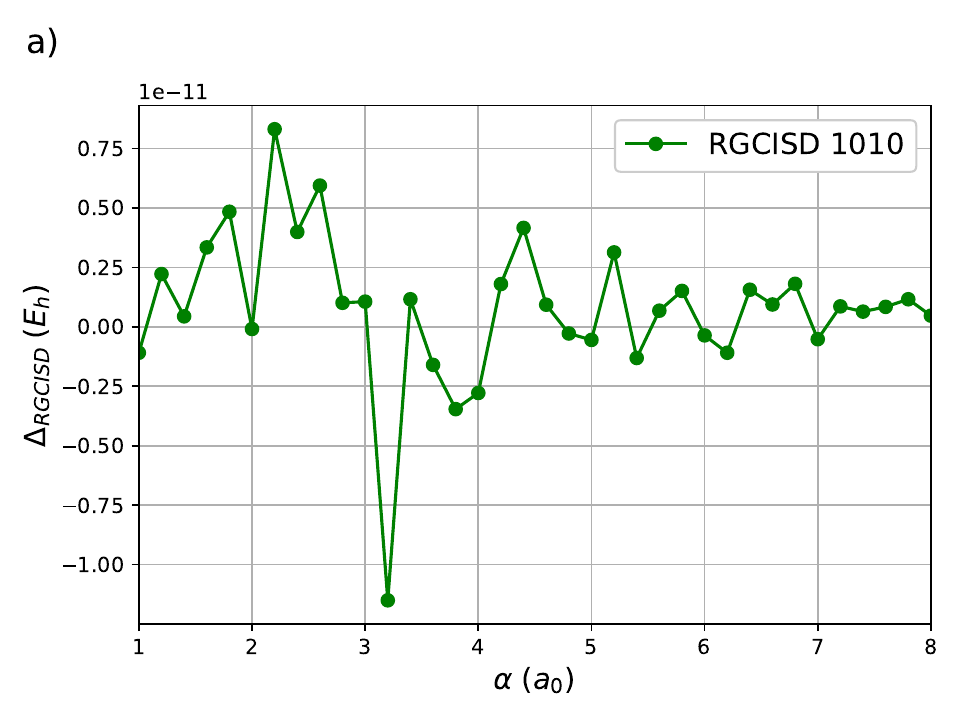} \hfill
	\includegraphics[width=0.475\textwidth]{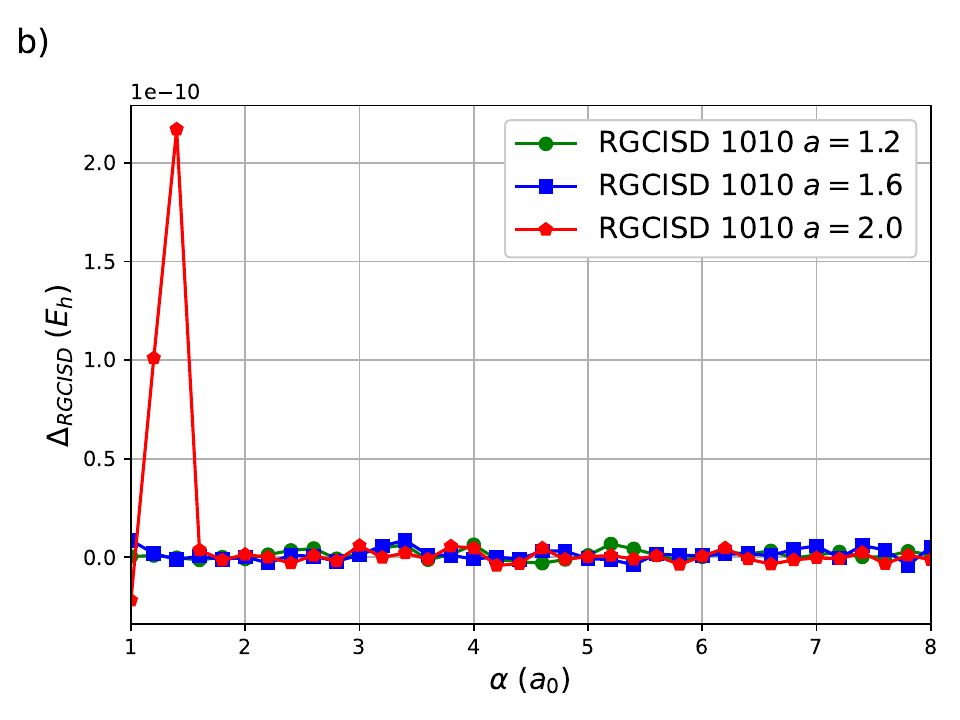} \\
	\includegraphics[width=0.475\textwidth]{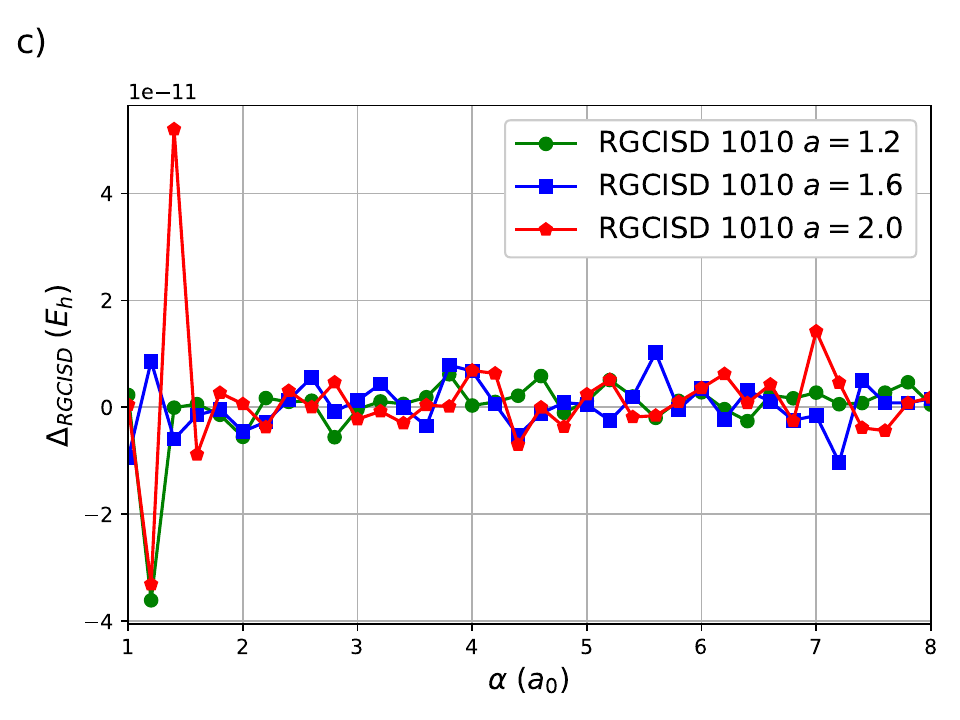} \hfill
	\includegraphics[width=0.475\textwidth]{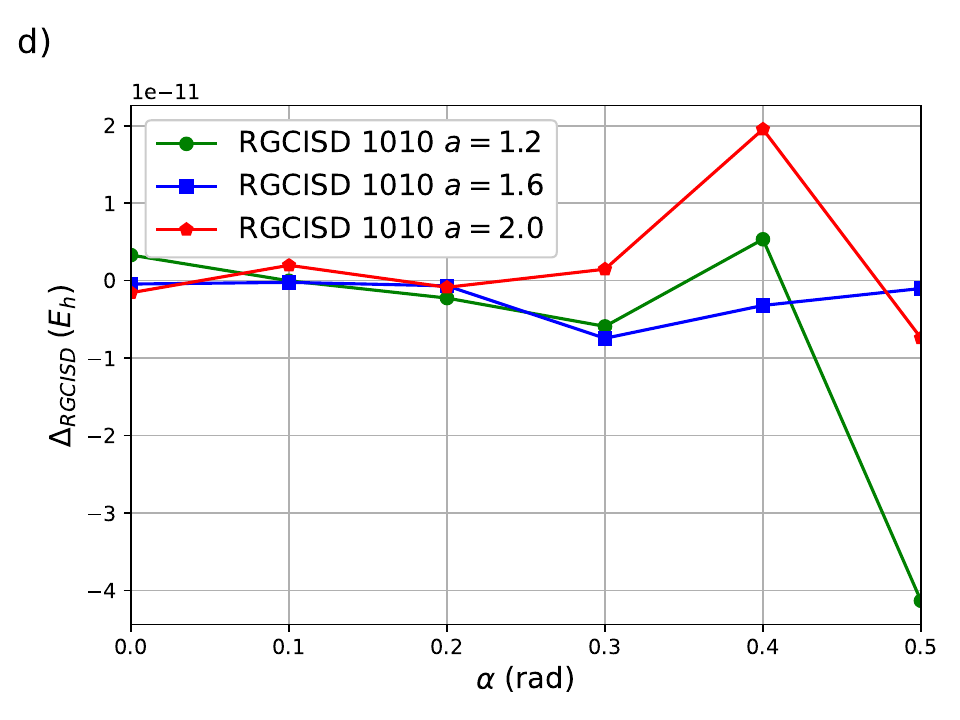} 
	\caption{RGCISD built from 1010 RG state for Paldus H$_{4}$ isomers: (a) S4, (b) P4, (c) D4, (d) H4. Results computed with the STO-6G basis in the OO-DOCI orbitals.} 
	\label{fig:H4_rgcisd}
\end{figure}

\begin{figure}[ht!] 
	\includegraphics[width=0.475\textwidth]{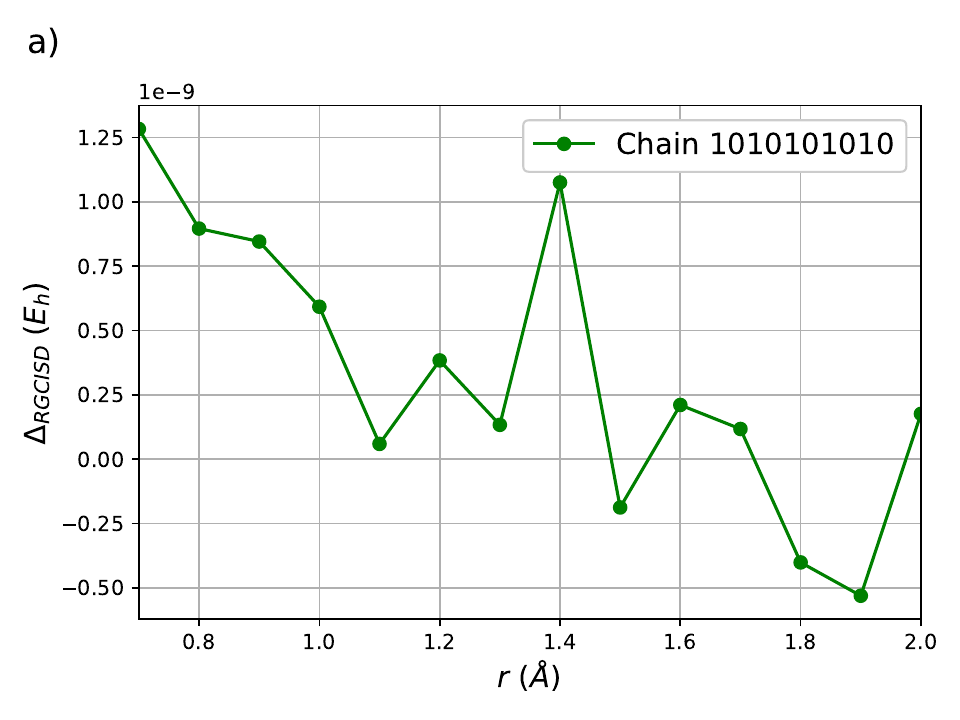} \hfill
	\includegraphics[width=0.475\textwidth]{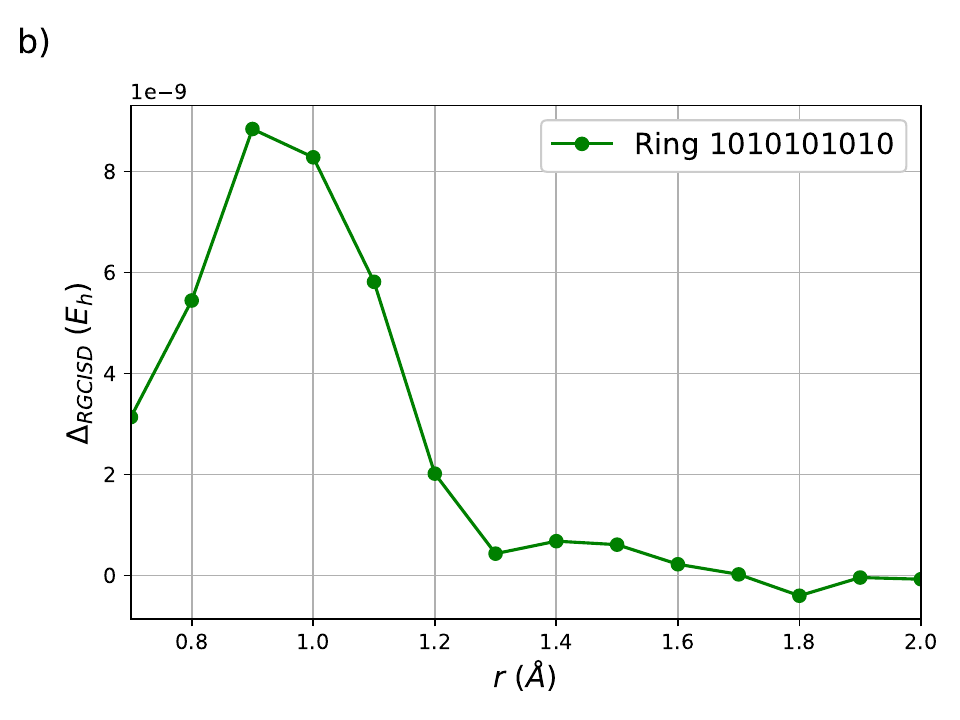} \\
	\includegraphics[width=0.475\textwidth]{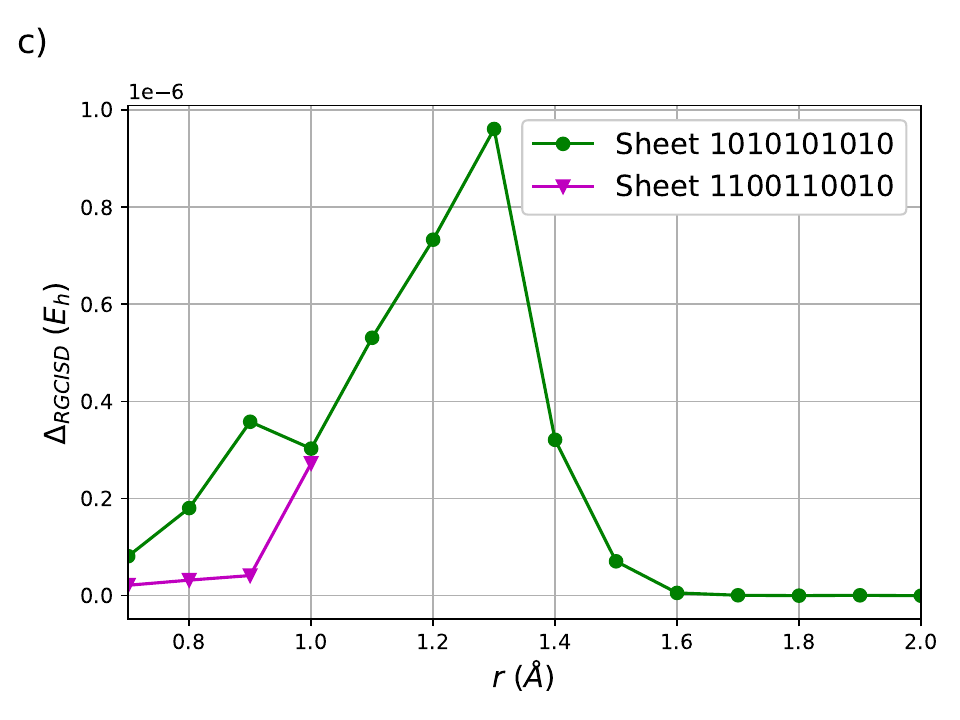} \hfill
	\includegraphics[width=0.475\textwidth]{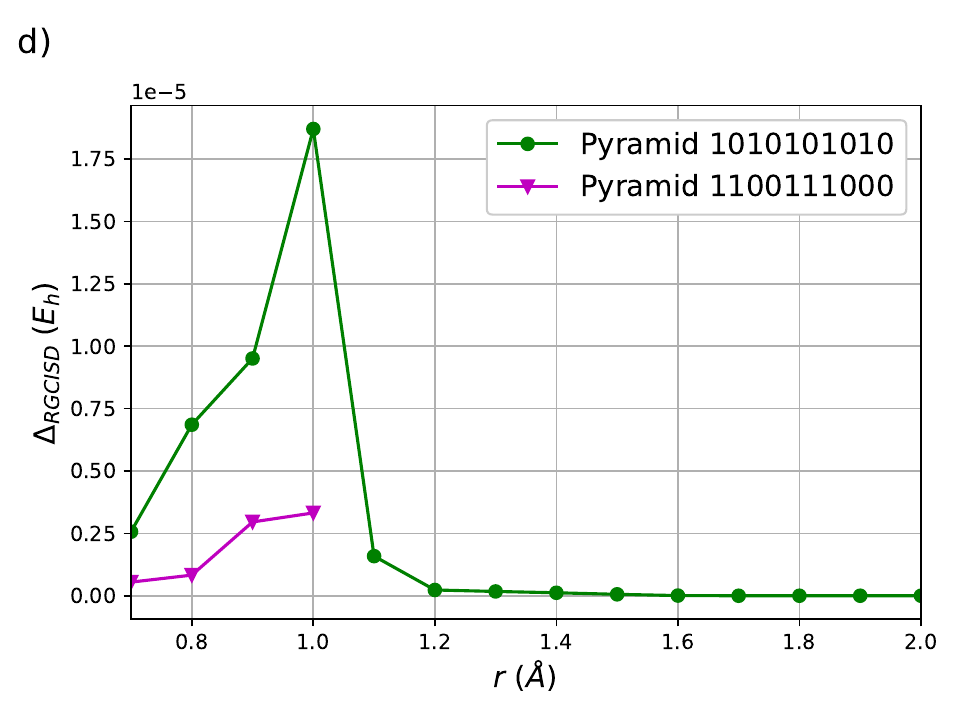} 
	\caption{RGCISD treatment of H$_{10}$ isomers: (a) 1D chain $\Delta_{RGCISD}$ of 1010101010 RG state. (b) 1d ring $\Delta_{RGCISD}$ of 1010101010 RG state. (c) 2D sheet $\Delta_{RGCISD}$ of 1010101010 and 1100110010 RG states. (d) 3D pyramid $\Delta_{RGCISD}$ of 1010101010 and 1100111000 RG states. Results computed with the STO-6G basis in the OO-DOCI orbitals.} 
	\label{fig:H10_rgcisd}
\end{figure}